\documentclass{article}
\usepackage{amsfonts}
\usepackage{amsmath}
\usepackage{graphicx}
\usepackage{amssymb}

\newtheorem{theorem}{Theorem}[section]

\newtheorem{claim}[theorem]{Claim}

\newtheorem{corollary}[theorem]{Corollary}

\newtheorem{definition}[theorem]{Definition}

\newtheorem{lemma}[theorem]{Lemma}

\newtheorem{problem}[theorem]{Problem}
\newtheorem{proposition}[theorem]{Proposition}
\newtheorem{remark}[theorem]{Remark}

\newenvironment{proof}[1][Proof]{\noindent\textbf{#1.} }{\ \rule{0.5em}{0.5em}}
\input{tcilatex}

\begin{document}

\title{Global geometry of 3-body motions with vanishing angular momentum, I}
\author{Wu-Yi Hsiang \\
Department of Mathematics \\
University of California, Berkeley \and Eldar Straume \\
Department of Mathematical Sciences\\
Norwegian University of Science and Technology\\
Trondheim, Norway}
\maketitle

\begin{abstract}
Following Jacobi's geometrization of Lagrange's least action principle,
trajectories of classical mechanics can be characterized as geodesics on the
configuration space $M$ with respect to a suitable metric which is the
conformal modification of the \emph{kinematic metric }by the factor ($U+h)$,
where $U$ and $h$ is the potential function and total energy, respectively.
In the special case of 3-body motions with zero angular momentum, the global
geometry of such trajectories can be reduced to that of their \emph{moduli
curves}, which\emph{\ }record the change of \ size and shape, in the moduli
space of oriented m-triangles, whose kinematic metric is, in fact, a
Riemannian cone over the shape space $M^{\ast}\simeq S^{2}(1/2).$

In this paper, we show that the moduli curve of such a motion is uniquely
determined by its \emph{shape curve} (which only records the change of
shape) in the case of $h\neq0$, while in the special case of $h=0$ it is
uniquely determined up to scaling. Thus, the study of the global geometry of
such motions can be further reduced to that of the shape curves, which are
time-parametrized curves on the 2-sphere characterized by a third order ODE
(cf. Theorem 3.9). Moreover, these curves have two remarkable properties,
namely the uniqueness of parametrization and the monotonicity, as stated in
Theorem 4.6 and Theorem 5.8, that constitute a solid foundation for a
systematic study of their global geometry and naturally lead to the
formulation of some pertinent problems such as those briefly discussed in \S %
7.
\end{abstract}

\tableofcontents

\section{Introduction}

\subsection{Local and global characterization of 3-body trajectories}

The classical 3-body problem of celestial mechanics studies the local and
global geometry of the trajectories of a 3-body system under the influence
of the gravitational forces, or equivalently a conservative system with
potential energy $-U$, where
\begin{equation}
U=\sum_{i<j}\frac{m_{i}m_{j}}{r_{ij}}   \label{1-0}
\end{equation}
is the Newtonian potential function. Thus, when the particles have position
vectors $\mathbf{a}_{i}=(x_{i},y_{i},z_{i})$ with respect to an inertial
frame, the trajectories are locally characterized by Newtons equation
\begin{equation}
m_{i}\mathbf{\ddot{a}}_{i}=\frac{\partial U}{\partial\mathbf{a}_{i}}=\frac{%
m_{i}m_{j}}{r_{ij}^{3}}(\mathbf{a}_{j}-\mathbf{a}_{i})+\frac{m_{i}m_{k}}{%
r_{ik}^{3}}(\mathbf{a}_{k}-\mathbf{a}_{i})\text{, \ }\left\{ i,j,k\right\}
=\left\{ 1,2,3\right\}   \label{1-1}
\end{equation}
where $r_{ij}=$ $\left\vert \mathbf{a}_{j}-\mathbf{a}_{i}\right\vert $ are
the mutual distances, and $(m_{1},m_{2},m_{3})$, $m_{i}>0$, is the given
mass distribution, assumed to be normalized so that $\sum m_{i}=1$. Since
the above equation is of order two, a trajectory is completely determined by
the initial position and velocity of the particles - often referred to as
the deterministic doctrine of classical mechanics.

We use the following notation
\begin{equation}
I=\sum m_{i}\left\vert \mathbf{a}_{i}\right\vert ^{2}\text{, \ }T=\frac{1}{2}%
\sum m_{i}\left\vert \mathbf{\dot{a}}_{i}\right\vert ^{2}\text{, \ \ \ }%
\mathbf{\Omega}=\sum m_{i}(\mathbf{a}_{i}\times\mathbf{\dot{a}}_{i})
\label{1-2}
\end{equation}
for the (polar) moment of inertia, kinetic energy and angular momentum,
respectively. These are the basic kinematic quantities, and their
interactions with the potential function $U$ play a major role in the
dynamics of the 3-body problem. For example, it is easy to deduce the
classical conservation laws from the system (\ref{1-1}), namely the
conservation of total energy
\begin{equation}
h=T-U,   \label{h}
\end{equation}
linear momentum $\sum m_{i}\mathbf{\dot{a}}_{i}$ and angular momentum $%
\mathbf{\Omega}$. As usual, the invariance of linear momentum allows us to
choose the inertial reference frame with the origin at the center of mass.
Moreover, by differentiation of $I$ twice with respect to time $t$ and using
(\ref{1-1}) we get
\begin{equation}
\ddot{I}=4T+2\sum\mathbf{a}_{i}\cdot\frac{\partial U}{\partial\mathbf{a}_{i}}%
=4T-2U=2(U+2h)   \label{L-J}
\end{equation}
where we have used the fact that $U$ is homogeneous of degree $-1$ as a
function of the vectors $\mathbf{a}_{i}$. This is the \emph{Lagrange-Jacobi}
equation.

On the other hand, trajectories can also be determined as solutions of a
suitable boundary value problem, and the simplest and most basic one is, for
example :
\begin{align*}
\text{"For a given pair of points }P,Q\text{, what are those trajectories }%
\gamma(t),t_{0}\leq t\leq t_{1}\text{,} & \text{ } \\
\text{w}\text{ith }\gamma(t_{0})=P\text{ and }\gamma(t_{1})=Q\text{ ?"} &
\end{align*}
Then solutions are found by applying an appropriate least action principle,
which characterizes solutions as extremals of an \emph{action integral }$%
J(\gamma)$\emph{\ }among those virtual motions $\gamma(t)$ with the given
pair of end points, together with some additional constraints.

Here we shall focus attention on the two least action principles due to
Lagrange and Hamilton, which are quite different but dual to each other :
\begin{align}
\text{Lagrange } & \text{: \ \ }J_{1}(\gamma)=\int_{\gamma}Tdt\text{, \
fixed energy }h\text{ }  \label{1-4a} \\
\text{Hamilton } & \text{: \ \ }J_{2}(\gamma)=\int_{\gamma}(T+U)dt\text{,
fixed time interval }[t_{0},t_{1}]   \label{1-4b}
\end{align}
The motions $t\rightarrow\gamma(t)$ are regarded as parametrized curves in
the Euclidean configuration space
\begin{equation}
M=\left\{ (\mathbf{a}_{1},\mathbf{a}_{2},\mathbf{a}_{3});\sum m_{i}\mathbf{a}%
_{i}=0\right\} \simeq\mathbb{R}^{6}   \label{M}
\end{equation}
Our aim, however, is to reduce the study of 3-body trajectories to a study
of associated curves in a lower dimensional space, namely the \emph{interior}
configuration space, i.e., the \emph{moduli space}
\begin{equation}
\bar{M}\simeq\mathbb{R}^{6}/SO(3)\approx\mathbb{R}_{+}^{3}\subset \mathbb{R}%
^{3}   \label{Mbar}
\end{equation}
With the appropriate assumptions, one expects that the least action
principles, as well as Newton's differential equation, can be pushed down to
the level of $\bar{M}$. In fact, one of our major results is that the study
of 3-body motions with vanishing angular momentum further reduces to the
analysis of specific curves on the shape space, which is the sphere $%
S^{2}\subset\mathbb{R}^{3}$.

The final step of our program is, of course, the lifting procedure from the
moduli curve $\bar{\gamma}(t)$ back to the actual trajectory $\gamma(t)$.
But this is a purely geometric construction which is well understood (cf.
e.g. \cite{HS2} or the following subsection) and it will not be a topic
here. Briefly, the curve in $\bar{M}$ determines the trajectory in $M$
uniquely up to a global congruence.

\subsection{Riemannian geometrization and reduction}

Classical mechanics up to present time is largely based upon developments
related to Hamilton's least action principle, involving Hamiltonian systems
and canonical transformations. Geometrically speaking, the underlying
structure is the symplectic geometry of the phase space. However, in this
paper we shall rather focus on the Riemannian geometric approach, based upon
Jacobi's reformulation of Lagrange's least action principle. In his famous
lectures \cite{Jacobi}, Jacobi introduced the concept of a \emph{kinematic
metric }
\begin{equation*}
ds^{2}=2Tdt^{2}
\end{equation*}
on the configuration space $M$, in terms of the kinetic energy $T$ of the
mechanical system. For example, in the case of an n-body system with total
mass $\sum m_{i}=1$
\begin{equation}
ds^{2}=2Tdt^{2}=\sum\limits_{i}m_{i}(dx_{i}^{2}+dy_{i}^{2}+dz_{i}^{2})
\label{1-5}
\end{equation}
Now, for a system with kinematic metric $ds^{2}$, potential function $U$ and
a given constant total energy $h$, set
\begin{align}
M_{(U,h)} & =\left\{ p\in M;h+U(p)\geq0\right\}  \label{1-6} \\
ds_{h}^{2} & =(h+U)ds^{2}  \notag
\end{align}
where $ds_{h}^{2}$ is the associated \emph{dynamical metric}. Then by
writing
\begin{equation*}
ds_{h}=\sqrt{h+U}ds=\sqrt{T}ds=\sqrt{2}Tdt
\end{equation*}
Jacobi transformed Lagrange's action integral (\ref{1-4a}) into an
arc-length integral
\begin{equation}
J_{1}(\gamma)=\frac{1}{\sqrt{2}}\int_{\gamma}ds_{h},   \label{1-7}
\end{equation}
and hence, in one stroke, the least action principle becomes the following
simple geometric statement :
\begin{align*}
& \text{" Trajectories with total energy }h\text{ are exactly those \emph{%
geodesic curves} in the } \\
& \text{space }M_{(U,h)}\text{ with the dynamical metric }ds_{h}^{2\text{ }}%
\text{"}
\end{align*}

Nowadays, such metric spaces are called $\emph{Riemannian}$ $\emph{manifolds}
$, and the global geometric study of geodesic curves is often referred to as
the Morse theory of geodesics. In particular, we note that the dynamical
metric $ds_{h}^{2}$ is a conformal modification of the underlying kinematic
metric $ds^{2}$.

In this geometric setting, the notion of "congruence class" is defined by
the action of the rotation group $SO(3)$, fixing the center of gravity (=
origin). It acts isometrically on the configuration space $(M,ds^{2})$ with
the kinematic metric, and also on the modified metric space ($%
M_{(U,h)},ds_{h}^{2})$ for any $SO(3)$-invariant potential function $U$. The
corresponding $SO(3)$-orbit spaces inherit the structure of a (stratified)
Riemannian manifold with the induced \emph{orbital distance metric}, which
we denote by
\begin{equation}
(\bar{M},d\bar{s}^{2})\text{, \ }(\bar{M}_{(U,h)},d\bar{s}_{h}^{2}\text{\ })%
\text{, \ }d\bar{s}_{h}^{2}=(h+U)d\bar{s}^{2},   \label{1-8}
\end{equation}
and similar to (\ref{1-6}), for $h$ negative the geodesics must stay inside
the Hill's region, namely the proper subset
\begin{equation}
\bar{M}_{h}=\bar{M}_{(U,h)}=\left\{ \bar{p}\in\bar{M};h+U(\bar{p}%
)\geq0\right\}   \label{Hills}
\end{equation}

\bigskip By definition, the projection map
\begin{equation*}
\pi:M\rightarrow\bar{M}=M/SO(3)
\end{equation*}
is a (stratified) Riemannian submersion, where the \emph{horizontal }tangent
vectors at $p\in$ $M$ are those perpendicular to the $SO(3)$-orbit. They are
mapped, by the tangent map $d\pi$, isometrically to the tangent space of $%
\bar{M}$ at $\bar{p}=\pi(p)$. Via the map $\pi$ there is a 1-1
correspondence between curves $\bar{\gamma}$ in $\bar{M}$ and their
horizontal lifting $\gamma$ in $M$ (resp. $M_{(U,h)})$, up to congruence. In
fact, for a (virtual) motion $\gamma(t)$, the property of being horizontal
is equivalent to the vanishing of the angular momentum vector $\mathbf{\Omega%
}$.

On the other hand, the above metric $d\bar{s}^{2}$ on $\bar{M}$ also has a
kinematic interpretation in analogy with (\ref{1-5}), namely
\begin{equation}
d\bar{s}^{2}=2\bar{T}dt^{2}\text{, \ }\bar{T}=T-T^{\omega}   \label{1-9}
\end{equation}
where $T^{\omega}$ is the purely rotational energy and hence the difference $%
\bar{T}$, representing that of the change of size and shape, is naturally
the kinetic energy at the level of $\bar{M}$. Therefore, we also refer to $d%
\bar{s}^{2}$ (resp. $d\bar{s}_{h}^{2})$ as the \emph{kinematic }(resp. \emph{%
dynamical}) metric on $\bar{M}$. Classical mechanics, indeed, tells us how
the term $T^{\omega}$ can be calculated from $\mathbf{\Omega}$ via the
socalled \emph{inertia operator} of the system; in particular, it follows
that $T^{\omega}=0$ if and only if $\mathbf{\Omega}=0$.

Now, assume $\mathbf{\Omega}=0$ and let $U$ be a nonnegative and $SO(3)$%
-invariant function on $M$. Then it is not difficult to see that both action
principles (\ref{1-4a}), (\ref{1-4b}) can be pushed down to $\bar{M}$. In
the first case, using Jacobi's reformulation (\ref{1-7}), we arrive
immediately at the following geometric statement similar to the one above :%
\begin{equation*}
\begin{array}{c}
\text{"Curves in }\bar{M}\text{ representing 3-body trajectories with total
energy }h \\
\text{(and }\mathbf{\Omega}=0\text{) are exactly those \emph{geodesic curves}
}\text{in the moduli } \\
\text{space }\bar{M}_{(U,h)}\text{ with the induced dynamical metric }d\bar {%
s}_{h}^{2}\text{."}%
\end{array}
\end{equation*}

In the case of (\ref{1-4b}), the Lagrange function $L=T+U$ is also defined
at the level of $\bar{M}$. Indeed, when $\mathbf{\Omega}=0$ we can view $%
\bar{M}$ as the configuration space for a simple conservative classical
mechanical system, namely with potential energy $-U$, kinetic energy $T$,
and conserved total energy $h=T-U$. It is easy to calculate the associated
Euler-Lagrange equations with respect to suitable coordinates in $\bar{M}$,
as demonstrated in Section 3.2. Finally, the reduced Newton's differential
equation on $\bar{M}$ can be calculated by the procedure described in \cite%
{HS2}, but we leave this topic here.

\subsection{A brief survey of the main results}

The results in this paper provide the foundations for the above Riemannian
geometric approach to the three-body problem with zero angular momentum. In
Section 2 we present the kinematic geometric framework for the reduction
method which we shall work out, consisting of the two successive reductions $%
M\rightarrow$ $\bar{M}\rightarrow M^{\ast}$, where $\bar{M}$ $\simeq \mathbb{%
R}^{3}$ is the (congruence) moduli space and $M^{\ast}\simeq S^{2}$ is the
shape space. The second reduction uses the cone structure of $\bar{M}$ over $%
M^{\ast}$ to eliminate the scaling variable $\rho=\sqrt{I}$ by radial
projection to the sphere.

A trajectory $\gamma(t)$ in the configuration space $M$ projects to its
moduli curve $\bar{\gamma}(t)$ in $\bar{M}$, and away from the base point $O$
the curve further projects to the shape curve $\gamma^{\ast}(t)$ on the
sphere. Conversely, whereas $\gamma(t)$ is determined up to congruence by $%
\bar {\gamma}(t)$, the real power of the above reduction method rests upon
the knowledge of the subtle relationship which, in fact, generally exists
between $\bar{\gamma}(t)$ and the geometric (i.e. unparametrized) curve $%
\gamma^{\ast }$.

In reality, the complete study of the three-body trajectories is hereby
reduced to the study of the relative geometry between the shape curve $%
\gamma^{\ast}$ and the gradient flow of $U^{\ast}$, namely the Newtonian
potential function restricted to the 2-sphere $M^{\ast}$. The major results
of the paper are divided into the following four main topics :

\begin{itemize}
\item The calculation of the reduced Newton's equations in $\bar{M}$ in
several ways, such as the geodesic equations of the Riemannian space ($%
\bar
{M}_{h},d\bar{s}_{h})$, and the reformulation of the geodesic
condition in terms of the curvature of the spherical shape curve. A suitable
combination of these equations also yields a separation of the radial
variable $\rho$ and hence a third order ODE on the 2-sphere which describes
all shape curves for any energy level $h$. The key step in this reduction is
Lemma \ref{split} which relates the intrinsic geometry of $%
(\gamma^{\ast},U^{\ast})$ to a kinematic quantity of $\bar{\gamma}(t)$.

\item The unique parametrization theorem (cf. Theorem \ref{unique}) asserts
that the time parametrized moduli curve $\bar{\gamma}(t)$ is (essentially)
determined by the oriented geometric shape curve\ $\gamma^{\ast}$.
Furthermore, the curve $\gamma^{\ast}$ is in fact uniquely determined by the
first two curvature coefficients at a generic point on the curve. The basic
technique used here is the local analysis of solution curves via power
series expansion.

\item The monotonicity theorem (cf. Theorem \ref{monotone}), which describes
a type of piecewise monotonic behavior of the shape curve $\gamma^{\ast}$.
Namely, the mass-modified latitude is a strictly monotonic function along $%
\gamma^{\ast}$ between two succeeding local maxima or minima, and they lie
on opposite hemispheres. In particular, the curve intersects the eclipse
circle at a unique point between two such local extrema.

\item Some initial applications to the study of triple collisions. In
particular, simple geometric proofs of the fundamental theorems of Sundman
and Siegel. Moreover, their asymptotic formulae for the derivatives of the
moment of inertia $I$ \ up to order two are extended to the derivatives of
any order.
\end{itemize}

The present exposition is based upon previous works of the authors (cf. \cite%
{Hsiang}, \cite{HS1}, \cite{HS2}) on the three-body problem, exploring its
"sphericality" as it manifests itself in various ways. The differential
equations which describe the moduli curves of the three-body trajectories,
are elaborated in Section 3 and 4, including a careful power series analysis
and comparison of the initial value problems at the moduli space and the
shape space level.

Section 5 is devoted to a geometric study of the gradient field of $U^{\ast}$%
, which also yields a simple and purely geometric proof of the monotonicity
theorem. A similar type of monotonicity for the shape curves was first
proved by Montgomery \cite{Mont} with his "infinitely many syzygies"
theorem, and later by Fujiwara et al \cite{Fuji}.

In Section 6 we recall the classical results and clarify some issues on the
work of Sundman and Siegel concerning triple collisions. Moreover, with the
results obtained so far, many challenging problems, for example in the study
of collisions and periodic motions, naturally present themselves for an in
depth study of the global geometry of shape curves. Some of these open
problems will be briefly discussed in Section 7.

\section{Kinematic geometry of m-triangles}

A three-body motion with vanishing angular momentum is always confined to a
fixed plane (for purely kinematic reasons), so the motions we shall study
are always planary. Therefore, we choose a plane $\mathbb{R}^{2}$ $\subset
\mathbb{R}^{3}$ with normal vectors $\pm\mathbf{n}$ and define an \emph{%
m-triangle} to be a triple $\delta=(\mathbf{a}_{1}\mathbf{,a}_{2}\mathbf{,a}%
_{3})$ of vectors $\mathbf{a}_{i}\in\mathbb{R}^{2}$ constrained by the
center of mass condition in (\ref{M}). Hence, for our purpose we shall
modify the definition (\ref{M}) of the \emph{configuration space} by taking
the subspace
\begin{equation}
M\simeq\mathbb{R}^{4}\subset\mathbb{R}^{6}:\sum_{i=1}^{3}m_{i}\mathbf{a}%
_{i}=0   \label{2-1}
\end{equation}
which consists of the above m-triangles in the fixed plane $\mathbb{R}^{2}$.
$M$ has the natural action of the rotation group $SO(2)$, and the \emph{%
moduli space}, representing the $SO(2)$-congruence classes of m-triangles,
is the orbit space
\begin{equation}
\bar{M}=M/SO(2)\simeq\mathbb{R}^{4}/SO(2)\supset\mathbb{R}^{6}/SO(3)
\label{2-1a}
\end{equation}

The degenerate (or collinear) $m$-triangles constitute the \emph{eclipse}
subvariety $E$ of $M$, and we say a nondegenerate $m$-triangle is \emph{%
positively} (resp. \emph{negatively}) oriented if $(\mathbf{a}_{1}\mathbf{,a}%
_{2}\mathbf{,n})$ is a positive (resp. negative) frame of $\mathbb{R}^{3}$.
Accordingly, we may write $M$ and $\bar{M}$ as the disjoint union of three
subsets

\begin{equation}
M=M_{+}\cup E\cup M_{-}\ ,\ \ \bar{M}=\bar{M}_{+}\cup\bar{E}\cup\bar{M}_{-}
\label{2-1b}
\end{equation}
and moreover, we observe that the moduli space of unoriented m-triangles
would be
\begin{equation*}
\bar{M}/\mathbb{Z}_{2}\simeq\mathbb{R}^{4}/O(2)\simeq\bar{M}_{\pm}\cup\bar
{%
E}\simeq\mathbb{R}^{6}/SO(3)=\mathbb{R}^{6}/O(3)
\end{equation*}

\begin{remark}
For a study of general (non-planary) 3-body motions, the natural
configuration space $\hat{M}$ consists of pairs ($\delta,\mathbf{n}$), where
$\delta$ is an m-triangle in $\mathbb{R}^{3}$ and $\mathbf{n}$ is a unit
vector perpendicular to all $\mathbf{a}_{i}$. $\hat{M}$ is a 6-dimensional
manifold (a 4-plane bundle over $S^{2})$ with the natural action of $SO(3)$,
and now the moduli space coincides with the above one (cf.\cite{HS2},
Section 2), namely%
\begin{equation*}
\hat{M}/SO(3)\simeq M/SO(2)=\bar{M}
\end{equation*}
\end{remark}

We shall describe in more detail the topology and induced Riemannian
structure of the above simple orbit spaces. Let $\delta$ (as above)\ and $%
\delta ^{\prime}=(\mathbf{b}_{1}\mathbf{,b}_{2}\mathbf{,b}_{3})$ be $m$%
-triangles. The following $SO(2)$-invariant, but mass dependent inner
product
\begin{equation}
\left\langle \delta,\delta^{\prime}\right\rangle =\sum m_{i}\mathbf{a}%
_{i}\cdot\mathbf{b}_{i}\text{ }   \label{2-2}
\end{equation}
is just the kinematic metric $ds^{2}$ of $M$ defined by (\ref{1-5}). In
particular, the squared norm is the moment of inertia, $\left\vert
\delta\right\vert ^{2}=I=\rho^{2}$. Let $S^{3}$ $\subset M$ be the unit
sphere $(\rho=1)$ and denote its spherical metric by $du^{2}$. Then we can
express $M$ as the Riemannian cone over $(S^{3},du^{2})$
\begin{equation}
M=\mathbb{R}^{4}=C(S^{3}):ds^{2}=d\rho^{2}+\rho^{2}du^{2}   \label{2-3}
\end{equation}
where $\rho$ measures the distance from the base point (origin) of the cone.

A description similar to (\ref{2-3}) applies to the moduli space
\begin{equation}
(\bar{M},d\bar{s}^{2})=(M,ds^{2})/SO(2)   \label{2-4}
\end{equation}
whose "unit sphere" $M^{\ast}$ $=(\rho=1)$ is called the \emph{shape space},
namely it is the image of $S^{3}$ in $\bar{M}$
\begin{equation}
(M^{\ast},d\sigma^{2})=(S^{3},du^{2})/SO(2)   \label{2-5}
\end{equation}
with the induced metric denoted by $d\sigma^{2}$. Thus, $\bar{M}$ also
inherits the structure of a Riemannian cone over its "unit sphere"
\begin{equation}
\bar{M}=C(S^{3})/SO(2)=C(S^{3}/SO(2))=C(M^{\ast}):d\bar{s}%
^{2}=d\rho^{2}+\rho^{2}d\sigma^{2}   \label{2-6}
\end{equation}
with $\rho$ still measuring the distance from the base point $O$. The shape
space is actually isometric to the 2-sphere of radius 1/2, as follows from
the well-known Hopf fibration construction $\ $ \ \emph{\ }
\begin{equation}
S^{3}\rightarrow\text{ }M^{\ast}=S^{3}/SO(2)=\mathbb{C}P^{1}\simeq
S^{2}(1/2)   \label{2-7}
\end{equation}

As a cone over $S^{2}$, $\bar{M}$ is clearly homeomorphic to $\mathbb{R}^{3}$
with the origin at the base point $O$. Away from $O$ they are even
diffeomorphic, when $\bar{M}$ has the induced smooth functional structure as
an orbit space of $\mathbb{R}^{4}$.

For the convenience of applying vector algebra we also recall the \emph{%
Euclidean model} of $\bar{M}$, where $\bar{M}$ is identified with $\mathbb{R}%
^{3}$, with Euclidean coordinates $(x,y,z)$ and associated spherical
coordinates $(I,\varphi,\theta)$, and the kinematic metric is expressed as
the following conformal modification of the standard Euclidean metric :
\begin{align}
d\bar{s}^{2} & =\frac{dx^{2}+dy^{2}+dz^{2}}{4\sqrt{x^{2}+y^{2}+z^{2}}}%
=d\rho^{2}+\frac{\rho^{2}}{4}(d\varphi^{2}+\sin^{2}\varphi\text{ }d\theta
^{2})  \label{2-8} \\
I^{2} & =\rho^{4}=x^{2}+y^{2}+z^{2}\text{, \ }0\leq\varphi\leq\pi\text{, \ }%
0\leq\theta\leq2\pi  \notag
\end{align}
Here $(\varphi,\theta)$ denotes any choice of spherical polar coordinates on
the sphere
\begin{equation*}
M^{\ast}=S^{2}:x^{2}+y^{2}+z^{2}=1
\end{equation*}
whose induced metric from the Euclidean 3-space is that of the round sphere
of radius 1
\begin{equation}
S^{2}(1):ds^{2}=d\varphi^{2}+\sin^{2}\varphi\text{ }d\theta^{2}
\label{round}
\end{equation}
whereas its induced (i.e. kinematic) metric as a submanifold of $(\bar
{M},d%
\bar{s}^{2})$ is
\begin{equation}
d\sigma^{2}=\frac{1}{4}ds^{2}   \label{scale}
\end{equation}
By (\ref{1-9}) and (\ref{2-8}) the total kinetic energy can be written as
\begin{equation}
T=\bar{T}+T^{\omega}=\frac{1}{2}\dot{\rho}^{2}+\frac{\rho^{2}}{8}(\dot {%
\varphi}^{2}+(\sin^{2}\varphi)\dot{\theta}^{2})+T^{\omega}   \label{T}
\end{equation}
where the rotational term $T^{\omega}$ vanishes precisely when $\mathbf{%
\Omega }=0$. Starting from Section 3 this is our standing assumption.

\begin{remark}
The radial distance function $\rho=\sqrt{I}$ is also referred to as the
\emph{hyper-radius} in the physics literature. For our purpose it is
generally more convenient to use $\rho$ rather than $I$ as the scaling
parameter, and we shall refer to $(\rho,\varphi,\theta)$ as a spherical
coordinate system of $\bar{M}$. We refer to \cite{HS2} for the relationships
between spherical coordinates, individual moments of inertia $%
(I_{1},I_{2},I_{3})$, or mutual distances $(r_{12},r_{23},r_{31})$.
\end{remark}

In the above Euclidean model of kinematic geometry the decomposition in (\ref%
{2-1b}) has a distinguished equator plane, namely the \emph{eclipse plane} $%
\bar{E}$ which divides $\mathbb{R}^{3}$ into the two half-spaces $\bar{M}%
_{\pm}=\mathbb{R}_{\pm}^{3}$. We choose the Euclidean coordinates so that $%
\bar{E}$ is the xy-plane and the half-space $z>0$ represents the congruence
classes $\bar{\delta}$ of the positively oriented m-triangles. Finally,
\begin{equation*}
E^{\ast}=\bar{E}\cap M^{\ast}:x^{2}+y^{2}=1,z=0
\end{equation*}
is the distinguished equator or \emph{eclipse circle} of the sphere $M^{\ast
}=S^{2}$.

On the other hand, the position of the various shapes $\delta^{\ast}$of
m-triangles on the sphere is uniquely determined by the position of the
three binary collision points $\mathbf{b}_{i}$, $i=1,2,3$, along the circle $%
E^{\ast}$, where $\mathbf{b}_{1}$ represents the shape of the degenerate
m-triangle with $\mathbf{a}_{2}=\mathbf{a}_{3}$ etc. We are still free to
choose the cyclic ordering $\mathbf{b}_{1}\rightarrow\mathbf{b}%
_{2}\rightarrow\mathbf{b}_{3}$ either in the eastward or westward direction.
Moreover, the mass distribution and the relative positions of the three
points $\mathbf{b}_{i}$ mutually determine each other. In fact, the angle $%
\beta_{1}$ between $\mathbf{b}_{2}$ and $\mathbf{b}_{3}$ is given by
\begin{equation}
\cos\beta_{1}=\frac{m_{2}m_{3}-m_{1}}{m_{2}m_{3}+m_{1}}\text{ \ etc. \ }
\label{2-9}
\end{equation}
and these formulae can be inverted.

Let $\left\{ N,S\right\} $ be the north and south pole of $S^{2}$. It is
often convenient to choose the spherical coordinate system $(\varphi,\theta)$
centered at $N$, namely $\varphi=0$ at the pole $N$. Then the poles
represent the m-triangles (congruent, but with opposite orientation) of
maximal area for a fixed size $\rho$, and more generally, the area of an
m-triangle is given by the formula
\begin{equation}
\Delta=\frac{\rho^{2}}{4\sqrt{m_{1}m_{2}m_{3}}}\left\vert \cos\varphi
\right\vert   \label{2-10}
\end{equation}

For a normalized (i.e. $\rho=1$) m-triangle of shape $\mathbf{p}\in S^{2}(1)
$ we also recall the formula for the mutual distances (cf. (\ref{1-1}))
\begin{equation}
r_{ij}=\frac{1}{2}\sqrt{\frac{1-m_{k}}{m_{i}m_{j}}}\left\vert \mathbf{p-b}%
_{k}\right\vert \text{ }=\sqrt{\frac{1-m_{k}}{m_{i}m_{j}}}\sin\sigma _{k}%
\text{\ }   \label{2-11}
\end{equation}
where $\left\vert \mathbf{p-b}_{k}\right\vert $ (resp. $2\sigma_{k}$) is the
Euclidean distance (resp. angle) between $\mathbf{p}$ and $\mathbf{b}_{k} $.
For proofs of (\ref{2-9}) - (\ref{2-11}) we refer to \cite{HS1} or \cite{HS2}%
.

\begin{remark}
\label{radius}In this paper we use both the kinematic and Euclidean model $%
S^{2}(r)$, $r=1/2$ or $1$, of the shape space $M^{\ast}$. Their arc-length
parameters are $\sigma$ and $s=2\sigma$, respectively, cf. (\ref{scale}). Of
course, the various geometric quantities, such as velocity, geodesic
curvature, gradient etc. must also be scaled appropriately when passing from
one model to the other.
\end{remark}

\section{Analysis on the moduli curves of 3-body motions with zero angular
momentum}

In this chapter we shall follow Jacobi's geometrization idea at the level of
the moduli space $\bar{M}=\mathbb{R}^{3}$. This enables us to reduce the
analysis of 3-body trajectories with zero angular momentum to that of the
corresponding moduli curves. According to Jacobi, for a given energy level $%
h $ the moduli curves can be interpreted as the geodesics of a specific
Riemannian metric $d\bar{s}_{h}^{2}$ on $\bar{M}$. Now, the standard
procedure for the calculation of the geodesic equations amounts to the
calculation of the Christoffel symbols of the metric, with respect to a
suitably chosen coordinate system suggested by the geometry of the space,
say. The resulting equations are ordinary differential equations whose
solutions are curves parametrized by the arc-length.

The kinematic geometry describes $\bar{M}$ with the scaling and rotational
symmetry of a Riemannian cone over a sphere, and therefore the spherical
coordinates $\rho,\varphi,\theta$ present themselves as the most natural
choice. In Section 3.1 we shall calculate the associated differential
equations. However, the natural parameter for 3-body trajectories is the
physical time $t$, and it is the effective usage of fixed energy that
enables us to express the equations in terms of $t$ as well, cf. (\ref{3-1}).

On the other hand, in the Hamiltonian least action principle the time
interval is fixed and $t$ is the natural parameter from the outset. The
Euler-Lagrange equations for the Lagrange function $L$ on $\bar{M}$ are
calculated in Section 3.2, and this approach, in fact, yields the same
system (\ref{3-1}) in a much simpler way.

However, we also seek a differential equation purely at the shape space
level, that is, with the scaling parameter $\rho$ eliminated. This demands a
deeper understanding of the relationship between $\rho$ and the geodesic
curvature of the shape curve. To this end we shall introduce an alternative
geometric approach to the study of geodesics in $\bar{M}$, which takes the
full advantage of the spherical symmetry and the cone structure of $\bar{M}$%
. This is the topic of Section 3.3.

Finally, in Section 3.4 we shall synthesize the results obtained in the
previous subsections and, in particular, we explain how the moduli curve can
be reconstructed from the shape curve.

\subsection{Calculation of the standard geodesic equations in $\bar{M}_{h}$}

We shall calculate the (standard) geodesic equations of $(\bar{M}_{h},d%
\bar
{s}_{h}^{2})$ relative to the spherical coordinates $%
(\rho,\varphi,\theta)$, where $(\varphi,\theta)$ denotes a choice of
spherical polar coordinates on the shape space $M^{\ast}=S^{2}$. The
homogeneity of the Newtonian potential function (\ref{1-0}) allows us to
write
\begin{equation}
U=\frac{1}{\rho}U^{\ast}(\varphi,\theta)   \label{U}
\end{equation}
where $U^{\ast}$ is the \emph{shape potential }function on $S^{2}$. By (\ref%
{1-8}) and (\ref{2-8}), the metric with the arc-length element $\bar
{s}_{h}
$ is

\begin{equation}
d\bar{s}_{h}^{2}=(\frac{U^{\ast}}{\rho}+h)d\bar{s}^{2}=(\frac{U^{\ast}}{\rho
}+h)\left( d\rho^{2}+\frac{\rho^{2}}{4}(d\varphi^{2}+\sin^{2}\varphi\text{ }%
d\theta^{2})\right)   \label{3-0}
\end{equation}
and the standard procedure for the calculation of the geodesic equations via
the Christoffel symbols, but expressed with respect to time $t$ as the
independent variable, yields the following system (cf. \cite{Str}, (5.11))
\begin{align}
(i)\text{ \ }0 & =\ddot{\rho}+\frac{\dot{\rho}^{2}}{\rho}-\frac{1}{\rho }(%
\frac{1}{\rho}U^{\ast}+2h)  \notag \\
(ii)\text{ \ }0 & =\text{\ }\ddot{\varphi}+2\frac{\dot{\rho}}{\rho}\dot{%
\varphi}-\frac{1}{2}\sin(2\varphi)\dot{\theta}^{2}-\frac{4}{\rho^{3}}%
U_{\varphi}^{\ast}  \label{3-1} \\
(iii)\text{ \ \ }0 & =\ddot{\theta}+2\frac{\dot{\rho}}{\rho}\dot{\theta }%
+2\cot(\varphi)\dot{\varphi}\dot{\theta}-\frac{4}{\rho^{3}}\frac{1}{\sin
^{2}\varphi}U_{\theta}^{\ast}  \notag
\end{align}
where equation (i) is just the Lagrange-Jacobi equation (\ref{L-J}).

The calculation of the above system goes as follows. For simplicity, let us
write%
\begin{equation*}
f^{2}=U+h=T\text{, \ }u=\bar{s}_{h}\text{, }
\end{equation*}
and $d\xi/du=\xi^{\prime}$, $d\xi/dt=\dot{\xi}$ for any function $\xi$. Then
by (\ref{1-9}) and (\ref{3-0})
\begin{equation*}
du^{2}=Td\bar{s}^{2}=2T^{2}dt^{2}=2f^{4}dt^{2}
\end{equation*}
and we deduce the useful identities relating arc-length and time derivatives
\begin{align}
\xi^{\prime} & =\frac{\dot{\xi}}{\dot{u}}\text{, \ }\xi^{\prime\prime}=\frac{%
\dot{u}\ddot{\xi}-\ddot{u}\dot{\xi}}{\dot{u}^{3}}\text{, \ \ }\frac{\ddot{u}%
}{\dot{u}}=2\frac{\dot{f}}{f}  \label{g1} \\
2f^{2} & =\dot{\rho}^{2}+\frac{\rho^{2}}{4}(\dot{\varphi}^{2}+(\sin
^{2}\varphi)\dot{\theta}^{2})\text{, \ }2f^{4}=\dot{u}^{2}   \label{g2}
\end{align}

Now, the first step is to calculate the geodesic equations with $u$ as the
independent variable, following the standard procedure, and the next step is
to transform the equations by changing to $t$ as the independent variable,
using the identities in (\ref{g1}) and (\ref{g2}). With the notation $%
(x_{1},x_{2},x_{3})=(\rho,\varphi,\theta)$ the equations in the first step
can be stated as%
\begin{equation}
\frac{d^{2}x_{k}}{du^{2}}+\dsum \limits_{i,j=1}^{3}\Gamma_{ij}^{k}\frac{%
dx_{i}}{du}\frac{dx_{j}}{du}=0\text{, \ }k=1,2,3   \label{g3}
\end{equation}
where the Christoffel sysmbols $\Gamma_{ij}^{k}$ are defined by
\begin{equation*}
\Gamma_{ij}^{k}=\frac{1}{2}\dsum \limits_{m}\left( \frac{\partial g_{jm}}{%
\partial x_{i}}+\frac{\partial g_{mi}}{\partial x_{j}}-\frac{\partial g_{ij}%
}{\partial x_{m}}\right) g^{km}
\end{equation*}
Here $(g^{ij})$ is the inverse of the matrix $(g_{ij})$ representing the
metric $du^{2}$, in the sense that
\begin{equation*}
du^{2}=f^{2}(d\rho^{2}+\frac{\rho^{2}}{4}(d\varphi^{2}+\sin^{2}\varphi\text{
}d\theta^{2})=\dsum g_{ij}dx_{i}dx_{j}
\end{equation*}
Thus the matrices are diagonal and
\begin{equation*}
g_{11}=f^{2},g_{22}=\frac{1}{4}f^{2}x_{1}^{2},g_{33}=\frac{1}{4}%
f^{2}x_{1}^{2}\sin^{2}x_{2}\text{, }g^{ii}=g_{ii}^{-1}
\end{equation*}

The first geodesic equation follows from (\ref{g3}) with $k=1$, so by
calculating the symbols $\Gamma_{ij}^{1}$ we obtain%
\begin{equation*}
\rho^{\prime\prime}-\frac{U^{\ast}}{2f^{2}\rho^{2}}\rho^{\prime}{}^{2}+(%
\frac{U^{\ast}}{2f^{2}}-\rho)\frac{1}{4}(\varphi^{\prime2}+(\sin^{2}\varphi)%
\theta^{\prime2})+\frac{\rho^{\prime}}{f^{2}\rho}(U_{\varphi}^{\ast
}\varphi^{\prime}+U_{\theta}^{\ast}\theta^{\prime})=0
\end{equation*}
Then, in the second step a straightforward calculation using (\ref{g1}) and (%
\ref{g2}) yields%
\begin{equation*}
\ddot{\rho}+\frac{\dot{\rho}^{2}}{\rho}+\left\{ -\frac{\dot{\rho}^{2}U^{\ast
}}{f^{2}\rho^{2}}-2\frac{f^{2}}{\rho}+\frac{U^{\ast}}{\rho^{2}}+(\frac{%
\dot
{U}^{\ast}}{f^{2}\rho}-2\frac{\dot{f}}{f})\dot{\rho}\right\} =0
\end{equation*}
where the bracket expression simplifies to $-\rho^{-1}(U+2h)$. The final
result is the first equation of (\ref{3-1}).

Next, let us consider the case $k=2$ where the first step yields the
equation
\begin{align*}
0 & =\varphi^{\prime\prime}-\frac{2U_{\varphi}^{\ast}}{f^{2}\rho^{3}}\left(
\rho^{\prime2}+\frac{\rho^{2}}{4}(\varphi^{\prime2}+(\sin^{2}\varphi
)\theta^{\prime2})\right) +\frac{U_{\varphi}^{\ast}}{f^{2}\rho}%
\varphi^{\prime2}-\frac{1}{2}(\sin2\varphi)\theta^{\prime2} \\
& +\frac{1}{f^{2}}(-\frac{U^{\ast}}{\rho^{2}}+\frac{2f^{2}}{\rho}%
)\rho^{\prime}\varphi^{\prime}+\frac{U_{\theta}^{\ast}}{f^{2}\rho}%
\varphi^{\prime}\theta^{\prime}
\end{align*}
The second step leads to the equation%
\begin{align*}
0 & =\ddot{\varphi}+2\frac{\dot{\rho}}{\rho}\dot{\varphi}-\frac{1}{2}%
\sin(2\varphi)\dot{\theta}^{2}-\frac{4}{\rho^{3}}U_{\varphi}^{\ast} \\
& +\left\{ -2\frac{\dot{f}}{f}\dot{\varphi}+\frac{U_{\varphi}^{\ast}}{%
f^{2}\rho}\dot{\varphi}^{2}-\frac{U^{\ast}}{f^{2}\rho^{2}}\dot{\rho}\dot{%
\varphi}+\frac{U_{\theta}^{\ast}}{f^{2}\rho}\dot{\varphi}\dot{\theta }%
\right\}
\end{align*}
where the bracket expression simplifies to%
\begin{equation*}
\frac{\dot{\varphi}}{f^{2}}(\frac{\dot{U}^{\ast}}{\rho}-\frac{U^{\ast}}{%
\rho^{2}}\dot{\rho}-\frac{d}{dt}f^{2})=0
\end{equation*}
This yields the second equation of (\ref{3-1}). The last case $k=3$ is
similar to the previous one, so we have omitted the calculations.

\subsection{An alternative derivation of the standard geodesic equations}

Another way of deriving the system (\ref{3-1}) is to calculate the
Euler-Lagrange equations for Hamilton's least action principle (\ref{1-4b}),
at the level of the moduli space $\bar{M}$. In fact, the Lagrange function $%
L=T+U$ descends to a function defined on the tangent bundle $T\bar{M}$ since
$\mathbf{\Omega}=0$, and
\begin{equation*}
T=\frac{1}{2}\left\vert \frac{d\bar{\gamma}}{dt}\right\vert ^{2}=\frac{1}{2}(%
\frac{d\bar{s}}{dt})^{2}
\end{equation*}
is actually the kinetic energy of a curve $\bar{\gamma}(t)$ in $\bar{M}$.
Thus we have a simple classical conservative mechanical system with
potential energy $-U$, kinetic energy $T$, and conserved total energy $h=T-U$%
. Therefore, in terms of the coordinates $(\rho,\varphi,\theta)$ the
associated Lagrange system is
\begin{equation}
\frac{d}{dt}\frac{\partial L}{\partial\dot{\rho}}=\frac{\partial L}{%
\partial\rho}\text{, \ \ }\frac{d}{dt}\frac{\partial L}{\partial\dot{\varphi
}}=\frac{\partial L}{\partial\varphi}\text{, \ \ }\frac{d}{dt}\frac{\partial
L}{\partial\dot{\theta}}=\frac{\partial L}{\partial\theta}   \label{3-20}
\end{equation}
where by (\ref{2-8})%
\begin{equation*}
L=T+U=\left( \frac{1}{2}\dot{\rho}^{2}+\frac{\rho^{2}}{8}(\dot{\varphi}%
^{2}+\sin^{2}\varphi\text{ }\dot{\theta}^{2})\right) +\frac{1}{\rho}U^{\ast
}(\varphi,\theta)
\end{equation*}

Now, the first equation of (\ref{3-20}) reads%
\begin{equation}
\frac{d}{dt}\dot{\rho}=\frac{\rho}{4}(\dot{\varphi}^{2}+\sin^{2}\varphi\text{
}\dot{\theta}^{2})-\frac{1}{\rho^{2}}U^{\ast}   \label{3-21}
\end{equation}
and by substituting the expression
\begin{equation*}
\frac{1}{4}(\dot{\varphi}^{2}+\sin^{2}\varphi\text{ }\dot{\theta}^{2})=\frac{%
1}{\rho^{2}}(2T-\dot{\rho}^{2})=\frac{1}{\rho^{2}}(2U+2h-\dot{\rho }^{2})
\end{equation*}
into (\ref{3-21}) the equation transforms to equation (i) of (\ref{3-1}).

Next, the second equation of (\ref{3-20}) reads%
\begin{equation*}
\frac{d}{dt}(\frac{\rho^{2}}{4}\dot{\varphi})=\frac{\rho^{2}}{4}\sin
\varphi\cos\varphi\text{ }\dot{\theta}^{2}+\frac{1}{\rho}U_{\varphi}^{\ast}
\end{equation*}
and after differentiation this becomes%
\begin{equation*}
\rho^{2}\ddot{\varphi}+2\rho\dot{\rho}\dot{\varphi}-\frac{\rho^{2}}{2}%
\sin2\varphi\text{ }\dot{\theta}^{2}-\frac{4}{\rho}U_{\varphi}^{\ast}=0
\end{equation*}
which is precisely equation (ii) of (\ref{3-1}). Similarly, the third
equation of (\ref{3-20}) reads
\begin{equation}
\frac{d}{dt}(\rho^{2}\sin^{2}\varphi\text{ }\dot{\theta})=\frac{4}{\rho }%
U_{\theta}^{\ast}   \label{3-22}
\end{equation}
where the left side becomes%
\begin{equation*}
2\rho\dot{\rho}\sin^{2}\varphi\text{ }\dot{\theta}+2\rho^{2}\sin\varphi
\cos\varphi\text{ }\dot{\varphi}\dot{\theta}+\rho^{2}\sin^{2}\varphi\text{ }%
\ddot{\theta}
\end{equation*}
Then it is easily seen that equation (\ref{3-22}) becomes equation (iii) of (%
\ref{3-1}).

\subsection{Cone surfaces and geodesics in $\bar{M}_{h}$}

To take full advantage of the cone structure of $\bar{M}$ and the scaling
property (\ref{U}) of the potential function $U$, one naturally seeks to
reduce the analysis of the moduli curve to that of the shape curve.
Therefore, in this subsection we shall study the geodesic equations with the
shape curve in the forefront.

\begin{definition}
Let $\bar{\gamma}$ be a curve in $\bar{M}$ not including the (triple
collision) base point $O$. The \emph{cone surface }spanned by $\bar{\gamma}$
consists of all rays emanating from $O$ and intersecting $\bar{\gamma}$. The
cone surface is denoted by $C(\bar{\gamma})$ or $C(\gamma^{\ast})$.
\end{definition}

The intersection of the cone surface with the shape space $M^{\ast}$ is the
associated shape curve $\gamma^{\ast}$, and conversely, the cone surface is
also uniquely determined by $\gamma^{\ast}$, which explains the notation $%
C(\gamma^{\ast})$.

Let $\bar{\gamma}(t)$ be the moduli curve of a 3-body motion with zero
angular momentum and $\gamma^{\ast}(t)$ the associated shape curve with
arc-length parameter $\sigma$ in $M^{\ast}\simeq S^{2}(1/2)$. Endowed with
the kinematic metric, the cone surface
\begin{equation}
C(\gamma^{\ast}):d\bar{s}^{2}=d\rho^{2}+\rho^{2}d\sigma^{2}\text{, \ }%
0\leq\sigma\leq\sigma_{1}   \label{3-2}
\end{equation}
is isometric to a (flat) Euclidean sector with polar coordinates $(\rho
,\sigma)$ and angular width $\sigma_{1}$.

Next, let the function $u(\sigma)$ denote the restriction of $U^{\ast}$
along $\gamma^{\ast}$. There is a pair of conformally related metrics on the
above sector, namely the flat metric (\ref{3-2}) and the dynamical metric
\begin{equation}
d\bar{s}_{h}^{2}=(h+\frac{u(\sigma)}{\rho})d\bar{s}^{2}   \label{3-3}
\end{equation}
defined on the subregion
\begin{equation}
C(\gamma^{\ast})_{h}=\left\{ (\rho,\sigma);\rho\geq0,0\leq\sigma\leq
\sigma_{1},h+\frac{u(\sigma)}{\rho}>0\right\}   \label{3-4}
\end{equation}

On the one hand, $\bar{\gamma}$ is a geodesic curve of the surface ($%
C(\gamma^{\ast})_{h}$ with the metric (\ref{3-3}), but on the other hand, it
is also a geodesic of the ambient space $(\bar{M}_{h},d\bar{s}_{h}^{2})$ and
hence the normal component of the \emph{curvature vector} of $\bar{\gamma}$
in $\bar{M}_{h}$ also vanishes. Accordingly, it is natural to write the
geodesic equations of $\bar{\gamma}$ in $\bar{M}_{h}$ as a pair of coupled
ODE's expressing, respectively, the vanishing of the tangential and normal
component of the curvature vector.

To analyze curvatures, let us fix some convention concerning orientation. We
assume $\bar{\gamma}$ (and hence also $\gamma^{\ast}$) is oriented; they are
curves in $\bar{M}\simeq\mathbb{R}^{3}$ and $M^{\ast}\simeq S^{2}$
respectively, and these spaces have their standard orientation. We choose a
positive orthonormal frame $(\mathbf{\tau,\eta,\nu})$ of $(\bar{M},d\bar
{s}%
^{2})$ along $\bar{\gamma}$, as follows. Let $\mathbf{\tau}$ (resp. $\mathbf{%
\tau}^{\ast})$ be the positive unit tangent field of $\bar{\gamma}$ (resp. $%
\gamma^{\ast}$), and choose the unit normal field $\mathbf{\nu}^{\ast }$ of $%
\gamma^{\ast}$ so that $(\mathbf{\tau}^{\ast},\mathbf{\nu}^{\ast})$ is a
positive frame of $M^{\ast}$. Then $\mathbf{\nu}=(1/\rho)\mathbf{\nu}^{\ast }
$ is a unit normal field of the cone surface $C(\bar{\gamma})$ and hence
orients the surface. Finally, $\mathbf{\eta}$ is the normal field of $\bar{%
\gamma}$ in $C(\bar{\gamma})$. For convenience, we write
\begin{equation}
\mathbf{\tau}=\cos\alpha\frac{\partial}{\partial\rho}+\sin\alpha\frac{1}{%
\rho }\frac{\partial}{\partial\sigma}\text{, \ }\mathbf{\eta}=-\sin\alpha
\frac{\partial}{\partial\rho}+\cos\alpha\frac{1}{\rho}\frac{\partial}{%
\partial\sigma}   \label{3-5a}
\end{equation}
where $\alpha$ denotes the angle between the radial and tangential
direction, that is, the angle between $\frac{\partial}{\partial\rho}$ and $%
\mathbf{\tau}$.

Let $\mathbf{n}$ be a vector normal to $\bar{\gamma}$, say $\mathbf{\eta}$
or $\mathbf{\nu}$ as above, and let $\mathcal{K}(\mathbf{n})$ and $\mathcal{K%
}_{h}(\mathbf{n})$ denote the associated geodesic curvatures of $\bar{\gamma}
$ with respect to the metrics $d\bar{s}^{2}$ and $d\bar{s}_{h}^{2}$,
respectively. It follows from the first variation formula of arc-length that
the curvatures of a given curve with respect to such a pair of conformally
related metrics are linked by the following formula
\begin{equation}
\mathcal{K}_{h}(\mathbf{n})=\mathcal{K}(\mathbf{n})\mathcal{-}\frac{1}{2}%
\frac{d}{d\mathbf{n}}\ln(h+U)   \label{3-5b}
\end{equation}
There are two cases to analyze, namely $\mathbf{n=\eta}$ or $\mathbf{\nu}$,
and for simplicity we write $\mathcal{K}=\mathcal{K(}\mathbf{\eta)}$, $%
\mathcal{K}^{\perp}=\mathcal{K}(\mathbf{\nu)}$ etc., and we assume $%
\gamma^{\ast}$ is not a single point.

First, the geodesic curvatures of $\bar{\gamma}$ in the surface $C(\gamma
^{\ast})$ with the two metrics are related by
\begin{equation}
\mathcal{K}_{h}=\mathcal{K}-\frac{1}{2}\frac{d}{d\mathbf{\eta}}\ln (h+U)=(%
\frac{d\alpha}{d\bar{s}}+\frac{d\sigma}{d\bar{s}})-\frac{1}{2}\frac {d}{d%
\mathbf{\eta}}\ln(h+\frac{u(\sigma)}{\rho})   \label{3-6}
\end{equation}
where we recognize $(\alpha+\sigma)$ as the angle between the tangent line
and a fixed reference ray in the Euclidean sector (\ref{3-2}). Similarly,
let $\mathcal{K}^{\perp}$ and $\mathcal{K}_{h}^{\perp}$ be the "surface
normal" geodesic curvatures of $\bar{\gamma}$ with respect to the metrics $d%
\bar
{s}^{2}$ and $d\bar{s}_{h}^{2}$, consequently
\begin{equation}
\mathcal{K}_{h}^{\perp}=\mathcal{K}^{\perp}\mathcal{-}\frac{1}{2}\frac {d}{d%
\mathbf{\nu}}\ln(h+\frac{U^{\ast}}{\rho})   \label{3-6a}
\end{equation}

To find an expression for $\mathcal{K}^{\perp}$, it is a key observation
that the \emph{principal curvatures} of $C(\gamma^{\ast})$ in $\bar{M}$ at a
given point $(\rho,\sigma)$ are the numbers $0$ and $\frac{1}{\rho}\mathcal{K%
}^{\ast}$, where $\mathcal{K}^{\ast}=\mathcal{K}^{\ast}(\sigma) $ denotes
the geodesic curvature of $\gamma^{\ast}$ on the sphere $M^{\ast}\simeq
S^{2}(1/2)$ at the point $\gamma^{\ast}(\sigma)$. Therefore, by the
classical Euler's formulas, the \emph{normal sectional curvature} of $%
C(\gamma^{\ast})$ in the tangential direction of $\bar{\gamma}$ is equal to
\begin{equation}
\mathcal{K}^{\perp}=\frac{1}{\rho}(\sin^{2}\alpha)\mathcal{K}^{\ast}(\sigma)
\label{3-7}
\end{equation}

Finally, the geodesic condition for $\bar{\gamma}$ in $\bar{M}_{h}$ reads
\begin{equation*}
\mathcal{K}_{h}=\mathcal{K}_{h}^{\perp}=0
\end{equation*}
which by (\ref{3-6})-(\ref{3-7}) is neatly expressed by two scalar ODEs. In
terms of the kinematic arc-length parameter $d\bar{s}$ of $\bar{\gamma}$,
regarded either as a curve in $C(\gamma^{\ast})$ or $\bar{M}$, we can state
the final result as the following theorem :

\begin{theorem}
\label{conesurface}Let $\bar{\gamma}(t)$ (resp. $\gamma^{\ast}(t)$) be the
moduli (resp. shape) curve of a given three-body motion with zero angular
momentum and total energy $h$. Set $\sigma$ to be the arc-length parameter
of $\gamma^{\ast}$ in the shape space $M^{\ast}\simeq S^{2}(1/2)$, $(\rho
,\sigma)$ the polar coordinate system of the associated cone surface $%
C(\gamma^{\ast})$, and $u(\sigma)$ the restriction of $U^{\ast}$ along $%
\gamma^{\ast}$. If $\bar{\gamma}$ is not a ray solution, then it is
characterized by the following pair of equations
\begin{align}
(i)\text{ } & \text{: }\frac{d\alpha}{d\bar{s}}+\frac{d\sigma}{d\bar{s}}-%
\frac{1}{2}(-\sin\alpha\frac{\partial}{\partial\rho}+\frac{\cos\alpha}{\rho }%
\frac{\partial}{\partial\sigma})\ln(h+\frac{u(\sigma)}{\rho})=0  \label{3-8a}
\\
(ii)\text{ } & \text{: \ }(\sin^{2}\alpha)\mathcal{K}^{\ast}-\frac{1}{2}%
\frac{\partial}{\partial\mathbf{\nu}^{\ast}}\ln(h+\frac{U^{\ast}}{\rho})=0
\label{3-8b}
\end{align}
where $\alpha$ is the angle between the radial direction and the tangential
direction and $\mathbf{\nu}^{\ast}$ is the positive unit normal vector field
of $\gamma^{\ast}$ in $M^{\ast}$.
\end{theorem}

\begin{remark}
\label{shape}The exceptional case of ray solutions, that is, $\gamma^{\ast} $
is a single point, can be settled directly from (\ref{3-5b}), where $%
\mathcal{K}(\mathbf{n})=\mathcal{K}_{h}(\mathbf{n})=0$ and hence for each $%
\mathbf{n}$
\begin{equation*}
\frac{d}{d\mathbf{n}}\ln(h+U)=0\text{ \ \ or \ \ }\nabla U\cdot\mathbf{n}=0
\end{equation*}
This is equivalent to $\nabla U^{\ast}=0$, so $\gamma^{\ast}$ is one of the
five critical points of $U^{\ast}$ on the 2-sphere, namely the two minima
(called Lagrange points) and the three saddle points (called Euler points)
lying on the equator (or eclipse) circle.

We also deduce the above result from the equations (ii), (iii) of (\ref{3-1}%
), namely $U_{\varphi}^{\ast}$ and $U_{\theta}^{\ast}$ must vanish if $%
\varphi$ and $\theta$ are set to be constant. Ray solutions yield the
simplest type of three-body motions, namely the shape invariant or so-called
homographic motions, which in the case of $\mathbf{\Omega}=0$ are confined
to a line. Then the only variable $\rho(t)$ is the solution of the
1-dimensional Kepler problem given by the Lagrange-Jacobi equation (i.e. (i)
of (\ref{3-1})).
\end{remark}

\begin{remark}
(i) For a given value of total energy $h$, the influence of $U^{\ast}$ on
the geometry of the associated cone surface $C(\gamma^{\ast})_{h}$ is via
the function $u=U^{\ast}|_{\gamma^{\ast}}$, and equation (\ref{3-8a}) is
exactly the geodesic equation of $C(\gamma^{\ast})_{h}$. Most of the
geodesic curves of $C(\gamma^{\ast})_{h}$ are, of course, not geodesics of
the ambient space $\bar{M}_{h}$ since they are not moduli curves of actual
three-body motions.

(ii) Equation (\ref{3-8b}), on the other hand, is expressed in terms of the
\emph{relative geometry} of the inclusion $\gamma^{\ast}\subset M^{\ast}$,
namely the geodesic curvature $\mathcal{K}^{\ast}$ and the normal derivative
of $U^{\ast}$along $\gamma^{\ast}$.
\end{remark}

Finally, recall from (\ref{3-5a}) that the scaling variable $\rho$ and the
angular variable $\alpha$, which measures the radial inclination of the
moduli curve in the cone $\bar{M}$, essentially determine each other via the
relations
\begin{equation}
\cos\alpha=\frac{d\rho}{d\bar{s}}\text{, \ }\sin\alpha=\rho\frac{d\sigma }{d%
\bar{s}}\text{,}   \label{3-10}
\end{equation}
which tell us, for example, how to calculate $\rho$ from the variation of $%
\alpha$ along the shape curve $\gamma^{\ast}\subset S^{2}(1/2)$ :
\begin{equation}
\rho(\sigma)=\rho(\sigma_{0})e^{\int_{\sigma_{0}}^{\sigma}\cot\alpha
(\sigma)d\sigma}\text{, \ \ for }\rho(\sigma_{0})\text{\ }\neq0
\label{3-11}
\end{equation}

\subsection{Synthesis of the analysis of the moduli curve and that of the
shape curve.}

In the previous sections we have used geometric ideas to obtain differential
equations in the moduli space $\bar{M}\simeq\mathbb{R}^{3}$ characterizing
3-body trajectories with zero angular momentum. Since $\bar{M}$ is a cone
over the 2-sphere $M^{\ast}$ defined by $\rho=1$, it is natural to project
the moduli curve $\bar{\gamma}$ down to its image curve $\gamma^{\ast}$ on
the sphere. However, unless one resolves the hidden interlocking between $%
\gamma^{\ast}$ and the scaling variable $\rho$, implicitly described by the
differential equations, one cannot reconstruct the moduli curve from its
shape curve and thus fully utilizing the reduction from $\bar{M}$ to $%
M^{\ast}$.

In order to separate the scaling variable $\rho$ from the spherical
variables $(\varphi,\theta)$ we shall proceed by combining the two systems
of geodesic equations, (\ref{3-1}) and (\ref{3-8a}) - (\ref{3-8b}), which we
derived in two different ways. Note that the natural parameter in mechanics
is the time $t$, whereas the arc-length parameter is the natural parameter
in metric geometry. This suggests a transformation of the latter system to
equations with $t$ as the independent variable, and as it turns out, this
also provides a remarkable simple solution of the above separation problem.

\subsubsection{Basic geometry of curves on the 2-sphere with a potential
function}

Since spherical curves play a crucial role in the present study, it is
convenient to collect some basic formulae concerning the differential
geometry of curves on $S^{2}(1)$, as well as the tangential and normal
derivatives of a given (potential) function $U^{\ast}$ on the sphere. We
shall express them in terms of a chosen spherical polar coordinate system $%
(\varphi,\theta)$.

For a given oriented curve $\gamma^{\ast}$, let $\mathbf{\tau}^{\ast}$
(resp. $\mathbf{\nu}^{\ast})$ be the unit tangent vector in the positive
direction (resp. unit normal vector) such that $(\mathbf{\tau}^{\ast},%
\mathbf{\nu}^{\ast})$ is a positively oriented frame of the sphere. We
consider a (regular) time parametrized curve $\gamma^{\ast}(t)=(\varphi(t),%
\theta(t))$ and set $s=s(t)\geq0$ to be the arc-length along the curve. As
before, differentiation of a function $f$ with respect to $t$ or $s $ are
denoted by $\dot{f}$ and $f^{\prime}$ respectively, and clearly $\dot{f}%
=f^{\prime}v$ where $v=\dot{s}$ is the speed of the curve. Then

\begin{equation}
\mathbf{\tau}^{\ast}=\frac{d\gamma^{\ast}}{ds}=\frac{1}{v}(\dot{\varphi}%
\frac{\partial}{\partial\varphi}+\dot{\theta}\frac{\partial}{\partial\theta }%
)\text{, \ }\mathbf{\nu}^{\ast}=\frac{1}{v}(-\dot{\theta}\sin\varphi \frac{%
\partial}{\partial\varphi}+\dot{\varphi}\frac{1}{\sin\varphi}\frac{\partial}{%
\partial\theta})   \label{C3}
\end{equation}
and the velocity vector field of the curve is
\begin{equation}
\frac{d\gamma^{\ast}}{dt}=v\mathbf{\tau}^{\ast}\text{, \ \ }v=\sqrt {\dot{%
\varphi}^{2}+(\sin^{2}\varphi)\dot{\theta}^{2}}   \label{speed}
\end{equation}
The scalar acceleration $\ $
\begin{equation}
\dot{v}=\frac{d}{dt}v=\frac{1}{v}[\dot{\varphi}\ddot{\varphi}+(\sin\varphi
\cos\varphi)\dot{\varphi}\dot{\theta}^{2}+\sin^{2}(\varphi)\dot{\theta}\ddot{%
\theta}]   \label{C9}
\end{equation}
and its higher time derivatives are needed to express time derivatives of a
function in terms of arc-length derivatives, using operators of increasing
order
\begin{equation}
\frac{d}{dt}=v\frac{d}{ds},\text{ \ }\frac{d^{2}}{dt^{2}}=v^{2}\frac{d^{2}}{%
ds^{2}}+\dot{v}\frac{d}{ds},\text{ etc.}   \label{C12}
\end{equation}

To calculate the geodesic curvature function $K^{\ast}$, let us first make
use of Euclidean coordinates
\begin{equation*}
x=\sin\varphi\cos\theta,\text{ \ }y=\sin\varphi\sin\theta,\text{ \ }%
z=\cos\varphi
\end{equation*}
and write $\mathbf{x}(s)=(x(s),y(s),z(s))$ and use the formula%
\begin{equation*}
K^{\ast}(s)=\mathbf{x}(s)\times\mathbf{x}^{\prime}(s)\cdot\mathbf{x}%
^{\prime\prime}(s)
\end{equation*}
where $\mathbf{x}^{\prime}(s)=\mathbf{\tau}^{\ast}$. This yields%
\begin{align}
K^{\ast} & =(\cos\varphi)\theta^{\prime}(1+\varphi^{\prime2})+\sin
\varphi(\varphi^{\prime}\theta^{\prime\prime}-\theta^{\prime}\varphi
^{\prime\prime})  \label{C6} \\
& =\frac{1}{v^{3}}\left\{ (\cos\varphi)\dot{\theta}(v^{2}+\dot{\varphi}%
^{2})+\sin\varphi(\dot{\varphi}\ddot{\theta}-\dot{\theta}\ddot{\varphi }%
)\right\}  \notag
\end{align}
and its intrinsic first derivative is
\begin{align}
K^{\ast\prime} & =\frac{d}{ds}K^{\ast}=(-(\sin\varphi)\varphi^{\prime}%
\theta^{\prime}+(\cos\varphi)\theta^{\prime\prime})(1+\varphi^{\prime2})+(%
\cos\varphi)\varphi^{\prime}(\varphi^{\prime}\theta^{\prime\prime}-\theta^{%
\prime}\varphi^{\prime\prime})  \notag \\
& +\sin\varphi(\varphi^{\prime}\theta^{\prime\prime\prime}-\theta^{\prime
}\varphi^{\prime\prime\prime})   \label{C7}
\end{align}

The gradient field of $U^{\ast}$ is the following vector field on the sphere
\begin{equation}
\nabla U^{\ast}=U_{\varphi}^{\ast}\frac{\partial}{\partial\varphi}+\frac{%
U_{\theta}^{\ast}}{\sin^{2}\varphi}\frac{\partial}{\partial\theta}
\label{C2}
\end{equation}
which allows us to calculate various derivatives of $U^{\ast}$. For example,
the tangential and normal derivatives along the curve are, respectively,%
\begin{align}
U_{\tau}^{\ast} & =\frac{\partial U^{\ast}}{\partial\mathbf{\tau}^{\ast}}%
=\nabla U^{\ast}\cdot\mathbf{\tau}^{\ast}=\frac{1}{v}(\dot{\varphi}%
U_{\varphi}^{\ast}+\dot{\theta}U_{\theta}^{\ast})  \label{C4} \\
U_{\nu}^{\ast} & =\frac{\partial U^{\ast}}{\partial\mathbf{\nu}^{\ast}}%
=\nabla U^{\ast}\cdot\mathbf{\nu}^{\ast}=\frac{1}{v}(-\dot{\theta}%
\sin\varphi U_{\varphi}^{\ast}+\dot{\varphi}\frac{1}{\sin\varphi}%
U_{\theta}^{\ast})   \label{C5}
\end{align}
and the intrinsic first derivative of $U_{\nu}^{\ast}$ is
\begin{align}
U_{\nu}^{\ast\prime} & =\frac{d}{ds}U_{\nu}^{\ast}=(\frac{U_{\theta}^{\ast}}{%
\sin\varphi})\varphi^{\prime\prime}+(-\sin\varphi
U_{\varphi}^{\ast})\theta^{\prime\prime}+(-\frac{\cos\varphi
U_{\theta}^{\ast}}{\sin^{2}\varphi })\varphi^{\prime2}  \label{C8} \\
& +(\frac{U_{\theta\theta}^{\ast}}{\sin\varphi}-U_{\varphi\theta}^{\ast}\sin%
\varphi)\theta^{\prime2}+(\frac{U_{\theta\theta}^{\ast}}{\sin\varphi}%
-\sin\varphi U_{\varphi\varphi}^{\ast}-\cos\varphi
U_{\varphi}^{\ast})\varphi^{\prime}\theta^{\prime}  \notag
\end{align}

Finally, for convenience and later reference let us introduce the following
definition :

\begin{definition}
For a given curve $\gamma^{\ast}$ and function $U^{\ast}$ on $S^{2}$, the
associated Siegel function along the curve is defined to be
\begin{equation}
\mathfrak{S}=\frac{U_{\nu}^{\ast}}{K^{\ast}}   \label{C10}
\end{equation}
\end{definition}

We regard the function as undefined along geodesic arcs. Moreover, the
function may have a singularity at isolated points where $K^{\ast}$
vanishes. Note that $\mathfrak{S}$ is independent of the orientation of $%
\gamma^{\ast}$. Observe the following formula for the logarithmic derivative
of $\mathfrak{S}$, as a function of $s$ (and similarly for $t$ as parameter)
\begin{equation}
\text{\ \ }\frac{\mathfrak{S}^{\prime}}{\mathfrak{S}}=\frac{%
U_{\nu}^{\ast\prime}}{U_{\nu}^{\ast}}-\frac{K^{\ast\prime}}{K^{\ast}}
\label{C11}
\end{equation}

\begin{remark}
We have named the above function after C.L. Siegel for the following reason.
In his study (cf. \cite{Sieg2}) of triple collisions in the three-body
problem, Siegel investigated the asymptotic behavior of the time derivatives
$\dot{I}$ , $\ddot{I}$ of the moment of inertia $I=\rho^{2}$. The major step
in his proof was, indeed, to show that the expression $\sqrt{I}(2T-\dot{I}%
^{2}/4I)$ tends to zero. It turns out that this expression equals $\mathfrak{%
S}$ whenever the latter is defined (see Lemma \ref{split}, where $%
\rho^{3}v^{2}$ equals the above expression). In particular, it is an
intrinsic quantity at the shape space level. We shall return to triple
collisions and Siegel's approach in Section 6.
\end{remark}

\subsubsection{Reformulation of the geodesic equations in terms of the shape
curve}

For convenience, let us write $V_{1}=\frac{\partial}{\partial\varphi}$ and $%
V_{2}=\frac{\partial}{\partial\theta}$ for the basic coordinate vector
fields of the spherical coordinate system on the unit sphere. By definition,
the \emph{acceleration }of the spherical curve $\gamma^{\ast}(t)=(\varphi
(t),\theta(t))$ is the expression
\begin{align*}
\ddot{\gamma}^{\ast} & =\frac{D}{dt}\gamma^{\ast}=\ddot{\varphi}V_{1}+\dot{%
\varphi}\nabla_{\dot{\gamma}^{\ast}}V_{1}+\ddot{\theta}V_{2}+\dot {\theta}%
\nabla_{\dot{\gamma}^{\ast}}V_{2} \\
& =\ddot{\varphi}V_{1}+\ddot{\theta}V_{2}+\dot{\varphi}^{2}%
\nabla_{V_{1}}V_{1}+\dot{\theta}^{2}\nabla_{V_{2}}V_{2}+\dot{\varphi}\dot{%
\theta}(\nabla_{V_{1}}V_{2}+\nabla_{V_{2}}V_{1})
\end{align*}
which is the covariant derivative of the velocity along the curve, with
respect to the metric (\ref{round}). By definition,
\begin{equation*}
\nabla_{V_{i}}V_{j}=\Gamma_{ij}^{1}V_{1}+\Gamma_{ij}^{2}V_{2}
\end{equation*}
and the only nonzero Christoffel symbols of the metric are $%
\Gamma_{12}^{2}=\Gamma_{21}^{2}=\cot\varphi$, $\Gamma_{22}^{1}=-\sin\varphi%
\cos\varphi$. Consequently,
\begin{equation}
\ddot{\gamma}^{\ast}=(\ddot{\varphi}-\dot{\theta}^{2}\sin\varphi\cos \varphi)%
\frac{\partial}{\partial\varphi}+(\ddot{\theta}+2\dot{\varphi}\dot{\theta}%
\cot\varphi)\frac{\partial}{\partial\theta}   \label{3-41}
\end{equation}

Now, take the above sphere and curve to be the shape space $M^{\ast}$ and a
shape curve $\gamma^{\ast}(t)$, respectively. Then it follows immediately
from (\ref{C2}) and (\ref{3-41}) that the equations (ii) and (iii) of the
system (\ref{3-1}) can be expressed neatly as the following coordinate-free
vector equation on the 2-sphere,
\begin{equation}
\ddot{\gamma}^{\ast}+(2\frac{\dot{\rho}}{\rho})\dot{\gamma}^{\ast}-\frac {4}{%
\rho^{3}}\nabla U^{\ast}=0   \label{3-42}
\end{equation}
Here the scaling variable $\rho=\sqrt{I}$ of the cone $\bar{M}$ plays the
role of an auxiliary function which couples equation (\ref{3-42}) to
equation (i) of (\ref{3-1}). The latter is the Lagrange-Jacobi equation (\ref%
{L-J}), which for a given shape curve $\gamma^{\ast}(t)$ is a second order
differential equation purely for $\rho(t)$. Another interpretation of the
coefficient of the velocity in (\ref{3-42}) follows from (\ref{3-10}),
namely we have
\begin{equation}
2\frac{\dot{\rho}}{\rho}=v\cot\alpha   \label{P}
\end{equation}

\begin{remark}
In fact, the differential equation (\ref{3-42}) can be completely decoupled
from the function $\rho(t)$ and hence it really becomes a differential
equation on the 2-sphere, as summarized in Theorem \ref{shape1} below.
\end{remark}

\subsubsection{Separation of the scaling function $\protect\rho(t)$ from the
shape space coordinates}

The system (\ref{3-8a})-(\ref{3-8b}) characterizes the moduli curves of
3-body trajectories with zero angular momentum, expressed in the language of
kinematic geometry and, in particular, for that reason the natural parameter
is the arc-length $\bar{s}$ of the moduli curve or the arc-length $\sigma$
of its image shape curve on the sphere of radius $1/2$. Although the
original mechanical system (\ref{1-1}) naturally involves the physical
parameter of time $t$, the latter is infinitesimally related to $\bar{s}$ by
the identity

\begin{equation}
\frac{d\bar{s}}{dt}=\sqrt{2T}=\sqrt{2}\sqrt{h+U}   \label{3-9}
\end{equation}
and this enables us to express the system (\ref{3-8a})-(\ref{3-8b}) in terms
of $t$ and hence combine it directly with the other system (\ref{3-1}).

To this end, let us consider a shape curve $\gamma^{\ast}(t)$ on the sphere $%
S^{2}(1/2)$. By (\ref{3-10}),
\begin{equation}
\sin\alpha=\rho\frac{d\sigma}{dt}/\frac{d\bar{s}}{dt}=\frac{\rho}{\sqrt{2}}%
(h+\frac{U^{\ast}}{\rho})^{-1/2}\left\vert \frac{d\gamma^{\ast}}{dt}%
\right\vert   \label{3-12}
\end{equation}
and then equation (\ref{3-8b}) becomes
\begin{equation}
\frac{\rho^{2}}{2}\left\vert \frac{d\gamma^{\ast}}{dt}\right\vert ^{2}%
\mathcal{K}^{\ast}=\frac{1}{2\rho}\frac{d}{d\mathbf{\nu}^{\ast}}U^{\ast}
\label{3-13}
\end{equation}
The latter is not only considerably simpler than equation (\ref{3-8b}), but
it also provides a simple formula to compute $\rho(t)=\rho(\bar{\gamma}(t))$
in terms of the geometry of the shape curve, namely
\begin{equation}
\rho^{3}=\frac{\frac{d}{d\mathbf{\nu}^{\ast}}U^{\ast}}{\mathcal{K}^{\ast
}\left\vert \frac{d\gamma^{\ast}}{dt}\right\vert ^{2}}   \label{3-14}
\end{equation}
In view of the integral formula (\ref{3-11}) this is, indeed, a pleasant
surprise which, in one stroke, shows how to reconstruct the moduli curve $%
\bar{\gamma}(t)$ from the shape curve $\gamma^{\ast}(t)$ by the simple
formula (\ref{3-14}).

The expression on the right hand side of (\ref{3-14}) refers to the
kinematic geometry with $M^{\ast}=S^{2}(1/2)$, and the whole product on this
side would change by the factor $4r^{2}$ if we had worked in the sphere $%
S^{2}(r)$. Henceforth, we shall return to the unit sphere $S^{2}(1)$, and
with the notation for speed, curvature and normal derivative from Section
3.4.1 we can restate (\ref{3-14}) in the following way :

\begin{lemma}
\label{split}Let $\mathfrak{S}$ be the Siegel function (\ref{C10}) of $%
(\gamma^{\ast},U^{\ast})$, which relates the intrinsic geometry of $%
\gamma^{\ast}$ with the gradient field $\nabla U^{\ast}$ on the unit
2-sphere. If $\bar{\gamma}(t)=(\rho(t),\gamma^{\ast}(t))$ is the
time-parametrized moduli curve of a three-body motion with zero angular
momentum, then the speed $v(t)$ of $\gamma^{\ast}(t)$ is related to $\rho(t)$
and $\mathfrak{S}(\gamma^{\ast}(t))$ by the identity $\mathcal{\ }$
\begin{equation}
\rho^{3}=\frac{4}{v^{2}}\mathfrak{S}\text{ \ \ \ \ or \ \ }\mathfrak{S}=%
\frac{1}{4}\rho^{3}v^{2}   \label{3-16}
\end{equation}
In particular, $\mathfrak{S}$ is always nonnegative!
\end{lemma}

Now, returning to equation (\ref{3-42}) we set
\begin{align}
P & =2\frac{\dot{\rho}}{\rho}=\frac{2}{3}\frac{\mathfrak{\dot{S}}}{\mathfrak{%
S}}-\frac{4}{3}\frac{\dot{v}}{v}  \label{3-17a} \\
Q & =-\frac{4}{\rho^{3}}=-\frac{v^{2}}{\mathfrak{S}}   \label{3-17b}
\end{align}
where the rightmost identity in (\ref{3-17a}) follows from (logarithmic)
differentiation of the identity (\ref{3-16}). Then we can state the
following result :

\begin{theorem}
\label{shape1}For 3-body motions with zero angular momentum and fixed total
energy, the associated shape curves $\gamma^{\ast}(t)$ on the unit 2-sphere
are characterized by the ODE
\begin{equation}
\frac{d^{2}}{dt^{2}}\gamma^{\ast}+P\frac{d}{dt}\gamma^{\ast}+Q\nabla U^{\ast
}=0   \label{3-43}
\end{equation}
where the first term is the (covariant) acceleration and the coefficients
are the functions $P,Q$ defined by (\ref{3-17a}), (\ref{3-17b}), which can
be expressed purely in terms of $\varphi$, $\theta$ and their derivatives up
to order 3.
\end{theorem}

\begin{remark}
The formula (\ref{3-16}), or equivalently
\begin{equation}
K^{\ast}=\frac{4U_{\nu}^{\ast}}{v^{2}\rho^{3}}   \label{K}
\end{equation}
expresses the geodesic curvature of the shape curve in terms of $\rho
,\varphi,\theta,\dot{\varphi},\dot{\theta}$. Therefore, since it only
involves first order derivatives in the moduli space, it is not surprising
to find that the same formula can, indeed, be derived more directly from the
general spherical curvature formula (\ref{C6}) by elimination of the second
order derivatives using equations (ii), (iii) in (\ref{3-1}).
\end{remark}

\subsubsection{Regular and irregular points and exceptional shape curves}

The formula for $\rho^{3}$ in (\ref{3-16}) involves the three quantities $%
v,U_{\nu}^{\ast}$ and $K^{\ast}$ and the product may become indefinite when
some of them vanish, namely we note the following implications
\begin{align}
\left[ v=0\right] & \Longrightarrow\left[ U_{\nu}^{\ast}=0\right]
\Longleftarrow\left[ K^{\ast}=0\right]  \label{implications} \\
\left[ U_{\nu}^{\ast}=0\right] & \Longrightarrow\left[ v=0\right] \text{ \
or }\left[ K^{\ast}=0\right]  \notag
\end{align}
It is worthwhile having a closer look at the geometric interpretation and
behavior of the shape curve due to the vanishing of any of these numbers,
and accordingly we shall make some definitions to distinguish the various
cases.

Formula (\ref{3-16}) expresses the Siegel function $\mathfrak{S}$ of $%
\gamma^{\ast}$ in two different ways, namely at a point $P_{0}=\gamma^{\ast
}(t_{0})$ its value is
\begin{equation}
\lim_{t\rightarrow t_{0}}\frac{1}{4}\rho^{3}v^{2}=\mathfrak{S}%
_{0}=\lim_{t\rightarrow t_{0}}\frac{U_{\nu}^{\ast}}{K^{\ast}}   \label{2lim}
\end{equation}
whenever any of the limits are defined, including $+\infty$ as a limiting
value.

\begin{definition}
\label{regular}$P_{0}$ is a \emph{regular }point if $0<\mathfrak{S}%
_{0}<\infty$, and otherwise it is \emph{irregular. }
\end{definition}

The irregular points consist of the \emph{cusps}, the \emph{collision} \emph{%
points} (binary or triple), and for completeness we also include \emph{%
escape points :} \emph{\ }
\begin{align}
\text{ \ \ cusp at }P_{0} & :\text{\ }\mathfrak{S}_{0}=0\text{, \ }v_{0}=0,%
\text{ }\nabla U^{\ast}(P_{0})\neq0  \notag \\
\text{triple collision at }P_{0} & :\mathfrak{S}_{0}=0\text{, \ }\rho _{0}=0%
\text{, \ }\nabla U^{\ast}(P_{0})=0  \label{reg/irreg} \\
\text{binary collision at }P_{0} & :\mathfrak{S}_{0}=\infty\text{, \ }%
v_{0}=\infty,\text{ }P_{0}=\mathbf{b}_{i}  \notag \\
\text{escape at }P_{0} & :\mathfrak{S}_{0}=?,\text{ \ \ }\rho_{0}=\infty,%
\text{ \ }v_{0}=0\   \notag
\end{align}
In Section 6 we shall return to triple collisions, but binary collisions and
escape to infinity behavior will not be a topic in this paper.

The simplest curves on the 2-sphere are the geodesic circles, characterized
by $K^{\ast}=0$ at each point. If such a curve is the shape curve of a
three-body motion, then it follows from (\ref{2lim}) that $U_{\nu}^{\ast}$
also vanishes. In fact, $\gamma^{\ast}$ coincides with a gradient line
segment if and only if it is a geodesic.

\begin{definition}
\label{exceptional}The shape curve $\gamma^{\ast}$ is called \emph{%
exceptional} if it is confined to a gradient line (or a geodesic circle), or
it consists of a single point.
\end{definition}

Clearly, a single point shape curve must be a fixpoint of the gradient flow,
see Remark \ref{shape}. Thus, apart from the exceptional shape curves, being
a regular or irregular point is an intrinsic property, that is, it depends
only on the geometric curve.

\begin{remark}
\label{exc}We omit the proof here, so we rather claim that the only
exceptional shape curves (of length \TEXTsymbol{>}0) are those representing
collinear motions or \emph{isosceles} \emph{triangle} motions. Their crucial
property is that the shape curve lies on a circle fixed by an isometry
(reflection) of the sphere which leaves $U^{\ast}$ invariant. It also
follows that a non-exceptional shape curve $\gamma^{\ast}$ can only
intersect an exceptional curve transversely, that is, neither tangentially
nor with zero speed.
\end{remark}

The equator circle $E^{\ast}$ represents the collinear motions, of course,
but isosceles motions exist only for special mass distributions, as follows.
An isosceles m-triangle has (at least) two equal masses, say $m_{1}=m_{2}$,
and the mass $m_{3}$ lies on the symmetry axis of the triangle. Their shapes
constitute the meridian through the north pole, the Euler point $\mathbf{e}%
_{3}$ and its antipodal point $-\mathbf{e}_{3}=\mathbf{b}_{3}$. The latter
point represents the collision of the two symmetric mass points somewhere on
the symmetry axis. Thus, an isosceles triangle motion arises when the
initial position and velocity have the above isosceles symmetry.

Henceforth, we shall assume the shape curve $\gamma^{\ast}$ is not of
exceptional type, unless otherwise stated. Consider the power series
expansions

\begin{equation*}
K^{\ast}=\sum\limits_{i=0}^{\infty}K_{i}s^{i},\text{ \ \ \ }U_{\nu}^{\ast
}=\sum\limits_{i=0}^{\infty}\omega_{i}s^{i}\text{\ }
\end{equation*}
at a regular point $P_{0}=\gamma^{\ast}(t_{0})$, where $s$ is the arc-length
measured from $P_{0}$. The value $\mathfrak{S}_{0}$ of $\mathfrak{S}$ at $s=0
$ can be calculated in two ways, possibly by the aid of l'Hospitals rule,%
\begin{equation}
\frac{1}{4}\rho_{0}^{3}v_{0}^{2}=\mathfrak{S}_{0}=\lim_{s\rightarrow0}\frac{%
U_{\nu}^{\ast}}{K^{\ast}}=\frac{\lim\frac{d}{ds}U_{\nu}^{\ast}}{\lim\frac{d}{%
ds}K^{\ast}}=...=\frac{\lim\frac{d^{k}}{ds^{k}}U_{\nu}^{\ast}}{\lim\frac{%
d^{k}}{ds^{k}}K^{\ast}}=\frac{\omega_{k}}{K_{k}}>0   \label{lim}
\end{equation}
where $k\geq0$ is the smallest integer such that $K_{k}\neq0$. Then $%
K_{i}=\omega_{i}=0$ for $i<k$, $\omega_{k}\neq0$, and we say $P_{0}$ is a
\emph{regular point} of order $k$. It is a finite number since otherwise $%
U_{\nu}^{\ast}$ and $K^{\ast}$ vanishes identically and $\gamma^{\ast}$
would be exceptional. In particular, we see that $\omega_{0}\neq0$ means $%
\gamma^{\ast}(t)$ is transversal to the gradient flow at $t=t_{0}.$

The \emph{order} $k$ of a cusp at $P_{0}$ $=(\varphi_{0},\theta_{0})$ can be
defined similarly by considering the limits in (\ref{lim}). The only
difference is that $\omega_{k}=0$ and $K_{k}\neq0$ at the last step. Cusps
arise when the moduli curve $\bar{\gamma}(t)$ in $\bar{M}$ \ is tangent to
the ray at the point $(\rho_{0},P_{0})$ and hence the projected curve $%
\gamma^{\ast}(t)$ on the 2-sphere "halts" at $t=t_{0}$. Geometrically, the
curve $\gamma^{\ast}$ near $P_{0}$ is a cusp consisting of two diverging
branches which emanate from $P_{0}$, both with the initial direction of $%
\nabla U^{\ast}(P_{0})$ and the initial curvature
\begin{equation}
K_{0}=\frac{1}{3}\frac{d}{d\mathbf{\nu}}\ln\left\vert \nabla
U^{\ast}(P_{0})\right\vert   \label{K0cusp}
\end{equation}
In Remark \ref{cusp} we further describe the totality of cusps at $P_{0}$,
at a fixed energy level $h$, as a specific family of curves $\gamma^{\ast}$
parametrized by a number $c>0$ (resp. $c\geq0$) for $h\geq0$ (resp. $h<0)$.
In the case $c=0$ the two branches coincide completely and $\bar{\gamma}%
(t_{0})$ $=(\rho_{0},P_{0})$ lies on the Hill's surface $\partial\bar{M}_{h}$%
, cf. (\ref{Hills}).

\subsubsection{On the initial value problem for the moduli curve and the
shape curve}

In the spherical coordinates $(\rho,\varphi,\theta)$ on $\bar{M}\simeq%
\mathbb{R}^{3}$, the time parametrized moduli curve and its associated shape
curve are simply related by%
\begin{equation*}
\bar{\gamma}(t)=(\rho(t),\gamma^{\ast}(t)),\ \ \ \
\gamma^{\ast}(t)=(\varphi(t),\theta(t)),
\end{equation*}
and conversely, formula (\ref{3-16}) is the key to the lifting procedure,
namely the reconstruction of $\bar{\gamma}(t)$ from its projection on the
2-sphere.

We choose an initial point $P_{0}=(\varphi_{0},\theta_{0})$ on the shape
curve, and assume (for simplicity) it is regular in the sense of Definition %
\ref{regular}. In particular, the (local) lifting procedure is well defined,
and the corresponding initial value problems for the ODE's (\ref{3-1}) and (%
\ref{3-43}) respectively, are equivalent. Thus, on the one hand, the
solution $\bar{\gamma}(t)$ is uniquely determined by its initial position
and velocity
\begin{equation}
\bar{\gamma}(t_{0})=(\rho_{0},\varphi_{0},\theta_{0})\text{, \ \ }\frac{d}{dt%
}\bar{\gamma}(t_{0})=(\rho_{1},\varphi_{1},\theta_{1}),   \label{3-44}
\end{equation}
but on the other hand it is also determined by the corresponding initial
data
\begin{equation}
(\varphi_{0},\theta_{0}),(\varphi_{1},\theta_{1}),(\varphi_{2},\theta _{2}),%
\text{ \ \ where }\varphi_{2}=\frac{1}{2}\ddot{\varphi}|_{t_{0}},\text{\ }%
\theta_{2}=\frac{1}{2}\ddot{\theta}|_{t_{0}},   \label{3-44b}
\end{equation}
at the shape space level.

However, the above chosen initial data determine an energy level $h$, and
conversely, if $h$ is already given, the initial data (\ref{3-44}) or (\ref%
{3-44b}) are acceptable only if the resulting energy level is $h$. To make
this more transparent, let us rather state the initial value problem in the
moduli space as a system with 4 equations :%
\begin{align}
(i)\text{ \ }0 & =\ddot{\rho}+\frac{\dot{\rho}^{2}}{\rho}-\frac{1}{\rho }(%
\frac{1}{\rho}U^{\ast}+2h)  \notag \\
(ii)\text{ \ }0 & =\text{\ }\ddot{\varphi}+2\frac{\dot{\rho}}{\rho}\dot{%
\varphi}-\frac{1}{2}\sin(2\varphi)\dot{\theta}^{2}-\frac{4}{\rho^{3}}%
U_{\varphi}^{\ast}  \label{3-45} \\
(iii)\text{ \ }0 & =\ddot{\theta}+2\frac{\dot{\rho}}{\rho}\dot{\theta}%
+2\cot(\varphi)\dot{\varphi}\dot{\theta}-\frac{4}{\rho^{3}}\frac{1}{\sin
^{2}\varphi}U_{\theta}^{\ast}  \notag \\
(iv)\text{ \ }0 & =\frac{1}{2}\dot{\rho}^{2}+\frac{\rho^{2}}{8}(\dot {\varphi%
}^{2}+\sin^{2}\varphi\text{ }\dot{\theta}^{2})-\frac{1}{\rho}U^{\ast
}(\varphi,\theta)-h  \notag
\end{align}
where the first order equation (iv) is the energy integral (\ref{h}) for the
fixed value $h$. In fact, any one of the four equations is redundant and can
be deduced from the other three. For example, with $h$ calculated from the
initial data (\ref{3-44}), the solution of (i)-(iii) also satisfies (iv).

More geometrically, if the initial data set (\ref{3-44}) is a point on a
specific level surface of type (iv) in the tangent bundle of $\bar{M}$, then
the solution $\bar{\gamma}(t)$ of (i)-(iii) in (\ref{3-45}) must lie on the
surface (iv) for all $t.$ But we can also determine the same solution $\bar{%
\gamma}(t)$ from (ii)-(iv). All this amounts to saying that for a given
value of $h$ the whole system (\ref{3-45}) is of total order $5.$

On the other hand, the ODE (\ref{3-43}) on the 2-sphere is independent of $h$
and its total order is 6. Its solutions are the projections $\gamma^{\ast}$
of all solutions of the system (\ref{3-45}) for any value of $h$. In
general, they are divided into three disjoint classes, distinguished by the
sign of $h$ (positive, zero or negative). The exceptions to this subdivision
are precisely the exceptional shape curves (cf. Definition \ref{exceptional}%
), which can represent three-body motions at any energy level. See Section
4.2.

Now, let us consider the problem of how to translate the initial data (\ref%
{3-44}), at a given energy level $h$, into a set of suitable initial data
consisting of five numbers depending only on the shape curve. A natural
first choice would be
\begin{equation}
\left( \varphi_{0},\theta_{0}\right) ,(\psi_{0},v_{0}),\mathfrak{S}_{0}
\label{data}
\end{equation}
where the angle $\psi_{0}\in\lbrack0,2\pi)$ specifies the initial direction
of $\gamma^{\ast}$, $v_{0}>0$ is the inital speed, and $\mathfrak{S}%
_{0}=(U_{\nu }^{\ast}/K^{\ast})_{0}>0$ is the initial value of the Siegel
function.

\begin{remark}
The generic points of $\gamma^{\ast}$ are the regular points of order $k=0$,
and hence $\gamma^{\ast}$ is transversal to the gradient flow almost
everywhere. For such points it is, perhaps, more natural to replace $%
\mathfrak{S}_{0}$ by $K_{0}$ in (\ref{data}). In fact, the two choices -
either $K_{0}$ or $\mathfrak{S}_{0}$ - are equivalent since the direction $%
\psi_{0}$ determines $\omega_{0}$ when $k=0$.
\end{remark}

The pairs $(\psi_{0},v_{0})$ and ($\varphi_{1},\theta_{1})$ evidently
determine each other, namely they specify the initial velocity of $%
\gamma^{\ast}$. Clearly, the 5-tuple (\ref{data}) is merely the data (\ref%
{3-44}) with the pair ($\rho_{0},\rho_{1})$ replaced by the single number $%
\mathfrak{S}_{0}$, which is a second order quantity at the shape space
level. Indeed, we recover $\rho_{0}$ immediately from (\ref{data}) by making
use of (\ref{lim}).

Thus, it is clear that the initial data information given by (\ref{3-44}) or
(\ref{data}) would be equivalent if we could also recover the radial speed $%
\rho_{1}$ from (\ref{data}). However, the energy integral (iv) of (\ref{3-45}%
) determines only $\rho_{1}^{2}$; in fact, there is no solution at all if $h$
is below the critical value%
\begin{equation}
h_{\min}=\frac{1}{8}\rho_{0}^{2}v_{0}^{2}-\frac{1}{\rho_{0}}%
U^{\ast}(\varphi_{0},\theta_{0})\text{, \ \ where\ }\rho_{0}=(\frac{4%
\mathfrak{S}_{0}}{v_{0}^{2}})^{1/3}   \label{hmin}
\end{equation}
There is a unique solution if $h=h_{\min}$, with $\rho_{1}=0$, and for $%
h>h_{\min}$ there are two solutions which are distinguished by the sign of $%
\rho_{1}$.

A slightly different approach is to combine the identities (\ref{P}) and (%
\ref{3-12}), which yields a value of $\sin\alpha$, where $0<\alpha<\pi$. In
fact, $\alpha<\pi/2$ means $\rho_{1}<0$ and $\alpha>\pi/2$ means $\rho_{1}>0$%
. But, we are still left with the problem of how to determine the sign of $%
\rho_{1}$ from the initial data (\ref{data}). Anyhow, up to now we have the
following result as a summary of the above local analysis at a regular point.

\begin{proposition}
\label{initalspeed copy(1)}Consider the three-body motions with zero angular
momentum and a given total energy $h$, whose oriented shape curve at a given
regular point have the same initial direction, speed, and curvature (or
Siegel number $\mathfrak{S}_{0}$, if the curvature $K_{0}$ vanishes). Then
the number of solutions is, up to congruence, equal to $0,1$ or $2$
depending on whether $h<h_{\min}$ (resp. $h=h_{\min}$ or $h>h_{\min})$,
where $h_{\min}$ is calculated from the given initial data by the formula (%
\ref{hmin}).
\end{proposition}

A three-body motion is said to be \emph{expanding }when $\dot{\rho}(t)>0$,
and it is \emph{contracting }when $\dot{\rho}(t)<0$. Accordingly, we say the
\emph{expansion index} at time $t_{0}$ is the sign $e(t_{0})=0,\pm1$ of $%
\dot{\rho}(t_{0})=\rho_{1}.$

\begin{corollary}
\label{initalspeed}A three-body motion $\gamma(t)$ with zero angular
momentum and a given total energy $h$ is uniquely determined up to
congruence by the oriented shape curve $\gamma^{\ast}$ $\subset S^{2}$ as a
subset (i.e. non-parametrized), together with the initial speed $v_{0}$ and
expansion index $e(t_{0})$ at a regular point $\gamma^{\ast}(t_{0})$.
\end{corollary}

\begin{remark}
\label{cusp}The above corollary is, in fact, also true when the point $%
P_{0}=\gamma^{\ast}(t_{0})$ is a cusp. Indeed, for each fixed $h$, the
family of possible cusps at $P_{0}$ is parametrized by the nonnegative
numbers
\begin{equation}
c=\rho_{1}^{2}=2(\frac{U^{\ast}(P_{0})}{\rho_{0}}+h)   \label{c}
\end{equation}
as follows from the energy integral (iv) in (\ref{3-45}).

For $c=0$ there is a unique moduli curve $\bar{\gamma}(t)$ starting out from
the "rest point" $(\rho_{0},P_{0})$ on the Hill's boundary $\partial\bar
{M}%
_{h}$, and the corresponding shape curve $\gamma^{\ast}$ is the simple cusp
(i.e. with one branch) emanating from $P_{0}$. In general, the local
geometry of $\gamma^{\ast}$ at $P_{0}$ determines the number $c$. In fact,
one can determine $c$ from the first curvature coefficients $K_{i},i>0$, but
note that $K_{0}$ is independent of $c$, by (\ref{K0cusp}). Finally, one
solves the initial value problem in $\bar{M}$ at $(\rho_{0},P_{0})$, with
the velocity component $\rho_{1}=\pm\sqrt{c}$ selected according to our
choice of $e(t_{0})$.
\end{remark}

For a given non-parametrized curve $\gamma^{\ast}$ in $M^{\ast}$, the
geodesic curves of the associated cone surface $C(\gamma^{\ast})_{h}$ are
characterized by the ODE of (\ref{3-8a}). The squared speed of such a curve
is, of course, given by $2T=2(U+h)$ and hence it is determined by the
position. A geodesic $\bar{\gamma}$ is therefore uniquely determined by the
initial position and direction at $t=t_{0}$, namely $\rho(t_{0})$, $%
\alpha(t_{0})$ and the point $\gamma^{\ast}(t_{0})$. By (\ref{3-14}) or (\ref%
{3-16}), and elimination of the speed in the shape space using (\ref{3-12}),
it follows that the validity of the identity
\begin{equation}
\sin^{2}\alpha=\frac{2\mathfrak{S}}{\rho h+U^{\ast}}   \label{3-15}
\end{equation}
along the entire curve $\bar{\gamma}$, where $\mathfrak{S}$ is the Siegel
function of $\gamma^{\ast}$ in $S^{2}(1)$, is a necessary and sufficient
condition for a geodesic of the cone surface\ to be a geodesic curve of $%
\bar{M}_{h}$ as well. This proves the following statement :

\begin{corollary}
A non-parametrized (i.e. geometric) curve $\gamma^{\ast}\subset S^{2}(1)$
can be suitably parametrized as the shape curve of a three-body motion with
zero angular momentum and total energy $h$ if and only if its cone surface $%
C(\gamma^{\ast})_{h}$ has a geodesic curve satisfying (\ref{3-15}) along the
entire curve.
\end{corollary}

\begin{problem}
\label{param}As indicated by the condition (\ref{3-15}), only a very special
kind of geometric curves on the 2-sphere $S^{2}$ can be suitably
parametrized as the shape curve of a three-body motion as above. How can
they be characterized in a neat way? Can such a curve have different time
parametrizations as the shape curve of three-body motions?
\end{problem}

\begin{problem}
\label{index}Is the initial speed and expansion index also determined by the
geometric shape curve in Corollary \ref{initalspeed}?\ (The case $h=0$ turns
out to be special.)
\end{problem}

\section{On the analysis of moduli and shape curves via power series}

We continue to use the notation and terminology from Section 3. Consider a
time parametrized moduli curve $t\rightarrow\bar{\gamma}(t)$ in $\bar{M}$ $%
\simeq\mathbb{R}^{3}$ which represents a three-body motion with vanishing
angular momentum, and let $t\rightarrow\gamma^{\ast}(t)$ be the associated
shape curve, namely its projection in the 2-sphere $S^{2}$. In this section
we shall investigate the possibility of reconstructing the parametrized
curve $\bar{\gamma}(t)$ solely from the oriented geometric shape curve.
Moreover, there is the question of how much geometric information about the
curve $\gamma^{\ast}$ is really needed for such a lifting procedure. At the
end we shall also answer the question concerning the uniqueness of the time
parametrization.

\subsection{Generation of recursive relations and intrinsic geometric
invariants}

In the local analysis of the moduli and the shape curve, and their
interaction with the potential function $U^{\ast}$, we shall distinguish
between two types of variables or quantities. Namely, on the one hand there
are the \emph{intrinsic} quantities which depend only on $\gamma^{\ast}$ as
an oriented geometric (i.e. unparametrized) curve and $U^{\ast}$ as a
function on $S^{2}$, and on the other hand there are the \emph{variable}
quantities, defined along the curve $\bar{\gamma}$ or $\gamma^{\ast}$, which
depend on the scaling function $\rho$ in the moduli space $\bar{M}$ or the
time parametrization of the curves. The basic intrinsic quantities are the
gradient field $\nabla U^{\ast}$ (or the tangential and normal derivatives $%
U_{\tau }^{\ast}$, $U_{\nu}^{\ast}$), the unit tangent field of $%
\gamma^{\ast}$, and the geodesic curvature function $K^{\ast}$ of $%
\gamma^{\ast}$. Moreover, we shall assume that $\gamma^{\ast}$ is not
exceptional and hence the linkage between $\gamma^{\ast}$ and $U^{\ast}$ is
also neatly encoded into the intrinsic Siegel function $\mathfrak{S}%
=U_{\nu}^{\ast}/K^{\ast}$, see (\ref{C10}) and Section 3.4.4.

Let $s$ be the arc-length parameter of $\gamma^{\ast}$ measured in the
positive direction from a given regular point $P_{0}$ of order $k\geq0$.
Then the coefficients of the power series expansions of the above functions,
such as $\ $
\begin{align}
K^{\ast} & =K_{0}+K_{1}s+K_{2}s^{2}+...  \notag \\
U^{\ast} & =u_{0}+\bar{u}_{1}s+\bar{u}_{2}s^{2}+...  \label{starseries} \\
U_{\nu}^{\ast} & =\omega_{0}+\omega_{1}s+\omega_{2}s^{2}+....  \notag \\
\mathfrak{S} & =\mathfrak{S}_{0}+\mathfrak{S}_{1}s+\mathfrak{S}_{2}s^{2}+...
\notag
\end{align}
yield intrinsic quantities localized at the point $P_{0}.$ Note the
expansion of the tangential derivative of $U^{\ast}$ is
\begin{equation}
U_{\tau}^{\ast}=\nabla U^{\ast}\cdot\mathbf{\tau}=\frac{d}{ds}U^{\ast}=\bar {%
u}_{1}+2\bar{u}_{2}s+3\bar{u}_{3}s^{2}+...   \label{tangderiv}
\end{equation}
and the coefficients $\mathfrak{S}_{n}$ are expressible as rational
functions of $K_{i}$ and $\omega_{i},k\leq i\leq n+k.$ Let us say the \emph{%
order} of a coefficient is the highest order of derivatives of local
coordinates in its expression. Thus, the pair $(\varphi_{0},\theta_{0})$ and
$u_{0}$ are the intrinsic (geometric) data of order 0 at $P_{0}$. Next, the
triple $\bar
{u}_{1}$, $\omega_{0}$, and the unit tangent vector at $P_{0}$
represent the intrinsic data of order $1$ at $P_{0}$, and $\omega_{n}$ and $%
\bar{u}_{n+1}$ (resp. $K_{n}$ and $\mathfrak{S}_{n}$) has order $n+1$ (resp.
$n+2)$.

We choose a spherical polar coordinate system $(\varphi,\theta)$, with $P_{0}
$ different from any of the "poles" $\varphi=0$ or $\pi$, and for a given
moduli curve $\bar{\gamma}(t)=(\rho(t),\varphi(t),\theta(t))$ we shall
expand the coordinate functions, as well as $U^{\ast}$ and its partial
derivatives, as power series with respect to $t:$
\begin{align}
\rho & =\rho_{0}+\rho_{1}t+\rho_{2}t^{2}+\rho_{3}t^{3}+....  \notag \\
\varphi & =\varphi_{0}+\varphi_{1}t+\varphi_{2}t^{2}+\varphi_{3}t^{3}+....
\notag \\
\theta & =\theta_{0}+\theta_{1}t+\theta_{2}t^{2}+\theta_{3}t^{3}+.....
\notag \\
v & =v_{0}+v_{1}+v_{2}t+v_{2}t^{2}+.....  \label{expansions} \\
U^{\ast} & =u_{0}+u_{1}t+u_{2}t^{2}+u_{3}t^{3}+....  \notag \\
U_{\varphi}^{\ast} & =\mu_{0}+\mu_{1}t+\mu_{2}t^{2}+\mu_{3}t^{3}+....  \notag
\\
U_{\theta}^{\ast} & =\eta_{0}+\eta_{1}t+\eta_{2}t^{2}+\eta_{3}t^{3}+...
\notag
\end{align}
For convenience, we also write%
\begin{align}
\sin(2\varphi) & =f_{0}+f_{1}t+f_{2}t^{2}+....  \notag \\
\sin^{2}(\varphi) & =g_{0}+g_{1}t+g_{2}t^{2}+....  \notag
\end{align}
and list some of the initial coefficiens :%
\begin{align}
u_{0} & =U^{\ast}(\varphi_{0},\theta_{0})\text{, \ \ \ }u_{1}=\mu_{0}%
\varphi_{1}+\eta_{0}\theta_{1}\text{, etc.}  \notag \\
f_{0} & =\sin(2\varphi_{0}),\text{ }f_{1}=2\cos(2\varphi_{0})\varphi _{1}%
\text{, etc.}  \label{coeff} \\
\text{\ }g_{0} & =\sin^{2}(\varphi_{0}),\text{ \ }g_{1}=f_{0}\varphi _{1}%
\text{, etc.}  \notag \\
v_{1} & =\frac{1}{v_{0}}[2\varphi_{1}\varphi_{2}+\sin(\varphi_{0})\cos(%
\varphi_{0})\varphi_{1}\theta_{1}^{2}+2\sin^{2}(\varphi_{0})\theta
_{1}\theta_{2}]  \notag
\end{align}
where the expression for $v_{1}$ follows from (\ref{C9}). We shall regard $%
\mu_{0}$, $\eta_{0}$ as intrinsic data, but they depend on the coordinate
system, of course.

Below we shall investigate dependence relations among the coefficients $%
\rho_{i}$, $\varphi_{j}$, $\theta_{k}$ of the coordinate functions in (\ref%
{expansions}) and various other coefficients. Some of them are directly
expressible in terms of the intrinsic data and hence regarded as constants,
whereas the others are the \emph{variables}.

\begin{definition}
\label{vari}The following list of coefficients from (\ref{expansions})
\begin{equation}
\rho_{0},v_{0};\rho_{1},\varphi_{1},\theta_{1};\rho_{2},\varphi_{2},%
\theta_{2}   \label{8variables}
\end{equation}
will be referred to as the variables of order $\leq2$. The variables of
order $n$ are $\rho_{n},\varphi_{n},\theta_{n}$ when $n>0$, and $%
\rho_{0},v_{0}$ are the only variables of order zero.
\end{definition}

Henceforth, assume the above moduli curve $\bar{\gamma}(t)$ is a solution of
the ODE system (\ref{3-45}). By inserting the power series into the
equations (i)-(iv) of (\ref{3-45}) and applying the method of undetermined
coefficients, we arrive at the following scheme of recursive relations for
the variables of increasing order $0,1,2..:$\qquad%
\begin{align}
E_{10} & :0=2\rho_{0}^{2}\rho_{2}+\rho_{0}\rho_{1}^{2}-2h\rho_{0}-u_{0}
\notag \\
E_{20} & :0=2\rho_{0}^{3}\varphi_{2}+2\rho_{0}^{2}\rho_{1}\varphi_{1}-\frac{1%
}{2}\rho_{0}^{3}f_{0}\theta_{1}^{2}-4\mu_{0}  \label{E0} \\
E_{30} & :0=2g_{0}\rho_{0}^{3}\theta_{2}+2g_{0}\rho_{0}^{2}\rho_{1}\theta
_{1}+\rho_{0}^{3}f_{0}\varphi_{1}\theta_{1}-4\eta_{0}  \notag \\
E_{40} & :0=\rho_{0}\rho_{1}^{2}+\frac{1}{4}\rho_{0}^{3}(%
\varphi_{1}^{2}+g_{0}\theta_{1}^{2})-2u_{0}-2h\rho_{0}  \notag
\end{align}%
\begin{align}
E_{11} &
:0=6\rho_{0}^{2}\rho_{3}+8\rho_{0}\rho_{1}\rho_{2}+\rho_{1}^{3}-2h%
\rho_{1}-u_{1}  \notag \\
E_{21} &
:0=6\rho_{0}^{3}\varphi_{3}+10\rho_{0}^{2}\rho_{1}\varphi_{2}+4(\rho_{0}^{2}%
\rho_{2}+\rho_{0}\rho_{1}^{2})\varphi_{1}-2f_{0}\rho_{0}^{3}\theta_{1}%
\theta_{2}  \notag \\
& -\frac{1}{2}(f_{1}\rho_{0}^{3}+3f_{0}\rho_{0}^{2}\rho_{1})\theta_{1}^{2}-4%
\mu_{1}  \label{E1} \\
E_{31} &
:0=6g_{0}\rho_{0}^{3}\theta_{3}+(10g_{0}\rho_{0}^{2}\rho_{1}+2g_{1}%
\rho_{0}^{3}+2f_{0}\rho_{0}^{3}\varphi_{1})\theta_{2}+2f_{0}\rho
_{0}^{3}\theta_{1}\varphi_{2}+  \notag \\
&
(f_{1}\rho_{0}^{3}+3f_{0}\rho_{0}^{2}\rho_{1})\varphi_{1}\theta_{1}+(4g_{0}%
\rho_{0}^{2}\rho_{2}+4g_{0}\rho_{0}\rho_{1}^{2}+2g_{1}\rho_{0}^{2}\rho_{1})%
\theta_{1}-4\eta_{1}  \notag
\end{align}
and in general%
\begin{align}
E_{1n} & :0=(n+2)(n+1)\rho_{0}^{2}\rho_{n+2}+.....  \notag \\
E_{2n} & :0=(n+2)(n+1)\rho_{0}^{3}\varphi_{n+2}+.....  \label{Ek} \\
E_{3n} & :0=(n+2)(n+1)\rho_{0}^{3}\theta_{n+2}+......  \notag
\end{align}
where the remaining terms are of less order since they involve $\rho
_{i},\varphi_{i},\theta_{i}$ for $i<n+2$. For example, the coefficients $%
u_{n},$ $\mu_{n},\eta_{n}$ occur in (\ref{Ek}) and their order is $n$. The
equations $E_{4n}$ for $n>0$ are omitted since they do not lead to
additional (algebraic independent) relations.

Now, let us select some independent and recursive relations from the above
ones, but first we take the basic identity (\ref{3-16}) and the expression (%
\ref{speed}) for the speed in the spherical metric, whose zero order terms
yield the two identities :
\begin{align}
E_{0} & :\rho_{0}^{3}v_{0}^{2}=4\mathfrak{S}_{0},\text{ \ \ }\mathfrak{S}%
_{0}=\frac{\omega_{0}}{K_{0}}  \label{7a} \\
E_{0}^{\prime} & :v_{0}=\sqrt{\varphi_{1}^{2}+g_{0}\theta_{1}^{2}}
\label{7b}
\end{align}
We shall use the symbols $J_{1},J_{2}$ etc. to denote various expressions
which are of intrinsic type. By using (\ref{7a}) the identities $E_{10}$ and
$E_{40}$ can be restated as
\begin{align}
E_{1} & :\rho_{0}(\rho_{1}^{2}-2h)=J_{1},\text{ \ \ \ }J_{1}=2u_{0}-%
\mathfrak{S}_{0}  \label{list1-4} \\
E_{4} & :\rho_{0}^{2}\rho_{2}=J_{4},\text{ \ \ \ \ \ \ \ \ \ \ \ \ \ \ }%
J_{4}=\frac{1}{2}(-u_{0}+\mathfrak{S}_{0})  \notag
\end{align}
Next, the direction $\psi_{0}$ of $\gamma^{\ast}$ at the point $(\varphi
_{0},\theta_{0})$ is intrinsic; it is also conveniently represented by the
unit tangent vector

\begin{equation*}
\mathbf{\tau}^{\ast}=\frac{1}{v_{0}}(\varphi_{1}\frac{\partial}{\partial
\varphi}+\theta_{1}\frac{\partial}{\partial\theta})=J_{\varphi}\frac{%
\partial }{\partial\varphi}+J_{\theta}\frac{\partial}{\partial\theta}
\end{equation*}
The coefficients $J_{\varphi}$, $J_{\theta}$ are intrinsic functions,
depending on the coordinate system, and they are related by the identity
\begin{equation}
J_{\varphi}^{2}+g_{0}J_{\theta}^{2}=1   \label{J-identity}
\end{equation}
Therefore, we adjoin to our list (\ref{list1-4}) the two identities
\begin{align}
E_{2} & :\varphi_{1}=J_{\varphi}v_{0}\text{ \ }  \label{list2-3} \\
E_{3} & :\theta_{1}=J_{\theta}v_{0}\text{\ }  \notag
\end{align}

Still, we have not used all zero order relations, namely $E_{20}$ and $E_{30}
$, and now we state them as%
\begin{align}
E_{5} & :\rho_{0}^{3}\varphi_{2}+\rho_{0}^{2}\rho_{1}\varphi_{1}=J_{5},\text{
\ \ \ }J_{5}=2\mu_{0}+f_{0}J_{\theta}^{2}\mathfrak{S}_{0}  \label{list5-6} \\
E_{6} & :\rho_{0}^{3}\theta_{2}+\rho_{0}^{2}\rho_{1}\theta_{1}=J_{6},\text{
\ \ \ }J_{6}=\frac{2\eta_{0}}{g_{0}}-2\frac{f_{0}}{g_{0}}J_{\varphi}J_{%
\theta }\mathfrak{S}_{0}  \notag
\end{align}
By continuing this way, we obtain for each $n>0$ three new relations%
\begin{align}
E_{3n+1} & :0=\rho_{0}^{2}\rho_{n+2}+.....  \notag \\
E_{3n+2} & :0=\rho_{0}^{3}\varphi_{n+2}+.....  \label{list7-9} \\
E_{3n+3} & :0=\rho_{0}^{3}\theta_{n+2}+......  \notag
\end{align}
involving at each step the new triple $\rho_{n+2},\varphi_{n+2},\theta_{n+2}$
of variables of order $n+2$. \

\begin{claim}
\label{possible}It is possible to solve the above recursive relations for
the variables (\ref{8variables}) completely in terms of the intrinsic local
geometric data in the shape space.
\end{claim}

This will be finally settled at the end of the subsection. At this point we
have altogether 3n+8 variables
\begin{equation*}
\rho_{0},v_{0};\rho_{1},\varphi_{1},\theta_{1};\rho_{2},\varphi_{2},\theta
_{2};...;\rho_{n+2},\varphi_{n+2},\theta_{n+2};
\end{equation*}
involved in 3n+8 recursive relations, and the first eight involve only the
variables up to order 2. However, $E_{0}^{\prime},E_{2}$ and $E_{3}$ are
obviously algebraic dependent due to the identity (\ref{J-identity}), so let
us search for one more independent relation among the variables of order $%
\leq2$. Since we expect such a relation to involve local intrinsic
quantities of order (at least) 3, a natural approach is to differentiate the
basic identity (\ref{3-16}) involving the Siegel function. Thus evaluation
of the resulting identity (\ref{3-17a}) at $t=t_{0}$ yields
\begin{equation}
3\frac{\rho_{1}}{\rho_{0}v_{0}}+2\frac{v_{1}}{v_{0}^{2}}=J_{7},\text{ \ \ \
\ \ }J_{7}=\frac{\mathfrak{S}_{1}}{\mathfrak{S}_{0}}=(\frac{\omega _{k+1}}{%
\omega_{k}}-\frac{K_{k+1}}{K_{k}})   \label{E1prime}
\end{equation}
Using the expression in (\ref{coeff}) for $v_{1}$ we can restate the above
identity as
\begin{equation}
3\frac{\rho_{1}}{\rho_{0}v_{0}}+\frac{4}{v_{0}^{3}}\left[ \varphi_{1}%
\varphi_{2}+\frac{1}{4}f_{0}\varphi_{1}\theta_{1}^{2}+g_{0}\theta_{1}%
\theta_{2}\right] =J_{7}   \label{10}
\end{equation}
By simple calculation and substitution using some of the previous relations $%
E_{i}$,
\begin{align*}
& \rho_{0}^{3}\left[ \varphi_{1}\varphi_{2}+\frac{1}{4}f_{0}\varphi
_{1}\theta_{1}^{2}+g_{0}\theta_{1}\theta_{2}\right] \\
& =\varphi_{1}(J_{5}-\rho_{0}^{2}\rho_{1}\varphi_{1})+\varphi_{1}(\frac{1}{4}%
f_{0}J_{\theta}^{2}\rho_{0}^{3}v_{0}^{2})+\theta_{1}(g_{0}J_{6}-g_{0}%
\rho_{0}^{2}\rho_{1}\theta_{1}) \\
& =-\frac{\rho_{1}}{\rho_{0}}\rho_{0}^{3}(\varphi_{1}^{2}+g_{0}%
\theta_{1}^{2})+\varphi_{1}(J_{5}+f_{0}J_{\theta}^{2}\mathfrak{S}%
_{0})+\theta_{1}g_{0}J_{6} \\
& =-4\frac{\rho_{1}}{\rho_{0}}\mathfrak{S}_{0}+v_{0}\left[
J_{\varphi}J_{5}+f_{0}J_{\varphi}J_{\theta}^{2}\mathfrak{S}%
_{0}+g_{0}J_{\theta}J_{6}\right]
\end{align*}
and by substitution into (\ref{10}), using the identity $%
\rho_{0}^{3}v_{0}^{2}$ $=4\mathfrak{S}_{0}$ and the expressions for $%
J_{5},J_{6}$ in (\ref{list5-6}), this leads to our new identity
\begin{equation}
E_{1}^{\prime}:\frac{\rho_{1}}{\rho_{0}v_{0}}=J_{8},\ \text{ \ \ \ \ \ \ }%
J_{8}=2\mathfrak{S}_{0}^{-1}(J_{\varphi}\mu_{0}+J_{\theta}\eta_{0})-J_{7}=%
\text{\ }\frac{1}{\mathfrak{S}_{0}}(2\bar{u}_{1}-\mathfrak{S}_{1})\text{\ \
\ }   \label{11}
\end{equation}
where $\bar{u}_{1}$ is the tangential derivative of $U^{\ast}$ at $P_{0},$
cf. (\ref{tangderiv}).

From the system of algebraic equations%
\begin{equation*}
E_{0},E_{0}^{\prime},E_{1},E_{1}^{\prime},E_{2},E_{3},E_{4},.....
\end{equation*}
we can now solve recursively and thus determine the variables%
\begin{equation*}
\rho_{0},v_{0,}\rho_{1},\varphi_{1},\theta_{1},\rho_{2},\varphi_{2},\theta
_{2},.....
\end{equation*}
successively in terms of the intrinsic data. In fact, this is obvious from
the structure of the equations, once we have determined $\rho_{0},v_{0},%
\rho_{1}$, namely using the three equations $E_{0},E_{1},E_{1}^{\prime}$:
\begin{equation}
\rho_{0}^{3}v_{0}^{2}=4\mathfrak{S}_{0},\text{ \ }\rho_{0}(%
\rho_{1}^{2}-2h)=J_{1},\text{ \ }\frac{\rho_{1}}{\rho_{0}v_{0}}=J_{8}
\label{3rel}
\end{equation}
It follows that%
\begin{equation}
2h\rho_{0}=-J_{1}+J_{8}^{2}\rho_{0}^{3}v_{0}^{2}=-J_{1}+4J_{8}^{2}\mathfrak{S%
}_{0}   \label{4-20}
\end{equation}
and consequently, for $h\neq0$,
\begin{equation}
\rho_{0}=\frac{1}{h}\left[ \frac{\mathfrak{S}_{0}}{2}(4J_{8}^{2}+1)-u_{0}%
\right] \text{, \ \ \ }v_{0}=\frac{2}{\rho_{0}^{3/2}}\sqrt {\mathfrak{S}_{0}}%
\text{, \ \ }\rho_{1}=2J_{8}\sqrt{\frac{\mathfrak{S}_{0}}{\rho_{0}}}
\label{3id}
\end{equation}

In the case $h=0$ the identity (\ref{4-20}) merely tells us that
\begin{equation}
u_{0}=\frac{1}{2}(4J_{8}^{2}+1)\mathfrak{S}_{0},   \label{4-21}
\end{equation}
and we can freely choose any initial size $\rho_{0}$ of $\rho$ and then
calculate $v_{0}$ and $\rho_{1}$ from (\ref{3id}). In particular, we
calculate the pair $(\varphi_{1},\theta_{1})$ using (\ref{list2-3}), where
the pair $(J_{\varphi},J_{\theta})$ represents the initial direction and
hence is intrinsic. Now, we are able to calculate successively each new
triple $(\rho_{n},\varphi_{n},\theta_{n})$, $n=2,3,4...$, expressed in terms
of the intrinsic data, as claimed above.

\begin{remark}
The initial direction, $(J_{\varphi},J_{\theta})$, is the only basic
intrinsic data with no invariant description, that is, independent of the
coordinate frame. However, from the recursive procedure it follows that $%
u_{n},v_{n},\rho_{n},n\geq0$, come out with coordinate free expressions
involving only the coefficients in (\ref{starseries}). In fact, we can
calculate $v_{n-1}$ in terms of $\rho_{i}$ $(i<n)$ and $v_{j}$ $(j<n-1)$, by
repeated differentiation of (\ref{3-16}), next we calculate $u_{n}$ by
applying differential operators such as (\ref{C12}) to $U^{\ast}$, and
finally $\rho_{n+2}$ is calculated using equation $E_{1n}$. The beginning
terms are
\begin{equation*}
u_{1}=v_{0}\bar{u}_{1},\text{ \ \ \ }u_{2}=v_{0}^{2}\bar{u}_{2}+\frac{1}{2}%
v_{1}\bar{u}_{1},\text{ \ \ }v_{1}=\frac{4\mathfrak{S}_{1}-3%
\rho_{0}^{2}v_{0}\rho_{1}}{2\rho_{0}^{3}}\text{\ \ \ }
\end{equation*}
\end{remark}

\subsection{Some basic results on the shape curves of three-body motions
with vanishing angular momentum}

Let us first review some of the above facts from the local analysis and then
draw a few immediate but important consequences. The above power series
developments amount to the explicit calculation of the solution $%
t\rightarrow \bar{\gamma}(t)$ of the system (\ref{3-45}) in the moduli space
with the initial data (\ref{3-44}). In doing so we started from the
following 5-tuple
\begin{equation}
(\varphi_{0},\theta_{0}),(\psi_{0},\mathfrak{S}_{0},\mathfrak{S}_{1}),
\label{5-data}
\end{equation}
where $\mathfrak{S}_{0},\mathfrak{S}_{1}$ may be replaced by $K_{0},K_{1}$
(or $K_{k},K_{k+1}$, for the smallest $k$ with $K_{k}\neq0)$ which consists
of three specific intrinsic local geometric invariants at the point $%
(\varphi _{0},\theta_{0})$ on the shape curve. In particular, we also
recover the time parametrized shape curve $\gamma^{\ast}(t)$ by projecting $%
\bar{\gamma}(t)$ to the 2-sphere.

Actually, since $\bar{\gamma}(t)$ is uniquely determined by the initial
value problem, (\ref{3-44}) and (\ref{3-45}), it suffices to recover (\ref%
{3-44}) from (\ref{5-data}), namely the missing information in (\ref{3-44})
is $\rho_{0}$ and $\rho_{1}$. This turns out to be possible when $h\neq0$,
but for a "good" reason (see below) it is impossible when $h=0$ since in
this case the shape curve only controls the product $\rho_{0}^{3}v_{0}^{2}$ $%
=4\mathfrak{S}_{0}$. In any case, with the quantity $\mathfrak{S}_{1}$ we
can actually determine $\rho_{1}$ and, in particular, the question in
Problem \ref{index} concerning the expansion index is settled.

The general three-body problem has the 10 classical conservation laws
(linear and angular momentum, and energy) due to its invariance under the
Galilean symmetry group. All of them have been used and, in particular, the
set of solutions is invariant under time translation, $t\rightarrow t+t_{0}$%
, as well as reversal of time ($t\rightarrow-t$) which reverses the
direction of the trajectory. However, there is also an additional
1-parameter size/time scaling symmetry group, whose induced action on
parametrized moduli curves sends $\bar{\gamma}(t)=(\rho(t),\gamma^{\ast}(t))$
to
\begin{equation}
\bar{\gamma}_{(r)}(t)=(\rho_{(r)}(t),\gamma_{(r)}^{\ast}(t))=(r\rho (r^{-%
\frac{3}{2}}t),\gamma^{\ast}(r^{-\frac{3}{2}}t))\text{, \ }\forall r>0,
\label{scaling}
\end{equation}
and changes the energy from $h$ to $h_{(r)}=r^{-1}h$. In particular,
although scaling and time translation leaves the oriented shape curve
geometrically unchanged, its time parametrization is subject to an affine
transformation%
\begin{equation}
t\rightarrow at+t_{0},\text{ \ \ }a=r^{-3/2}>0   \label{aff}
\end{equation}
Since the energy level $h=0$ is scaling invariant, this also explains why
the reconstruction of a unique initial size $\rho_{0}$ fails when $h=0$.

\begin{remark}
For any $n\geq1$ the Newtonian n-body problem has the above 1-parameter
symmetry group $\left\{ \Phi_{r}\right\} $, acting on size and time but
leaves the shape invariant. For example, for a periodic motion with period $%
P_{(1)}$ and average (or initial) size $\rho_{(1)}$, the group sweeps out a
periodic motion with the same shape, and the ratio $P_{(r)}^{2}/%
\rho_{(r)}^{3}$ is independent of $r$. The case $n=1$ means the restricted
case $n=2$ with one of the masses (e.g. a planet) infinitesimal small, in
which case there is only one shape (a point) and the above ratio depends
only on the large mass (the sun). This gives Kepler's third law, so the
above symmetry group is essentially the generalization of this law.

On the other hand, for three-body motions with vanishing angular momentum,
the identity $\rho^{3}v^{2}$ $=4\mathfrak{S}$ of (\ref{3-16}) gives another
quantity, $\rho^{3}v^{2}$, which is invariant under the above symmetry group.
\end{remark}

\begin{definition}
A time reparametrization of $\bar{\gamma}(t)$ or $\gamma^{\ast}(t)$ by an
affine transformation (\ref{aff}), for any $a\neq0$, is called \emph{%
canonical}, otherwise it is called \emph{exceptional}.\emph{\ }
\end{definition}

Of course, in order to stay at a given nonzero energy level a canonical
reparametrization must have $a=\pm1$, and moreover, the orientation of the
curve is reversed if $a<0$. Now we can state the following basic \emph{%
unique parametrization theorem} :

\begin{theorem}
\label{unique} A three-body motion with zero angular momentum is, up to
congruence and canonical reparametrization, uniquely determined by its
oriented shape curve on the 2-sphere. In fact, it suffices to know the
direction and the first two Siegel numbers $\mathfrak{S}_{0},\mathfrak{S}%
_{1} $ at any regular point on the shape curve. In particular, there are no
exceptional reparametrizations.
\end{theorem}

As shown before, the theorem still holds with $(\mathfrak{S}_{0},\mathfrak{S}%
_{1})$ replaced by the curvature numbers $(K_{0},K_{1})$ if the point is
regular of order 0, and this is, indeed, the generic type of points.

A curve on the 2-sphere which is the shape curve of a motion with total
energy $h$ can also be the shape curve for some motion with any other energy
$h^{\prime}$ of the same sign as $h$. Indeed, we find the other motions by
suitable canonical reparametrizations of the given motion, and by (\ref{3id}%
) it also follows that the sign of $h$ (viewed as a number $0,\pm1)$ is an
intrinsic invariant at the shape space level. More precisely, we have the
following quantitative measurement of the energy type :

\begin{theorem}
Let $\gamma^{\ast}$ be a geometric curve on the 2-sphere, with the Siegel
function $\mathfrak{S}$ (with respect to $U^{\ast}$, as usual), and consider
the function
\begin{equation*}
\Delta=\frac{\mathfrak{S}}{2}(4\left[ \frac{1}{\mathfrak{S}}\frac {d}{d%
\mathbf{\tau}}(2U^{\ast}-\mathfrak{S)}\right] ^{2}+1)-U^{\ast}
\end{equation*}
along the curve, where $\frac{d}{d\mathbf{\tau}}$ denotes the tangential
derivative. If $\gamma^{\ast}$ can be realized as the shape curve of a
three-body motion with vanishing angular momentum, then the sign of $\Delta$
is constant along the curve (whenever $\Delta$ is defined), namely equal to
the sign of the total energy $h$ of the motion.
\end{theorem}

\begin{corollary}
A given oriented (geometric) curve on the 2-sphere can be time parametrized
in at most one way, up to canonical reparametrization, as the shape curve $%
t\rightarrow\gamma^{\ast}(t)$ of a three-body motion with zero angular
momentum. Moreover, the sign of the total energy of such a motion is
determined by the local relative geometry of $(\gamma^{\ast},U^{\ast})$ at a
(regular) point.
\end{corollary}

The above uniqueness property of time parametrization of geometric shape
curves, the minimal amount of geometric information needed to determine the
shape curve, and the monotonicity theorem which we shall discuss in Section
5, are our basic tools for the understanding of both the local and global
picture of shape curves representing three-body motions with vanishing
angular momentum. The monotonicity property tells us the \emph{m-latitude }%
function is monotonic increasing or decreasing until the curve turns back
somewhere in the opposite hemisphere. Thus the curve resembles an
"oscillating motion" between the upper and lower hemisphere which never
stops, unless it ends at a triple collision or escapes to infinity. The
curve crosses the equator circle transversely, or it goes to a binary
collision and bounces back (via regularization) to the same hemisphere. In
Section 7.5 we describe the problem of how to construct such a curve by
linking together its maximal monotonic segments.

\section{The monotonicity theorem for shape curves}

\subsection{A closer look at the gradient vector field of $U^{\ast}$}

The analysis of trajectories, moduli curves or shape curves describing
three-body motions depends, of course, ultimately on the function $U^{\ast}$%
, whose behavior is largely reflected by the geometry of its gradient field.
In this subsection some useful facts are established which are beyond those
simpler statements concerning the critical or singular points of $U^{\ast}$.

We shall apply vector algebra in the Euclidean model for the moduli space,
namely with $\bar{M}=\mathbb{R}^{3}$ as the Euclidean space (cf. Section 2)
and vectors denoted by boldface letters. Thus the shape space $M^{\ast}$ $%
=S^{2}(1)$ consists of unit vectors $\mathbf{p}=(x,y,z),\left\vert \mathbf{p}%
\right\vert ^{2}=x^{2}+y^{2}+z^{2}$, and $\mathbf{p\cdot q}$ denotes the
usual inner product. Set
\begin{align}
\hat{m} & =\sum\hat{m}_{i},\text{ \ }\bar{m}=m_{1}m_{2}m_{3},\text{ \ }\hat{m%
}_{1}=m_{2}m_{3}\text{ etc. \ and }\sum m_{i}=1  \notag \\
\text{\ }k_{i} & =2\frac{\hat{m}_{i}^{\frac{3}{2}}}{\sqrt{1-m_{i}}}\text{, \
}\beta_{i}=\cos^{-1}(\frac{\hat{m}_{i}-m_{i}}{\hat{m}_{i}+m_{i}})=\sin ^{-1}(%
\frac{2\sqrt{\bar{m}}}{\hat{m}_{i}+m_{i}})  \label{4-1} \\
\mathbf{b}_{1} & =(1,0,0)\text{, }\mathbf{b}_{2}=(\cos\beta_{3},\sin
\beta_{3},0)\text{, }\mathbf{b}_{3}=(\cos\beta_{2},-\sin\beta_{2},0)  \notag
\end{align}
where $0<\beta_{i}<\pi$ is the angle between the binary collision points $%
\mathbf{b}_{j}$ and $\mathbf{b}_{k},\left\{ i,j,k\right\} =\left\{
1,2,3\right\} $. The Newtonian shape potential function is the restriction
of $U$ to the above 2-sphere,
\begin{equation}
U^{\ast}=\sum_{i=1}^{3}\frac{\hat{m}_{i}}{s_{i}}=\sum\limits_{i=1}^{3}\frac{%
k_{i}}{\left\vert \mathbf{p-b}_{i}\right\vert }\text{, \ cf. (\ref{2-11})}
\label{4-2}
\end{equation}
where the mutual distances $s_{i}=r_{jk}$ are normalized to $I=1$, and hence
by a formula of Lagrange
\begin{equation}
I=\sum\hat{m}_{i}s_{i}^{2}=1   \label{4-2a}
\end{equation}

The basic behavior of $U^{\ast}$ is, of course, given by the 8 special
points on the 2-sphere, namely
\begin{equation}
\mathbf{b}_{1},\mathbf{e}_{3},\mathbf{b}_{2},\mathbf{e}_{1},\mathbf{b}_{3},%
\mathbf{e}_{2};\text{ }\mathbf{p}_{0},\mathbf{p}_{0}^{\prime}
\label{8points}
\end{equation}
where the first six are cyclically ordered (eastward) along the equator
circle $E^{\ast}$ representing degenerate m-triangles. The $\mathbf{b}_{i}$
are poles where $U^{\ast}$ tends to $\infty$, and the Euler points $\mathbf{e%
}_{i}$ are the saddle points. Finally, the remaining two are the minima, say
$\mathbf{p}_{0}$ lies on the northern hemisphere and $\mathbf{p}_{0}^{\prime}
$ is the symmetric (mirror) image with respect to the equator plane.

We can also use the points $\mathbf{b}_{i}$ to describe the gradient field,
as follows. Let $d\mathbf{x}$ be an arbitrary infinitesimal vector
perpendicular to $\mathbf{p}$ (i.e. $d\mathbf{x\in}T_{\mathbf{p}}S^{2})$.
Then, on the one hand
\begin{equation}
\nabla U^{\ast}(\mathbf{p})\cdot d\mathbf{x}\equiv U^{\ast}(\mathbf{p}+d%
\mathbf{x)}-U^{\ast}(\mathbf{p})\text{ \ }(\func{mod}\left\vert d\mathbf{x}%
\right\vert ^{2})   \label{4-3}
\end{equation}
and on the other hand,
\begin{align}
U^{\ast}(\mathbf{p}+d\mathbf{x)}-U^{\ast}(\mathbf{p}) &
=\sum_{i=1}^{3}k_{i}\left( (2-2(\mathbf{p}+d\mathbf{x)\cdot b}_{i})^{-\frac{1%
}{2}}-(2-2\mathbf{p}\cdot\mathbf{b}_{i})^{-\frac{1}{2}}\right)  \notag \\
& \equiv\left( \sum_{i=1}^{3}\frac{k_{i}\mathbf{b}_{i}}{\left\vert \mathbf{%
p-b}_{i}\right\vert ^{3}}\right) \cdot d\mathbf{x}\text{ \ \ }(\func{mod}%
\text{ }\left\vert d\mathbf{x}\right\vert ^{2})   \label{4-4}
\end{align}
Set
\begin{equation}
\mathbf{B}=\mathbf{B(p)=}\sum_{i=1}^{3}\frac{k_{i}\mathbf{b}_{i}}{\left\vert
\mathbf{p-b}_{i}\right\vert ^{3}}   \label{4-5}
\end{equation}
Then it follows from (\ref{4-3}) and (\ref{4-4}) that the gradient of the
function $U^{\ast}$ is the orthogonal projection of the vector $\mathbf{B}$
to the tangent plane of the sphere at $\mathbf{p}$, namely
\begin{equation}
\nabla U^{\ast}=\mathbf{B-(B\cdot p)p}   \label{4-5a}
\end{equation}

The characterization of the critical points of $U^{\ast}$ is, of course,
well known. However, with the following lemma we also like to establish the
identity (\ref{4-6}).

\begin{lemma}
Let $\mathbf{p}_{0}$ and $\mathbf{p}_{0}^{\prime}$ represent the pair of
equilateral m-triangles with $I=1$ and with opposite orientations. Then $%
\mathbf{p}_{0}$ and $\mathbf{p}_{0}^{\prime}$ are the minima of $U^{\ast}$,
and moreover
\begin{equation}
\mathbf{b}_{i}\cdot(\mathbf{b}_{i}-\mathbf{p}_{0})=\mathbf{b}_{i}\cdot(%
\mathbf{b}_{i}-\mathbf{p}_{0}^{\prime})=\frac{2m_{j}m_{k}}{(1-m_{i})\hat{m}}
\label{4-6}
\end{equation}
\end{lemma}

\begin{proof}
The determination of the critical points away from the equator follows
readily by Lagrange's multiplier method in $\bar{M}$ with the constraint $%
I=1 $. As coordinates we can, for example, use the individual moments of
inertia $I_{j}$, but the calculations are simplest in terms of the mutual
distances $r_{ij}=s_{k}$ using (\ref{4-2}) and (\ref{4-2a}). This shows the
minimum $(s_{1}^{0},s_{2}^{0},s_{3}^{0})$ of $U^{\ast}$ (on any of the
hemispheres) satisfies the following set of equations with a multiplier $%
\lambda$
\begin{equation}
\frac{1}{2(s_{i}^{0})^{3}}=\lambda\text{, \ }i=1,2,3   \label{4-8}
\end{equation}
and hence, by (\ref{4-2a}), all the sides $s_{i}^{0}$ are equal to $1/\sqrt{%
\hat{m}}$ . Moreover, by (\ref{4-2})
\begin{align}
\frac{k_{i}}{\left\vert \mathbf{p}_{0}\mathbf{-b}_{i}\right\vert } & =\frac{%
\hat{m}_{i}}{s_{i}^{0}}=m_{j}m_{k}\sqrt{\hat{m}}  \notag \\
& \Longrightarrow2-2\mathbf{p}_{0}\cdot\mathbf{b}_{i}=\left\vert \mathbf{p}%
_{0}\mathbf{-b}_{i}\right\vert ^{2}=\frac{4m_{j}m_{k}}{(1-m_{i})\hat{m}}
\label{4-9} \\
& \Longrightarrow\mathbf{p}_{0}\cdot\mathbf{b}_{i}=1-\frac{2m_{j}m_{k}}{%
(1-m_{i})\hat{m}}  \notag
\end{align}
and this gives (\ref{4-6}).
\end{proof}

\begin{remark}
It is also straightforward to check the identities
\begin{align}
\frac{k_{i}}{\left\vert \mathbf{p}_{0}-\mathbf{b}_{i}\right\vert ^{3}} & =%
\frac{m_{j}m_{k}\sqrt{\hat{m}}}{\left\vert \mathbf{p}_{0}-\mathbf{b}%
_{i}\right\vert ^{2}}=\frac{1}{4}\hat{m}^{\frac{3}{2}}(1-m_{i})  \label{4-10}
\\
\sum\limits_{i=1}^{3}(1-m_{i})\mathbf{b}_{i} & =0   \label{4-11}
\end{align}
\end{remark}

To simplify the notation below, let us write
\begin{align}
\psi_{i}(t) & =(1-\frac{t}{c_{i}})^{-\frac{3}{2}},\text{ \ }c_{i}=\frac{%
m_{j}m_{k}}{(1-m_{i})\hat{m}}\text{ , \ cf. (\ref{4-6})}  \label{4-11b} \\
\Psi_{i}(t) & =\frac{1}{4}\hat{m}^{\frac{3}{2}}(1-m_{i})\psi_{i}(t)
\label{4-11c}
\end{align}
where $\psi_{i}$ is defined for $t<c_{i},i=1,2,3$, and observe that the
derivative of $\psi_{i}$ is strictly positive. The following two lemmas will
be useful.

\begin{lemma}
The expression (\ref{4-5}) of $B$ \ can be restated as%
\begin{equation}
B(\mathbf{p})=\sum_{i=1}^{3}\Psi_{i}(\xi_{i})\mathbf{b}_{i}   \label{4-11d}
\end{equation}
where%
\begin{equation}
\xi_{i}=\mathbf{b}_{i}\cdot\mathbf{(p-p}_{0})   \label{4-11e}
\end{equation}
\end{lemma}

\begin{proof}
By (\ref{4-10})
\begin{align}
\mathbf{B} & =\sum_{i=1}^{3}\frac{k_{i}\mathbf{b}_{i}}{\left\vert \mathbf{p}-%
\mathbf{b}_{i}\right\vert ^{3}}=\sum_{i=1}^{3}\frac{k_{i}}{\left\vert
\mathbf{p}_{0}-\mathbf{b}_{i}\right\vert ^{3}}\left( \frac{\left\vert
\mathbf{p}_{0}-\mathbf{b}_{i}\right\vert }{\left\vert \mathbf{p}-\mathbf{b}%
_{i}\right\vert }\right) ^{3}\mathbf{b}_{i}  \notag \\
& =\sum_{i=1}^{3}\frac{\hat{m}^{\frac{3}{2}}}{4}(1-m_{i})\left( \frac{%
\mathbf{b}_{i}\cdot(\mathbf{b}_{i}-\mathbf{p}_{0})}{\mathbf{b}_{i}\cdot(%
\mathbf{b}_{i}-\mathbf{p)\ }}\right) ^{\frac{3}{2}}\mathbf{b}_{i}
\label{4-15} \\
& =\frac{\hat{m}^{\frac{3}{2}}}{4}\sum_{i=1}^{3}(1-m_{i})\left( 1-\frac{%
\mathbf{b}_{i}\cdot(\mathbf{p}-\mathbf{p}_{0})}{\mathbf{b}_{i}\cdot(\mathbf{b%
}_{i}-\mathbf{p}_{0}\mathbf{)}}\right) ^{-\frac{3}{2}}\mathbf{b}_{i}  \notag
\\
& =\frac{\hat{m}^{\frac{3}{2}}}{4}\sum_{i=1}^{3}(1-m_{i})\psi_{i}(\xi _{i})%
\mathbf{b}_{i}=\sum_{i=1}^{3}\Psi_{i}(\xi_{i})\mathbf{b}_{i}  \notag
\end{align}
\end{proof}

\begin{lemma}
\label{positiv}Let $\mathbf{p}$ be a unit vector different from $\mathbf{p}%
_{0}$ and $\mathbf{p}_{0}^{\prime}$. Then $\mathbf{B\cdot(p-p}_{0})$ is
strictly positive.\ \ \ \ \ \ \ \ \
\end{lemma}

\begin{proof}
By the mean value theorem, there exists $0<\varepsilon_{i}<1$%
\begin{equation*}
\Psi_{i}(\xi_{i})=\Psi_{i}(\xi_{i})-\Psi_{i}(0)=\Psi_{i}^{\prime}(%
\varepsilon_{i}\xi_{i})\xi_{i},\text{ \ }1\leq i\leq3,
\end{equation*}
and we recall that the derivative of $\Psi_{i}$ is strictly positive. By (%
\ref{4-11d}) and (\ref{4-11e}) \
\begin{align*}
\mathbf{B(p})\cdot(\mathbf{p-p}_{0}) & =\left(
\sum_{i=1}^{3}\Psi_{i}(\xi_{i})\mathbf{b}_{i}\right) \cdot(\mathbf{p-p}_{0})
\\
& =\sum_{i=1}^{3}\Psi_{i}(\xi_{i})\xi_{i}=\sum_{i=1}^{3}\Psi_{i}^{\prime
}(\varepsilon_{i}\xi_{i})\xi_{i}^{2}>0
\end{align*}
\end{proof}

Note that $\mathbf{B}$ is, by definition, a linear combination of $\{\mathbf{%
b}_{1}\mathbf{,b}_{2}\mathbf{,b}_{3}\}$, thus lying in the $xy$-plane and
consequently
\begin{equation*}
\mathbf{B\cdot(p-p}_{0})=\mathbf{B\cdot(\bar{p}-\bar{p}}_{0})
\end{equation*}
where $\mathbf{\bar{p}}$ (resp. $\mathbf{\bar{p}}_{0})$ is the orthogonal
projection of $\mathbf{p}$ (resp. $\mathbf{p}_{0})$ in the $xy$-plane.
Clearly, the geometric meaning of the positivity of the inner product in the
lemma is that the angle between $\mathbf{B}$ and the vector from $\mathbf{%
\bar{p}}_{0}$ to $\mathbf{\bar{p}}$ is strictly less than $\pi/2$.

\subsection{The relative geometry between $\protect\nabla U^{\ast}$ and the
conjugate pair of co-axial families of circles associated to \{p$_{0}$,p$%
_{0}^{\prime}$\}}

In spherical geometry, circles are the simplest kind of curves and they are
characterized by the constancy of their geodesic curvature. Associated to a
given pair of points, such as the minima $\{\mathbf{p}_{0},\mathbf{p}%
_{0}^{\prime}\}$ of $U^{\ast}$, there are two co-axial families of circles,
namely the family of circles passing through the two given points and its
dual family consisting of those circles which are orthogonal to all circles
of the former family. We shall denote the conjugate pair of coaxial families
of circles associated to $\{\mathbf{p}_{0},\mathbf{p}_{0}^{\prime}\}$ by $%
\mathcal{F}$ and $\mathcal{F}^{\prime}$.

To a given $\mathbf{p}\in S^{2}(1)$ other than $\mathbf{p}_{0},\mathbf{p}%
_{0}^{\prime}$, let us denote the unique circle of $\mathcal{F}$ (resp. $%
\mathcal{F}^{\prime}$) passing through $\mathbf{p}$ by $\Gamma_{p}$ (resp.%
\textbf{\ }$\Gamma_{p}^{\prime}$). In fact, $\Gamma_{p}$ is simply the
intersection of $S^{2}(1)$ and the plane spanned by the triple $\{\mathbf{p}%
_{0},\mathbf{p}_{0}^{\prime},\mathbf{p}\}$. Therefore, the tangent line of $%
\Gamma_{p}$ at $\mathbf{p}$ is the intersection of the tangent plane $T_{%
\mathbf{p}}S^{2}(1)$ and the above plane. In fact, the following pair of
vectors
\begin{equation}
\mathbf{T}=\mathbf{p}\times\left[ (\mathbf{p-p}_{0})\times(\mathbf{p-p}%
_{0}^{\prime})\right] ,\text{ \ \ }\mathbf{T}^{\prime}\mathbf{=p}\times%
\mathbf{T}   \label{4-12}
\end{equation}
constitutes a positively oriented orthogonal basis $\left\{ \mathbf{T,T}%
^{\prime}\right\} $ for the tangent plane at $\mathbf{p}$ such that $\mathbf{%
T}$ (resp. $\mathbf{T}^{\prime}\mathbf{)}$ is tangent to $\Gamma_{p}$ (resp.
$\Gamma_{p}^{\prime})$. The vector $\mathbf{T}$ defines the \emph{southward }%
direction, that is, away from $\mathbf{p}_{0}$, whereas $\mathbf{T}^{\prime}$
defines the \emph{eastward} direction along $\Gamma _{p}^{\prime}$. \

For example, in the case of uniform mass distribution, $\{\mathbf{p}_{0},%
\mathbf{p}_{0}^{\prime}\}$ $=\left\{ N,S\right\} $ are the north and south
pole of the sphere, whose conjugate pair of coaxial families of circles are
the usual \emph{longitude circles} (or \emph{meridians}) and \emph{latitude
circles. }In this case
\begin{equation*}
\mathbf{T}=2\left[ (\mathbf{p\cdot k)p-k}\right] \text{ , \ }\mathbf{T}%
^{\prime}=-2\mathbf{p\times k}\text{\ }
\end{equation*}
and these are positive multiples of the coordinate vectors $\frac{\partial }{%
\partial\varphi}$ and $\frac{\partial}{\partial\theta}$ associated with
spherical polar coordinates $(\varphi,\theta)$ centered at the pole $N$.

\begin{proposition}
\label{south}The inner product between the "southward" vector $\mathbf{T}$
and the gradient vector $\nabla U^{\ast}$ at a point $\mathbf{p}$ on the
sphere is
\begin{equation}
\mathbf{T\cdot}\nabla U^{\ast}=\mathbf{T\cdot B}=\left[ \mathbf{p\cdot }(%
\mathbf{p}_{0}\mathbf{-p}_{0}^{\prime})\right] \left[ \mathbf{B\cdot (p-p}%
_{0})\right]   \label{4-13}
\end{equation}
In particular, on the northern (resp. southern) hemisphere the angle $\beta$
between $\mathbf{T}$ and $\nabla U^{\ast}$ is in the range $0\leq\beta<\pi/2$
\ (resp. $\pi/2<\beta\leq\pi$).
\end{proposition}

\begin{proof}
By (\ref{4-5a}), $\mathbf{T\cdot}\nabla U^{\ast}=\mathbf{T\cdot B}$, and
clearly $\mathbf{B\cdot p}_{0}=\mathbf{B\cdot p}_{0}^{\prime}$, consequently

\begin{align}
\mathbf{T\cdot B} & =\mathbf{B}\cdot\{\mathbf{p\times\lbrack p\times(p}_{0}-%
\mathbf{p}_{0}^{\prime})]+\mathbf{p\times(p}_{0}\times\mathbf{p}%
_{0}^{\prime})\}  \notag \\
& =\left\vert
\begin{array}{cc}
\mathbf{B\cdot p} & 1 \\
\mathbf{B\cdot(p}_{0}-\mathbf{p}_{0}^{\prime}) & \mathbf{p\cdot(p}_{0}-%
\mathbf{p}_{0}^{\prime})%
\end{array}
\right\vert +\left\vert
\begin{array}{cc}
\mathbf{B\cdot p}_{0} & \mathbf{p\cdot p}_{0} \\
\mathbf{B\cdot p}_{0} & \mathbf{p\cdot p}_{0}^{\prime}%
\end{array}
\right\vert  \label{4-14} \\
& =(\mathbf{B\cdot p)(p\cdot(p}_{0}-\mathbf{p}_{0}^{\prime}))+(\mathbf{%
B\cdot p}_{0})(\mathbf{p\cdot(p}_{0}^{\prime}-\mathbf{p}_{0}))  \notag \\
& =\left[ \mathbf{p\cdot(p}_{0}-\mathbf{p}_{0}^{\prime})\right] \left[
\mathbf{B\cdot(p-p}_{0})\right]  \notag
\end{align}

Finally, by Lemma \ref{positiv} the second factor in (\ref{4-13}) is always
positive, whereas the first factor is positive on the northern hemisphere
and vanishes precisely on the equator circle.
\end{proof}

\subsection{The monotone m-latitude theorem}

Spherical polar coordinates $(\varphi,\theta)$ centered at the north pole $N$
parametrize, of course, the latitude and longitude (meridian) circles, which
constitute the pair of coaxial families of circles associated to the pair $%
\left\{ N,S\right\} $ of geometric centers of the two hemispheres. However,
instead of using the colatitude $\varphi$ let us rather parametrize the
latitude circles by the \emph{latitude in radians}, $-\pi/2\leq\lambda\leq
\pi/2$, namely $\lambda=$ $\pi/2-\varphi$ and hence $\lambda$ is positive on
the northern hemisphere.

For equal masses the pair of minima $\left\{ \mathbf{p}_{0},\mathbf{p}%
_{0}^{\prime}\right\} $ of $U^{\ast}$ happens to coincide with the pair $%
\left\{ N,S\right\} $, but this does not hold for non-equal masses. However,
there exists a unique M\"{o}bius transformation which maps $\mathbf{p}_{0}$
to $N$, $\mathbf{p}_{0}^{\prime}$ to $S$ and the equator circle $E^{\ast}$
to itself. Such a M\"{o}bius transformation maps $\mathcal{F}$ (resp. $%
\mathcal{F}^{\prime}$) to the family of meridians (resp. latitude circles).

\begin{definition}
For a given mass distribution $\left\{ m_{1},m_{2},m_{3}\right\} $, the
\emph{m-modified latitude} of $\mathbf{p\in S}^{2}(1)$ is defined to be the
latitude in radians of the image of $\mathbf{p}$ under the above M\"{o}bius
transformation, and it is denoted by $\lambda(\mathbf{p}).$
\end{definition}

For example, $\lambda(\mathbf{p}_{0})=\pi/2,\lambda(\mathbf{p}_{0}^{\prime
})=-\pi/2$, and $\lambda(\mathbf{p})=0$ if and only if $\mathbf{p\in}E^{\ast}
$. Moreover, $\lambda(\mathbf{p})=-\lambda(\mathbf{p}^{\prime})$ for any
pair $\left\{ \mathbf{p,p}^{\prime}\right\} $ representing similar
m-triangles of opposite orientations.

For a given (smooth) curve $\gamma^{\ast}(t)$ on the sphere we shall
consider the associated function%
\begin{equation}
\lambda_{\gamma^{\ast}}(t)=\lambda(\gamma^{\ast}(t))   \label{lambda}
\end{equation}
which records the m-modified latitude along the curve. It turns out that for
shape curves representing three-body motions with zero angular momentum this
function has a remarkable monotonicity property. Namely, it oscillates
between local maxima where it is positive and local minima where it is
negative, and between two such extremals it is monotonic. The only
exceptions arise when the function is a constant, as described by the
following lemma.

\begin{lemma}
If the m-latitude function $\lambda_{\gamma^{\ast}}$ is constant along $%
\gamma^{\ast}$, then $\gamma^{\ast}$ is an exceptional shape curve which is
either a single point or is confined to the equator circle (cf. Definition %
\ref{exceptional}).
\end{lemma}

\begin{proof}
Assume $\gamma^{\ast}$ is not a single point and is confined to an
m-modified latitude circle $\Gamma^{\prime}$ different from the equator. We
choose a spherical polar coordinate system $(\varphi,\theta)$ centered at
the geometric center of $\Gamma^{\prime}$; hence $\varphi$ is also constant
along $\gamma^{\ast}$. By equation (ii) of the system (\ref{3-1}), $%
U_{\varphi }^{\ast}$ must be negative along $\Gamma^{\prime}$, that is, the
gradient $\nabla U^{\ast}$ is pointing inward along the circle. However, $%
\Gamma ^{\prime}$ encloses $\mathbf{p}_{0}$ (or $\mathbf{p}_{0}^{\prime}$)
and Proposition \ref{south} tells us that $\nabla U^{\ast}$ is pointing
outward, so this is a contradiction.
\end{proof}

Now, let us assume $\gamma^{\ast}$ is not exceptional as in the above lemma.
We shall state and prove the \emph{Monotone m-latitude theorem} :

\begin{theorem}
\label{monotone}Let $\gamma^{\ast}(t),a\leq t\leq b,$ be a segment of the
associated shape curve of a 3-body trajectory with vanishing angular
momentum, and let $\lambda_{\gamma^{\ast}}$ as in (\ref{lambda}) be the
function recording the m-modified latitude along $\gamma^{\ast}[a,b]$.
Suppose that $a\leq t_{0}\leq b$ is a critical point of $\lambda_{\gamma^{%
\ast}}$ (i.e. $\lambda_{\gamma^{\ast}}^{\prime}(t_{0})=0)$ or is possibly a
singularity. Then $\lambda_{\gamma^{\ast}}(t_{0})$ must \ be a local maximum
(resp. minimum) when $\gamma^{\ast}(t_{0})$ lies on the northern (resp.
southern) hemisphere.
\end{theorem}

\begin{proof}
We may assume the point $\mathbf{q}_{0}=\gamma^{\ast}(t_{0})$ on the
m-modified latitude circle $\Gamma^{\prime}=\Gamma_{q_{0}}^{\prime}$ is
strictly inside either the northern or southern hemisphere, since $%
\gamma^{\ast}$ crosses the equator transversely or it hits a binary
collision point and bounces back into the same hemisphere (by
regularization). Moreover, by using the reflectional symmetry which reverses
orientation we may reduce the proof to the case that $\mathbf{q}_{0}$ lies
on the northern hemisphere.

There are two cases to consider; either $\mathbf{q}_{0}$ is a cusp, that is,
the speed $v_{0}$ of $\gamma^{\ast}$ vanishes, or $\mathbf{q}_{0}$ is a
regular point ($v_{0}\neq0)$ and hence the curvature function of $\gamma
^{\ast}(t)$ is smooth at $t_{0}.$

If a cusp is encountered at $\mathbf{q}_{0}$, it cannot be the critical
point $\mathbf{p}_{0}$ of $U^{\ast}$. In general, the nonzero vector $\nabla
U^{\ast}(\mathbf{q}_{0})$ actually gives the outgoing direction of the cusp,
which by Proposition \ref{south} is directed "southward" and hence $%
\lambda_{\gamma^{\ast}}$ is strictly increasing (resp. decreasing) when $%
\gamma^{\ast}$ approaches (resp. leaves) $\mathbf{q}_{0}$.

In the other case, $\gamma^{\ast}$ and $\Gamma^{\prime}$ are tangent to each
other at $\mathbf{q}_{0}$. By reversal of time if necessary, we may assume
that the velocity vector of $\gamma^{\ast}$ at $\mathbf{q}_{0}$ points in
the "eastward" direction of $\Gamma^{\prime}$, that is, the positive
direction of $\Gamma^{\prime}$ as the oriented boundary of the circular cap
containing $\mathbf{p}_{0}$. Geometrically speaking, a local maximum of $%
\lambda _{\gamma^{\ast}}$ at $t_{0}$ means exactly that the geodesic
curvature $K_{0}$ of $\gamma^{\ast}$ at $\mathbf{q}_{0}$ is strictly less
than that of $\Gamma^{\prime}$, which is a positive constant $k_{0}$. Thus,
it suffices to show $K_{0}<k_{0}$, and we claim, in fact, that $K_{0}\leq0$.

Suppose to the contrary that $K_{0}$ is at least positive, and recall
Theorem \ref{conesurface} and its identity (\ref{3-8b}), but scaled with $%
M^{\ast}$ as the unit sphere. Then, on the one hand
\begin{equation*}
(\sin^{2}\alpha)K_{0}\geq0,
\end{equation*}
and on the other hand, the positive normal vector to $\gamma^{\ast}$ is
\begin{equation*}
\mathbf{\nu}^{\ast}=\frac{-\mathbf{T(\mathbf{q}_{0})}}{\left\vert \mathbf{T(%
\mathbf{q}_{0})}\right\vert }
\end{equation*}
and, by Proposition \ref{south}, evaluation at $\mathbf{\mathbf{q}_{0}}$
yields
\begin{align*}
-\frac{d}{d\mathbf{\nu}^{\ast}}\ln(U+h) & =-\frac{1}{U+h}\nabla U^{\ast
}\cdot\mathbf{\nu}^{\ast} \\
& =\frac{1}{\ (U+h)}\nabla U^{\ast}\mathbf{\cdot}\frac{\mathbf{T}}{%
\left\vert \mathbf{T}\right\vert }>0
\end{align*}
This implies that
\begin{equation*}
2(\sin^{2}\alpha)K_{0}-\frac{d}{d\mathbf{\nu}^{\ast}}\ln(U+h)>0
\end{equation*}
and hence contradicts the identity (\ref{3-8b}). Consequently, $K_{0}$
cannot be positive and, in particular, we conclude that $\lambda_{\gamma^{%
\ast}}(t_{0})$ is a local maximum.
\end{proof}

\begin{corollary}
Suppose that $\gamma^{\ast}(t)$ is the associated shape curve of a 3-body
motion with vanishing angular momentum, without ever encountering a triple
collision or escape to infinity. Then it contains infinitely many eclipse
points (resp. local maxima and local minima for $\lambda_{\gamma^{\ast}}$),
and they occur at alternating sequences of times.
\end{corollary}

\section{The asymptotic behavior at a triple collision}

Three-body motions leading to a triple collision have vanishing angular
momenum, and their moduli curves $\bar{\gamma}(t)$ are exactly those
geodesic curves in $(\bar{M}_{h},d\bar{s}_{h}^{2})$ leading to the base
point $O$ $=(\rho=0)$ as the limit$.$Therefore, it is also natural to review
and study their basic asymtotic properties at the triple collision.

The classical works of Sundman and Siegel tell us that the triple collision
is the only essential singularity of three-body motions, and their \emph{%
asymptotic theorem}, briefly stated as Theorem \ref{asymptotic} below,\emph{%
\ }gives a qualitative description of the behavior at the singularity. We
mention here some major works in the classical literature which have
contributed to the understanding of the collision motions, namely Sundman%
\cite{Sund1}, \cite{Sund2}, Levi-Civita\cite{Lev}, Siegel\cite{Sieg1}, \cite%
{Sieg2}, Siegel-Moser \cite{SM}, Wintner\cite{Win}. Unfortunately, the
proofs one finds in the above literature are rather long and difficult, and
thus it is worthwhile to provide simpler proofs, as well as improvements of
their results, in the setting of kinematic geometry.

The asymptotic theorem is, in fact, a direct consequence of the asymptotic
estimates of $I$ and its lowest derivatives $\dot{I}$ and $\ddot{I}$, and
since $\rho=\sqrt{I}$ is the kinematic distance to the triple collision
(base) point, these lower order asymptotic estimates are needed somehow for
any proof of the above theorem. On the other hand, the theorem can actually
be regarded as the geometric interpretation of such estimates.

Along the way we shall also give remarks on the works of Sundman and Siegel,
and in the final subsection we shall apply Wintner's idea of using a
logarithmic time scale to deduce the asymptotic formulae for the time
derivatives of $I$ of any order, cf. Theorem \ref{higher}.

\subsection{Ray solutions as a model for the asymptotic behavior at a triple
collision}

We begin with some vector algebra in the Euclidean space $\mathbb{R}%
_{(1)}^{3}\oplus$ $\mathbb{R}_{(2)}^{3}\oplus$ $\mathbb{R}_{(3)}^{3}$ $=%
\mathbb{R}^{9}$ of all triples $\delta=(\mathbf{a}_{1}\mathbf{,a}_{2}\mathbf{%
,a}_{3})$, or rather in the subspace of m-triangles defined by $\sum m_{i}%
\mathbf{a}_{i}=0$, equipped with the Jacobi metric (\ref{2-2}), cross
product and exterior product
\begin{equation*}
\delta\times\delta^{\prime}=\sum m_{i}\mathbf{a}_{i}\times\mathbf{b}_{i}\in%
\mathbb{R}^{3}\text{, \ }\delta\wedge\delta^{\prime}=\sum\mathbf{a}_{i}\wedge%
\mathbf{b}_{j}\in\wedge^{2}\mathbb{R}^{9}\text{\ \ }
\end{equation*}
where the standard basis vectors $\mathbf{e}_{r}\wedge\mathbf{e}_{s}\in%
\mathbb{R}_{(i)}^{3}\wedge\mathbb{R}_{(j)}^{3}$ has length $\sqrt {m_{i}m_{j}%
}$. Some useful relationships between these operations are expressed by
\begin{align*}
\left\vert \delta\wedge\delta^{\prime}\right\vert ^{2} & =\det\left\vert
\begin{array}{cc}
\delta\cdot\delta & \delta\cdot\delta^{\prime} \\
\delta\cdot\delta^{\prime} & \delta^{\prime}\cdot\delta^{\prime}%
\end{array}
\right\vert =\left\vert \delta\right\vert ^{2}\left\vert \delta^{\prime
}\right\vert ^{2}-(\delta\cdot\delta^{\prime})^{2}\geq\left\vert \delta
\times\delta^{\prime}\right\vert ^{2} \\
\left\vert \delta\wedge\delta^{\prime}\right\vert ^{2} & =\sum
m_{i}^{2}\left\vert \mathbf{a}_{i}\times\mathbf{b}_{i}\right\vert
^{2}+\sum\limits_{i<j}m_{i}m_{j}\mu_{ij}\mathbf{\ }\text{\ \ }(\mu_{ij}\geq0)
\end{align*}
In particular, for a motion $\delta(t)$ with velocity $\dot{\delta}(t)$ and
individual angular momenta $\mathbf{\Omega}_{i}$, we deduce the relations
\begin{equation}
\left\vert \delta\wedge\dot{\delta}\right\vert ^{2}=\sum\left\vert \mathbf{%
\Omega}_{i}\right\vert ^{2}+\left\vert \mathbf{\Omega}_{mix}\right\vert
^{2}=2IT-\frac{1}{4}\dot{I}^{2}\geq\left\vert \mathbf{\Omega }\right\vert
^{2}   \label{5-0}
\end{equation}
These can also be interpreted in tems of the splitting of kinetic energy%
\begin{equation*}
T=T^{\rho}+(T^{\sigma}+T^{\omega})=\frac{1}{8}\frac{\dot{I}^{2}}{I}+\frac{%
\left\vert \delta\wedge\dot{\delta}\right\vert ^{2}}{2I}
\end{equation*}
where $T^{\rho}$, $T^{\sigma}$or $T^{\omega}$ is due to radial motion,
change of shape, or (rigid) rotational motion, respectively. In the case of
planary motions, equality holds in (\ref{5-0}) since $T^{\omega}=$ $%
\left\vert \mathbf{\Omega}\right\vert ^{2}/2I$. In particular, equality
holds if $\mathbf{\Omega}=0$ since in that case the motion is planary, e.g.
by a simple geometric argument.

Ray solutions provide, of course, the simplest examples of triple collision
motions. Here each particle moves along a fixed line through the center of
gravity (origin) and hence $\mathbf{\Omega}_{i}=0$ for each $i$, and also $%
T^{\sigma}=T^{\omega}=0$. By Remark \ref{shape}, the shape of the ray is a
critical point of $U^{\ast}$ on the 2-sphere $M^{\ast}$, namely a Lagrange
point or an Euler point. In other words, the motion is either a homothetic
deformation of an equilateral triangle or a degenerate triangle which is an
Euler configuration. Let $\mu$ be the value of $U^{\ast}$ at the above
critical point. Then the Lagrange-Jacobi equation (\ref{L-J}) reads
\begin{equation}
\ddot{I}=2\mu\frac{1}{\sqrt{I}}+4h   \label{L-Jb}
\end{equation}
and hence the only variable of the problem, $I(t),$ is the solution of a
1-dimensional Kepler problem.

The equation (\ref{L-Jb}) can be solved explicitly, but we seek the
solutions with the (singular) initial condition $I(0)=0$. In the special
case of $h=0$,
\begin{equation}
I(t)=Kt^{\frac{4}{3}}\text{, \ \ \ }K=(\frac{9\mu}{2})^{\frac{2}{3}}\text{ \
\ }   \label{5-10a}
\end{equation}
and for general $h$ there is a formula $t=$ $F_{h}(I)$ which can be inverted
and, for example, this yields a series development of type%
\begin{equation*}
I(t)=t^{\frac{4}{3}}(K+\sum\limits_{i=1}^{\infty}k_{i}t^{\frac{2i}{3}})
\end{equation*}

To facilitate our study of asymptotic estimates in general, let us introduce
the commonly used notation
\begin{equation}
f\sim g\text{ \ \ }\Longleftrightarrow\lim\text{\ }f/g=1\text{ \ as }%
t\rightarrow0   \label{5-10c}
\end{equation}
Then all ray solution have the same \emph{asymptotic behavior} at $t=0$, in
the sense that their time derivatives of $I(t)$ at $t=0$ yield the same
asymptotic formulae, beginning with
\begin{equation}
I(t)\sim Kt^{\frac{4}{3}},\ \ \dot{I}(t)\sim\frac{4}{3}Kt^{\frac{1}{3}},\
\ddot{I}\ \sim\frac{4}{9}Kt^{-\frac{2}{3}},   \label{5-10b}
\end{equation}
and clearly the higher order asymtotic formulae follow the same pattern,
namely
\begin{equation}
\frac{d^{k}}{dt^{k}}I(t)\sim\frac{d^{k}}{dt^{k}}Kt^{\frac{4}{3}}\text{, \ }%
k=0,1,2,3,...   \label{5-11}
\end{equation}

For a general triple collision motion, it was first realized that $\mathbf{%
\Omega}$ must vanish, hence also $T^{\omega}$ vanishes. On the other hand,
although $T^{\rho}$ was found to be the dominating kinetic energy, $%
T^{\sigma}$ cannot vanish for a non-radial motion and may perhaps tend to
infinity at some lower order of magnitude. However, although the general
asymptotic behavior is certainly more involved due to the change of shape,
it turns out that the estimates (\ref{5-11}) still hold, by Theorem \ref%
{higher}. Moreover, a general triple collision motion has one of the above
simple ray solutions as its asymptotic limit, according to Theorem \ref%
{asymptotic}.

Exact information on the limiting behavior of the shape is not really needed
to derive the asymptotic formulae (\ref{5-11}) for the motion in the radial
direction. The theory of Sundman and Siegel establishes the formulae only up
to $k=2$, namely the asymptotic estimates (\ref{5-10b}). To proceed from $%
k=0,1$ to $k=2$ Siegel introduced the following function $g(t)$ and proved
its crucial property
\begin{equation*}
g(t)=(8IT-\dot{I}^{2})t^{-2/3}\rightarrow0\text{ \ as }t\rightarrow0\text{ \
\ \ (cf. Siegel\cite{Sieg2}, Chap. III, \S 1)}
\end{equation*}
which in our setting can be reformulated as
\begin{equation}
\frac{\left\vert \delta\wedge\dot{\delta}\right\vert ^{2}}{\rho}=2\rho
T^{\sigma}=\frac{1}{4}\rho^{3}v^{2}\rightarrow0\text{ \ as }t\rightarrow0
\label{Siegel2}
\end{equation}
where $v$ is the speed of the shape curve. From our viewpoint, we recognize
the expression in (\ref{Siegel2}) as the Siegel function $\mathfrak{S}$ of
the associated shape curve $\delta^{\ast}(t)$ on the 2-sphere, cf. (\ref%
{3-16}).

\subsection{The results of Sundman and Siegel}

The following basic fact on the vanishing of the angular momentum of
three-body motions leading to triple collision had already been stated by
Weierstrass when Sundman first proved the following classical statement at
the beginning of the 20th century.

\begin{lemma}
\label{Sundman}(Sundman) The angular momentum $\mathbf{\Omega}$ is
necessarily zero for a triple collision motion.
\end{lemma}

\begin{proof}
First, by translation and (possibly) reversal of time, we shall rather
assume (in all Section 6) there is a \emph{triple explosion} at $t=0$. Using
the Lagrange-Jacobi equation (\ref{L-J}), it follows from $I\rightarrow0$
that $\ddot{I}\rightarrow\infty$ and hence $\dot{I}>0$ for $t\in(0,t_{0})$
and $t_{0}$ suitably small. Sundman discovered and made use of the rightmost
inequality in (\ref{5-0}). Namely, for a given value $\left\vert \mathbf{%
\Omega}\right\vert >0$ it is not so difficult to see that $I(t)$ has a
positive lower bound.

However, we shall proceed with a slightly different proof since (\ref{5-0})
also involves the individual momenta $\mathbf{\Omega}_{i}$, and this will
enable us to prove a stronger version of the lemma (see Corollary \ref%
{individ}). For that purpose we set
\begin{equation*}
C(t)=\left\vert \delta\wedge\dot{\delta}\right\vert ^{2}\text{, \ }%
C_{0}=\inf C\text{ \ for }t\in(0,t_{0})\text{\ }
\end{equation*}
Now, multiplying the Lagrange-Jacobi equation by $\dot{I}$ gives the
inequality
\begin{equation*}
\dot{I}\ddot{I}=\dot{I}(2T+2h)=\frac{\dot{I}}{I}C+2h\dot{I}+\frac{1}{4I}\dot{%
I}^{3}\geq\frac{\dot{I}}{I}C+2h\dot{I}
\end{equation*}
which by integration yields
\begin{equation*}
\int\limits_{t}^{t_{0}}\dot{I}\ddot{I}dt\geq\int\limits_{t}^{t_{0}}\frac {%
\dot{I}}{I}Cdt+2h(I_{0}-I)
\end{equation*}%
\begin{equation*}
\frac{1}{2}\dot{I}_{0}^{2}\geq\frac{1}{2}(\dot{I}_{0}^{2}-\dot{I}^{2})\geq
C_{0}\ln(\frac{I_{0}}{I})+2h(I_{0}-I)
\end{equation*}%
\begin{equation*}
C_{0}\leq\frac{2hI-2hI_{0}+\frac{1}{2}\dot{I}_{0}^{2}}{\ln(I_{0}/I)}%
\rightarrow0\text{ \ as }t\rightarrow0
\end{equation*}
Hence, $C_{0}=0$ and the constant sum $\mathbf{\Omega=}\sum\mathbf{\Omega}%
_{i}$ is zero.
\end{proof}

\begin{remark}
Clearly, the above proof also gives $\inf$ $\left\vert \mathbf{\Omega}%
_{i}\right\vert =0$ for each $i$, but one cannot yet conclude that $\mathbf{%
\Omega}_{i}\rightarrow0$. See Corollary \ref{individ} for this last step.
\end{remark}

The major results of Sundman and Siegel concerning a general triple
collision motion $\delta(t)$ can be summarized as follows.

\begin{itemize}
\item \textbf{Sundman} : $t^{-\frac{2}{3}}\bar{\delta}(t)$ tends to a limit
whose shape is that of an equilateral triangle or an Euler configuration.

\item \textbf{Siegel} : The magnified or "big triangle" $t^{-\frac{2}{3}%
}\delta(t)$ approaches a fixed m-triangle $\tilde{\delta}_{0}$ in the
Euclidean configuration space $M.$
\end{itemize}

Standard references for proofs of these statements are Siegel\cite{Sieg2}
and Siegel-Moser\cite{SM}. One finds that results proved by Sundman are,
typically, seen to express properties at the moduli space level, that is,
statements about the moduli curve $\bar{\delta}(t)$. Siegel improved his
results by lifting them up to the configuration space level, where he
studied the motion of the "big triangle" in the Hamiltonian setting and
performed a series of successive canonical transformations to simplify the
analysis.

A major step was to establish the validity of the above asymptotic estimates
(\ref{5-10b}), and with the following proposition we shall provide a proof
of this - in the spirit of Sundman and Siegel. Moreover, for the sake of
completeness, in the last subsection we shall extend the proof to the higher
order asymptotic estimates, $k>2$ in (\ref{5-11}), using ideas due to
Wintner.

\begin{proposition}
\label{lower}For a three-body motion $\delta(t)$ with $I(t)\rightarrow0$ as $%
t\rightarrow0$, the asymptotic estimates in (\ref{5-10b}) hold, where $K$ is
the expression in (\ref{5-10a}) with $\mu=\lim U^{\ast}(\delta^{\ast}(t))$.
\end{proposition}

\begin{proof}
For a ray solution, $I=\rho^{2}\sim Kt^{\frac{4}{3}}$ and $\rho\dot{\rho}%
^{2}\sim\frac{4}{9}K^{\frac{3}{2}}$. This suggests a study of the asymptotic
behavior of $\rho\dot{\rho}^{2}$ for triple collisions in general, using
equation (\ref{L-J}) and kinematic geometry. Now, $\mathbf{\Omega}=0$ and
the kinetic energy has the splitting
\begin{equation}
T=\frac{1}{2}(\frac{d\bar{s}}{dt})^{2}=T^{\rho}+T^{\sigma}=\frac{1}{2}\dot{%
\rho}^{2}+\frac{1}{8}\rho^{2}v^{2}\text{, \ cf. (\ref{2-8})}   \label{5-12}
\end{equation}
and we set
\begin{equation}
\mathfrak{R}(t)=\rho T^{\rho}\text{, \ \ }\mathfrak{\tilde{S}}(t)=\frac{1}{2}%
\mathfrak{S(t)}=\rho T^{\sigma}   \label{5-13}
\end{equation}
where $\mathfrak{S}$ is the Siegel function, see (\ref{C10}), (\ref{3-16}).
By the energy integral $U^{\ast}=\rho T-\rho h$ we can also write
\begin{equation}
U^{\ast}\sim\mathfrak{R}(t)+\mathfrak{\tilde{S}}(t)\text{ as }t\rightarrow0
\label{5-14}
\end{equation}

Our first claim is that
\begin{equation}
\mathfrak{R}(t)\rightarrow\mu>0\text{ \ and \ }\mathfrak{\tilde{S}}%
(t)\rightarrow0   \label{5-15}
\end{equation}
as $t\rightarrow0$. Let us differentiate $\mathfrak{R}$ and substitute for $%
\ddot{I}$ using (\ref{L-J}), or equivalently (i) in (\ref{3-1}), to obtain
\begin{equation*}
\mathfrak{\dot{R}}=\frac{d}{dt}(\frac{\rho}{2}\dot{\rho}^{2})=-\frac{1}{2}%
\dot{\rho}^{3}+\dot{\rho}(U+2h)=\dot{\rho}(T^{\sigma}+h)
\end{equation*}%
\begin{equation}
\int_{t}^{t_{0}}\mathfrak{\dot{R}}dt=\mathfrak{R}(t_{0})-\mathfrak{R}%
(t)=\int_{t}^{t_{0}}\dot{\rho}T^{\sigma}dt+h(\rho(t_{0})-\rho(t))
\label{5-6}
\end{equation}
Since $\mathfrak{R}\geq0$ and the integral on the right side is $\geq0$, $%
\lim\mathfrak{R}(t)=$ $\mu\geq0$ must exist, that is, $\mathfrak{R}=\mu+o(1)$
for small $t$.

Suppose we had $\mu=0$. Since $\min U^{\ast}>0$, equation (i) in (\ref{3-1})
implies
\begin{equation*}
\ddot{\rho}\rho^{2}=U^{\ast}+2h\rho-2\mathfrak{R}=U^{\ast}+o(1)\geq C>0
\end{equation*}
and consequently
\begin{equation*}
\dot{\rho}\ddot{\rho}=\frac{\dot{\rho}}{\rho^{2}}(\ddot{\rho}\rho^{2})=\frac{%
\dot{\rho}}{\rho^{2}}(U^{\ast}+o(1))\geq C\frac{\dot{\rho}}{\rho^{2}}
\end{equation*}%
\begin{align*}
\dot{\rho}(t_{0})^{2}-\dot{\rho}(t)^{2} & =2\int_{t}^{t_{0}}\dot{\rho}\ddot{%
\rho}dt\geq2C\int_{t}^{t_{0}}\frac{\dot{\rho}}{\rho^{2}}dt \\
& =2C(\frac{1}{\rho(t)}-\frac{1}{\rho(t_{0})})\rightarrow\infty
\end{align*}
This is clearly impossible, so we conclude $\mu>0$.

Next, we deduce successively
\begin{align}
\frac{1}{2}\rho\dot{\rho}^{2} & =\mu+o(1)\text{ \ }\Longrightarrow\dot{\rho }%
^{2}=\frac{1}{\rho}(2\mu+o(1))\Longrightarrow  \notag \\
\dot{\rho} & =\sqrt{2\mu}\text{ }\rho^{-\frac{1}{2}}+o(\rho^{-\frac{1}{2}})
\label{5-2}
\end{align}%
\begin{align}
\frac{2}{3}\rho^{\frac{3}{2}} & =\int_{0}^{\rho}\sqrt{\rho}d\rho=\int
_{0}^{t}\dot{\rho}\sqrt{\rho}dt  \notag \\
& =\int_{0}^{t}(\sqrt{2\mu}+o(1))dt=\sqrt{2\mu}t+o(t)   \label{5-3}
\end{align}
and hence by (\ref{5-2}) and (\ref{5-3})
\begin{equation}
I=\rho^{2}\sim Kt^{\frac{4}{3}}\text{, \ \ \ }K=(\frac{9}{2}\mu)^{\frac{2}{3}%
}   \label{5-4}
\end{equation}%
\begin{equation}
\dot{I}=2\rho\dot{\rho}\sim\frac{4}{3}Kt^{\frac{1}{3}}\text{, \ \ }T^{\rho }=%
\frac{1}{2}\dot{\rho}^{2}\sim\frac{\mu}{\rho}   \label{5-5}
\end{equation}

Next we show $\mathfrak{\tilde{S}}(t)\rightarrow0$. The integral on the
right side of (\ref{5-6}) exists, but the integrand is
\begin{equation*}
\dot{\rho}T^{\sigma}=\frac{\dot{\rho}}{\rho}\mathfrak{\tilde{S}}\sim\frac {2%
}{3}\frac{\mathfrak{\tilde{S}}}{t}
\end{equation*}
and the integral of $1/t$ is divergent, hence $\lim\inf\mathfrak{\tilde{S}}%
(t)=0$.

It is, however, more difficult to show $\lim\sup\mathfrak{\tilde{S}}(t)=0$,
but let us apply an idea from Siegel\cite{Sieg2}. Namely, suppose to the
contrary, that $\lim\sup\mathfrak{\tilde{S}}(t)>0$. Then, for a given $%
\epsilon>0$ there is an infinite decreasing sequence of numbers in $(0,t_{0})
$, $\epsilon>t_{1}>t_{2}>..>t_{k}>0$, $\lim t_{i}=0$, so that
\begin{align}
& \epsilon\leq\mathfrak{\tilde{S}}(t)\leq3\epsilon\text{, \ for }t\in
J_{k}=[t_{2k},t_{2k-1}]  \notag \\
& \mathfrak{\tilde{S}}(t_{2k})=\epsilon\text{, \ }\mathfrak{\tilde{S}}%
(t_{2k-1})=3\epsilon  \label{5-7} \\
& \left\vert \mathfrak{R}(t_{2k})-\mathfrak{R}(t_{2k-1})\right\vert
\leq\epsilon  \notag
\end{align}
By (\ref{5-14}), in each interval $J_{k}$, $U^{\ast}$ and hence also the
norm of $\nabla U^{\ast}$ are bounded by the same constant $C_{1}$, and then
it is not difficult to show
\begin{equation*}
\dot{T}=\dot{U}=O(t^{-\frac{5}{3}})\text{, \ for }t\in J_{k}
\end{equation*}
But for small $t$ we also have
\begin{equation}
T=U+h=\frac{1}{\rho}(U^{\ast}+h\rho)\leq\frac{C_{2}}{\rho}   \label{5-16}
\end{equation}
for a suitable constant $C_{2}$, consequently
\begin{equation}
\left\vert \frac{d}{dt}(\rho T)\right\vert =\left\vert \rho\dot{T}+\dot{\rho
}T\right\vert \leq\frac{C_{2}}{t}\text{, \ for }t\in J_{k}\text{ }
\label{5-17}
\end{equation}

By (\ref{5-7}) and (\ref{5-17}),
\begin{align*}
2\epsilon & =\mathfrak{\tilde{S}}(t_{2k-1})-\mathfrak{\tilde{S}}%
(t_{2k})=(\rho T-\mathfrak{R})_{t=t_{2k-1}}-(\rho T-\mathfrak{R})_{t=t_{2k}}
\\
& \leq C_{2}\int_{t_{2k}}^{t_{2k-1}}\frac{dt}{t}+\epsilon
\end{align*}
which implies
\begin{equation}
\int_{t_{2k}}^{t_{2k-1}}\frac{\mathfrak{\tilde{S}}(t)}{t}dt\geq\epsilon
\int_{t_{2k}}^{t_{2k-1}}\frac{dt}{t}\geq\ \frac{\epsilon^{2}}{C_{2}}\text{ ,
\ for each }k   \label{5-18}
\end{equation}
and hence the sum of the integrals is infinite. On the other hand,
\begin{equation*}
\frac{\mathfrak{\tilde{S}}(t)}{t}=\frac{\rho T^{\sigma}}{t}\sim\frac{3}{2}%
\dot{\rho}T^{\sigma}
\end{equation*}
and the integral of $\dot{\rho}T^{\sigma}$ on $[0,t_{0}]$ exists by (\ref%
{5-6}), so this is a contradiction.

Having proved that $\mathfrak{\tilde{S}}(t)\rightarrow0$, it follows from (%
\ref{5-14}) that $U^{\ast}\rightarrow\mu$, and now the Lagrange-Jacobi
equation yields
\begin{equation}
\ddot{I}\sim2T\sim2T^{\rho}=\dot{\rho}^{2}\sim\frac{4}{9}Kt^{-\frac{2}{3}}
\label{5-19}
\end{equation}
where $K$ is the expression from (\ref{5-4}). This completes the proof.
\end{proof}

\begin{corollary}
\label{individ}The quantity $\left\vert \delta\wedge\dot{\delta}\right\vert
^{2}$ in (\ref{5-0}) tends to zero at the triple collision. In particular,
the individual angular momenta $\mathbf{\Omega}_{i}$ as well as the "mixed"
momentum term $\mathbf{\Omega}_{mix}$ tend to zero.
\end{corollary}

Since $\mathfrak{S}(t)\rightarrow0$, the above statement follows immediately
from
\begin{equation}
\left\vert \delta\wedge\dot{\delta}\right\vert ^{2}=2IT-\frac{1}{4}\dot{I}%
^{2}=2IT^{\sigma}=2\rho\mathfrak{\tilde{S}}\rightarrow0\text{, \ \ cf. (\ref%
{5-0})}   \label{5-20}
\end{equation}

By "infinite magnification" at the triple collision the solution $\delta(t)$
coincides with one of the ray solutions in Section 6.1. This is the idea
behind the classical \emph{asymptotic theorem}, and now we give a simple
proof of this in the setting of kinematic geometry.

\begin{theorem}
\label{asymptotic}(Sundman-Siegel) Any triple collision orbit is asymptotic
to one of the ray solutions.
\end{theorem}

\begin{proof}
By (\ref{3-10}) and (\ref{5-19})
\begin{equation*}
2T=(\frac{d\bar{s}}{dt})^{2}\sim\frac{4}{9}Kt^{-\frac{2}{3}}\sim(\frac{d\rho
}{dt})^{2}=2T^{\rho}
\end{equation*}%
\begin{equation}
\cos^{2}\alpha=(\frac{d\rho}{d\bar{s}})^{2}\rightarrow1\text{ \ as }%
t\rightarrow0   \label{5-21}
\end{equation}
which simply means that the moduli curve $\bar{\delta}(t)$ of the given
triple collision motion $\delta(t)$ is tangent to a ray or, equivalently,
the limit of its \emph{infinite magnification} exists. It also follows that
the limit ray must itself be a geodesic in $\bar{M}_{h}$, namely one of
those rays representing the shape of a Lagrange or Euler configuration.

However, the claim is also that $\delta(t)$ itself approaches a ray in $M$.
To see this, consider as above the angle $\tilde{\alpha}$ between the radial
and tangential direction in $M$, that is, the angle between the vectors $%
\delta(t)$ and $\dot{\delta}(t)$. It follows that
\begin{equation}
\cos\tilde{\alpha}=\frac{\delta\cdot\dot{\delta}}{\left\vert
\delta\right\vert \left\vert \dot{\delta}\right\vert }=\frac{\dot{\rho}}{%
\left\vert \dot{\delta }\right\vert }=\frac{\dot{\rho}}{\sqrt{2T}}%
\rightarrow1\text{ \ as }t\rightarrow0   \label{5-22}
\end{equation}

The limit ray in $M$ projects to a geodesic ray in $\bar{M}_{h}$, namely a
ray consisting of the homothetic images of either a fixed equilateral
triangle or a fixed degenerate triangle of Euler's type. As shown in Section
6.1, these are the rays which admit triple collision motions, and thus the
given motion $\delta(t)$ will be asymptotic to the corresponding limit ray
solution with the same energy $h$.
\end{proof}

\begin{corollary}
The shape curve $\delta^{\ast}(t)$ converges to a Lagrange or Euler point $%
\delta_{0}^{\ast}$ on the 2-sphere, and the "big triangle" $t^{-2/3}\delta(t)
$ converges to an m-triangle $\tilde{\delta}_{0}$ with the shape $%
\delta_{0}^{\ast}$ and moment of inertia
\begin{equation}
\tilde{I}_{0}=K=(\frac{9\mu}{2})^{\frac{2}{3}}\text{, }\mu=U^{\ast}(\delta
_{0}^{\ast})   \label{K2}
\end{equation}
\end{corollary}

\begin{remark}
\label{Gauss-Bonnet}Actually, a limiting shape of Euler's type cannot be
reached unless the whole three-body motion itself is collinear, see e.g. \S %
13 in Siegel-Moser\cite{SM}, which refines and improves the classical
Sundman-Siegel approach. The latter is described in detail in Siegel's
lectures \cite{Sieg2} of about 240 pages. On p.138 he writes :"The
difficulty of the problem consists in the fact that we cannot yet prove
(this will be proved only at the end) that the big triangle referred to a
\underline{fixed} coordinate system has a limiting position as $t\rightarrow0
$; all that we have proved so far is the existence of a limiting
configuration relative to a rotating coordinate system. The triangle itself
may go on rotating about its centre of gravity, ..."
\end{remark}

However, although finiteness of the rotation of the "big triangle" was
proved via the convergence of the "big triangle", neither an estimate of the
actual angle of rotation nor its precise definition was addressed in the
above studies of the triple collision. In reality, the "big triangle" is
approaching its final shape and position quite fast and in a monotonic way.
To make this precise, we propose to measure how much the equilateral
limiting triangle $\tilde{\delta}_{0}$ deviates in position from some
natural \emph{reference} equilateral m-triangle $\zeta$, depending on the
given collision motion $\delta(t)$, but also $\zeta=\zeta(t)$ will be a
function depending on the chosen time interval $\left[ 0,t\right] $ under
consideration. This goes as follows.

We may assume the shape curve $\delta^{\ast}(t)$ is on the northern
hemisphere and hence starts at the Lagrange point $\mathbf{p}%
_{0}=\delta_{0}^{\ast}$. Let $\delta_{1}=\delta(t_{1})$ be the m-triangle at
a given time $t_{1}>0$ and write $\delta_{1}^{\ast}=\delta^{\ast}(t_{1})$.
Then there is a unique \underline{linear} m-triangle motion
\begin{equation*}
Z(t)=\frac{(t_{1}-t)}{t_{1}}\zeta_{1}+\frac{t}{t_{1}}\delta_{1}\text{, \ }%
t\in\left[ 0,t_{1}\right]
\end{equation*}
with vanishing angular momentum, connecting $\delta_{1}$ to some equilateral
m-triangle $\zeta_{1}$ $=\zeta(t_{1})$ (cf. \cite{HS2} , Section 3.3). The
shape curve of this (virtual) motion is the geodesic arc on the sphere $%
S^{2}(1)$ from the Lagrange point $\mathbf{p}_{0}$ to the point $\delta
_{1}^{\ast}$, and together with the curve segment of $\delta^{\ast}(t)$ from
$\delta_{1}^{\ast}$ to $\mathbf{p}_{0}$ they constitute a closed curve $C_{1}
$ on the sphere. We define the \emph{rotation angle }$\psi(t_{1})$ of $%
\delta(t)$ at the triple collision, measured from time $t=t_{1},$ to be half
of the signed area
\begin{equation}
\psi(t_{1})=\frac{1}{2}Area(D_{1})=\int_{C_{1}}\omega   \label{devi}
\end{equation}
of the region $D_{1}$ enclosed by $C_{1}$. This is motivated by the
kinematic Gauss-Bonnet theorem (cf. \cite{HS2}) for three-body motions with
zero angular momentum, where traversal of a loop on the 2-sphere amounts to
a net rotation (i.e. a \emph{geometric phase} ) of the m-triangle in the
configuration space, which can be calculated as the line integral of a
kinematic 1-form $\omega$ (depending on the mass distribution and region of $%
S^{2}$). Moreover, $d\omega$ $=\frac{1}{2}dA$ where $dA$ is the area form of
the unit sphere.

Now, for $t_{1}$ not too large, the shape curve $\delta^{\ast}(t)$ will stay
on one side of the geodesic arc since its curvature will have a fixed sign.
So the rotation angle (\ref{devi}) decreases monotonically to zero as $%
t_{1}\rightarrow0$, and hence the kinematic geometric approach explains
Siegel's angle of rotation and yields as well a recipe for how to measure it
quantitatively.

\subsection{Higher order asymptotic estimates at a triple collision}

The asymptotic formulae for the energy functions $\Xi=T,T^{\rho},U$ and
their time derivatives up to order $k$ can be developed inductively together
with those formulae for $I$ up to order $k+2$. However, from the three
identities
\begin{equation}
T^{\rho}=\frac{\dot{I}^{2}}{8I}\text{, \ }T=U+h=\frac{1}{2}\ddot{I}-h
\label{6-0}
\end{equation}
it is easy to show that the three cases of $\Xi$, for a given order $k$,
yield the same asymptotic formula. Therefore, the final description of the
asymptotic behavior of the above quantities can be stated as follows.

\begin{theorem}
\label{higher}For a three-body motion $\delta(t),t\geq0$, with a triple
collision at $t=0$, the following asymptotic estimates hold as $t\rightarrow0
$ :
\begin{equation*}
\frac{d^{k}}{dt^{k}}I\sim\frac{d^{k}}{dt^{k}}(Kt^{\frac{4}{3}})\text{, \ \ }%
\frac{d^{k}}{dt^{k}}T\sim\frac{d^{k}}{dt^{k}}(\frac{\mu}{\rho})\sim \frac{%
d^{k}}{dt^{k}}(\frac{2}{9}Kt^{-\frac{2}{3}})\text{, \ for all }k\geq0
\end{equation*}
where $K=(\frac{9\mu}{2})^{\frac{2}{3}}$, $\mu=U^{\ast}(\delta_{0}^{\ast})$,
and $\delta_{0}^{\ast}$ is the limiting shape.
\end{theorem}

From the initial cases $k=0,1,2$, proved in the previous subsection, we
shall complete the proof of the above theorem for $k>2$ by deducing the
following equivalent formulae for the behavior of $\rho=\sqrt{I}$,
\begin{equation}
\frac{d^{k}}{dt^{k}}\rho\sim\frac{d^{k}}{dt^{k}}(\sqrt{K}t^{\frac{2}{3}})%
\text{, \ }k\geq3,   \label{5-23}
\end{equation}
In fact, they will follow inductively as a rather direct consequence of
Newton's equation (\ref{1-1}) and its energy integral, namely
\begin{equation}
\nabla U(\delta)=\frac{d^{2}}{dt^{2}}\delta\text{, \ \ \ }T=U+h,
\label{5-27}
\end{equation}
but only after an appropriate transformation of space and time. This is the
composition of a time dependent space transformation and a pure time
transformation, as follows :

\begin{itemize}
\item Magnification of the motion $\delta(t)=(\mathbf{a}_{1}(t),\mathbf{a}%
_{2}(t),\mathbf{a}_{3}(t))$ by the time factor $t^{-2/3}$, as in the works
of Sundman and Siegel, to assure convergence at $t=0$ of the magnified
motion. We use the notation \
\begin{align}
\tilde{\delta}(t) & =t^{-2/3}\delta(t)=(\mathbf{\tilde{a}}_{1,}\mathbf{%
\tilde{a}}_{2}\mathbf{,\tilde{a}}_{3})\text{, \ \ }\mathbf{\tilde{a}}%
_{i}(t)=t^{-2/3}\mathbf{a}_{i}(t)  \label{5-23a} \\
f(t) & =f(\delta(t))\text{, \ \ }\tilde{f}(t)=f(\tilde{\delta}(t))\text{, \
\ }\tilde{\delta}_{0}=\lim_{t\rightarrow0}\tilde{\delta}(t)   \label{5-23b}
\end{align}
where $f$ \ is any (homogeneous) function on\ $M$ or its tangent bundle
which we shall evaluate along the trajectory.

\item A logarithmic transformation of time; set
\begin{equation*}
u=-\log t\text{ \ \ (or \ }t=e^{-u})
\end{equation*}
and hence $t\rightarrow0$ means $u\rightarrow\infty$. This transforms a
function $g(t)$ to the function $\breve{g}(u)=g(e^{-u})$.
\end{itemize}

The composition of the two transformations yields the motion $u\rightarrow
\hat{\delta}(u)$ in $M$, and we write
\begin{align}
\hat{\delta}(u) & =\tilde{\delta}(e^{-u})=(\mathbf{\hat{a}}_{1},\mathbf{\hat{%
a}}_{2},\mathbf{\hat{a}}_{3})\text{, \ }\mathbf{\hat{a}}_{i}\mathbf{(}u)=e^{%
\frac{2}{3}u}\mathbf{a}_{i}(e^{-u})  \label{5-28} \\
\text{ \ \ }\hat{f}(u) & =f(\hat{\delta}(u))\text{, \ }\hat{\rho }%
(u)=\left\vert \hat{\delta}(u)\right\vert =e^{\frac{2}{3}u}\rho (e^{-u})%
\text{, \ \ \ }\hat{\delta}_{0}=\tilde{\delta}_{0}  \notag
\end{align}
This motion is, of course, a solution of the transformed equations in (\ref%
{5-27}), which can be stated as%
\begin{align}
\nabla\hat{U} & =\frac{d^{2}}{du^{2}}\hat{\delta}-\frac{1}{3}\frac{d}{du}%
\hat{\delta}-\frac{2}{9}\hat{\delta}  \label{5-32} \\
\hat{T} & =\hat{U}+he^{-\frac{2}{3}u}-\frac{2}{9}\hat{\rho}^{2}+\frac{1}{3}%
\hat{\rho}\frac{d}{du}\hat{\rho}   \label{5-33}
\end{align}
with the appropriate interpretation of $\hat{T}$, $\hat{U}$ and $\nabla%
\hat
{U}$, cf. (\ref{6-3}), (\ref{6-6}). For example, from the above
definitions
\begin{equation*}
\hat{U}(u)=U(\hat{\delta}(u))=e^{-\frac{2}{3}u}U(\delta(e^{-u}))=t^{\frac {2%
}{3}}U(t)
\end{equation*}

To derive the above equations and prepare for its usage, we shall make a few
more definitions and establish some useful identities for differential
operators generated by $\frac{d}{dt}$ and $\frac{d}{du}$. For functions of $%
t $ (or $u$) it is convenient to write \
\begin{equation*}
f_{1}\approx f_{2}\text{ }\Longleftrightarrow(f_{1}-f_{2})=o(1)\text{ as }%
t\rightarrow0\text{ \ (or }u\rightarrow\infty)
\end{equation*}
and, for example, since the magnified motion converges,
\begin{equation}
\hat{\delta}\approx\hat{\delta}_{0}\text{, \ }\hat{\rho}\approx\left\vert
\tilde{\delta}_{0}\right\vert =\sqrt{K}   \label{6-2}
\end{equation}
We say $f(t)$ has \emph{order} $q$ at $t=0$ if \
\begin{equation*}
\frac{f(t)}{t^{q}}\rightarrow f_{0}\neq0\text{ \ as }t\rightarrow0
\end{equation*}
and then the notation
\begin{equation}
\tilde{f}(t)=t^{-q}f(t)\approx f_{0}   \label{5-26}
\end{equation}
is consistent with (\ref{5-23b}) since a homogeneous function $g$ of degree $%
d$ on \ $M$, with $g(\tilde{\delta}_{0})$ $\neq0$, has order $q=2d/3$ at $t=0
$.

The transformed potential function, kinetic energy, and gradient are given
by
\begin{align}
\hat{U} & =U(\hat{\delta})=\sum\limits_{i<j}\frac{m_{i}m_{j}}{\left\vert
\mathbf{\hat{a}}_{i}-\mathbf{\hat{a}}_{j}\right\vert }\text{, \ }\hat{T}=%
\frac{1}{2}\sum m_{i}\left\vert \frac{d}{du}\mathbf{\hat{a}}_{i}\right\vert
^{2}  \label{6-3} \\
\nabla\hat{U} & =\left( \frac{1}{m_{1}}\frac{\partial\hat{U}}{\partial%
\mathbf{\hat{a}}_{1}},...\right) =t^{4/3}\nabla U   \label{6-6}
\end{align}
and by substituting these expressions together with
\begin{equation*}
\frac{d^{2}}{dt^{2}}\delta=\frac{d^{2}}{dt^{2}}(t^{\frac{2}{3}}\tilde{\delta
})=t^{-\frac{4}{3}}\left( -\frac{2}{9}\hat{\delta}-\frac{1}{3}\frac{d}{du}%
\hat{\delta}+\frac{d^{2}}{du^{2}}\hat{\delta}\right)
\end{equation*}
into the equations (\ref{5-27}) one obtains the system (\ref{5-32})-(\ref%
{5-33}).

\begin{lemma}
\label{equiv}If $f(t)$ has order $q$ at $t=0$, with $\tilde{f}(t)\approx
f_{0}$ and $q\notin\left\{ 0,1,2,..\right\} $, then there is the equivalence
\begin{align*}
& \left[ 1\leq k\leq m\text{, \ }t^{k}\frac{d^{k}}{dt^{k}}\tilde {f}%
(t)\approx0\right] \Longleftrightarrow \\
& \left[ 1\leq k\leq m\text{, }\frac{d^{k}}{dt^{k}}f(t)\sim\frac{d^{k}}{%
dt^{k}}(f_{0}t^{q})\right]
\end{align*}
\end{lemma}

\begin{proof}
By applying the Leibniz formula
\begin{equation*}
\frac{d^{m}}{dt^{m}}\tilde{f}(t)=\frac{d^{m}}{dt^{m}}(t^{-q}f(t))=\sum
_{i=0}^{m}\binom{m}{i}\frac{d^{m-i}}{dt^{m-i}}(t^{-q})\frac{d^{i}}{dt^{i}}%
f(t)
\end{equation*}
one proves the above equivalence by induction on $m$. We refer to Lemma 6.1
in \cite{Str} for a detailed proof.
\end{proof}

Moreover, using the operator identity
\begin{equation}
t^{k}\frac{d^{k}}{dt^{k}}=(-1)^{k}(n_{k,1}\frac{d}{du}+...+n_{k,k}\frac{d^{k}%
}{du^{k}})\text{, \ }n_{k,k}=1,n_{k,i}\in\mathbb{Z}   \label{5-24}
\end{equation}
associated with the logarithmic time change, $t\rightarrow u=-\log t$, one
can verify the following equivalence
\begin{equation}
\left[ 1\leq k\leq m\text{, }t^{k}\frac{d^{k}}{dt^{k}}g\approx0\right]
\Longleftrightarrow\left[ 1\leq k\leq m\text{, }\frac{d^{k}}{du^{k}}\breve {g%
}\approx0\right]   \label{5-25}
\end{equation}

The reason for introducing the change of variable $t\rightarrow u$ is the
following useful lemma of \emph{Tauberian type.}

\begin{lemma}
\label{Tauberian}(cf. \#363 in Wintner \cite{Win}) Let $f(u)$ be defined for
$u>0$ and assume $f(u)$ has a limit and $\frac{d^{2}}{du^{2}}f(u)$ is
bounded as $u\rightarrow\infty$. Then $\frac{d}{du}f(u)\rightarrow0$ as $%
u\rightarrow \infty$.
\end{lemma}

Finally, we turn to the proof of the asymptotic formulae (\ref{5-23}). The
"initial" data needed to start up are provided by Proposition \ref{lower}
and (\ref{6-2}), which by Lemma \ref{equiv} and (\ref{5-25}) can be restated
as
\begin{equation}
\hat{\rho}\approx\sqrt{K}\text{, \ }\hat{U}\approx\frac{\mu}{\sqrt{K}}\text{
, }\frac{d}{du}\hat{\rho}\approx\frac{d^{2}}{du^{2}}\hat{\rho}\text{\ }%
\approx0   \label{5-34}
\end{equation}
Then, by equation (\ref{5-33}), we first deduce $\hat{T}\approx0$, or
equivalently $\frac{d}{du}\mathbf{\hat{a}}_{i}\approx0$ for each $i$.
Moreover, by (\ref{5-34}) each $\mathbf{\hat{a}}_{i}$ is bounded, $\hat{U}$
is bounded and hence $\left\vert \mathbf{\hat{a}}_{i}-\mathbf{\hat{a}}%
_{j}\right\vert $ has a lower bound when $i\neq j$. It follows that all
partial derivatives of $\hat{U}$, with respect to components of $\mathbf{%
\hat
{a}}_{i}$ and of any order, are bounded (as functions of $u)$. In
particular, in equation (\ref{5-32}) $\nabla\hat{U}$ is bounded and hence
also $\frac{d^{2}}{du^{2}}\hat{\delta}$ is bounded.

Now, apply the operator $\frac{d}{du}$ repeatedly to the equation (\ref{5-32}%
) and deduce
\begin{equation*}
\frac{d^{k}}{du^{k}}\nabla\hat{U}\text{ is bounded }\Longrightarrow \frac{%
d^{k+1}}{du^{k+1}}\hat{\delta}\text{ is bounded, \ for all }k\geq1
\end{equation*}
By the above Tauberian lemma it follows that
\begin{equation*}
\frac{d^{k}}{du^{k}}\hat{\delta}\approx0\text{, \ for all }k\geq1
\end{equation*}
Similarly, apply $\frac{d}{du}$ successively to the equation (\ref{5-33})
and deduce that the highest derivative of $\hat{\rho}$ is always bounded,
hence by the Tauberian lemma
\begin{equation}
\frac{d^{k}}{du^{k}}\hat{\rho}\approx0\text{, for all }k\geq3   \label{5-35}
\end{equation}
By the equivalence (\ref{5-25}) and Lemma \ref{equiv}, the statement (\ref%
{5-35}) is equivalent to the statement of (\ref{5-23}), and this completes
the proof of Theorem \ref{higher}.

\begin{remark}
The asymptotic estimates in Theorem \ref{higher} are also valid for a
general collision (i.e. total collapse) of an n-body motion, for any $n\geq2$%
. The proof is esssentially the same as above and the previous subsection,
since we have used only the Riemannian cone structure of the moduli space $%
\bar{M}$ and, for example, the angle $\alpha$ is similarly defined for any $%
n>2$. In fact, the actual structure of the shape space is irrelevant as far
as the asymptotic behavior of the radial motion is concerned. Moreover, the
exponent $\nu$ of $t$ in the formula $I\sim Kt^{\nu}$ is independent of $n$,
but depends on the degree $-e$ of homogenity of the potential function, $%
U\sim\frac{1}{r^{e}}$, namely $\nu=4/(2+e)$ where we assume $0<e<2$, and $e=1
$ is the Newtonian case. We refer to \cite{Str}.\
\end{remark}

\section{A brief discussion of some open problems}

In this concluding section we shall formulate and explain some natural open
problems in the present geometric setting. Recall that the trajectories of
3-body motions with zero angular momentum are already uniquely determined up
to congruence by their associated moduli curves, which can be characterized
(geometrically) as geodesic curves in $(\bar{M}_{h},d\bar{s}_{h}^{2})$.
Furthermore, these geodesics together with their time evolution are
essentially determined by their shape curve on the 2-sphere. Finally, we
recall two major results concerning shape curves, namely the unique
parametrization theorem and the monotone m-latitude theorem (cf. Section 4.2
and 5.3). Therefore, from now on shape curve means geometric shape curve,
unless otherwise specified, and we also assume they are oriented.

The geometric behavior of these spherical curves raises many interesting
questions for an in depth understanding of 3-body motions. Here we propose a
few natural problems of basic importance.

\subsection{Shape curves of periodic motions with vanishing angular momentum}

The study of periodic orbits is naturally a central topic of the 3-body
problem as a whole. Clearly, the moduli curve and the shape curve of such a
motion are periodic, but the converse may not be true. Therefore, we say the
three-body motion is \emph{congruence periodic }or \emph{shape periodic} if
the moduli curve or shape curve, respectively, is periodic with respect to
time. However, it is an important consequence of the unique parametrization
property that the time parametrization of these curves is dictated by the
geometry of the shape curve, whenever the latter is non-exceptional. Then
the notion of congruence periodic is the same as shape periodic, and this
means the shape curve is \emph{periodic} in a geometric sense which we
explain as follows.

Since it is natural to allow binary collision points, periodic shape curves
can be characterized as the topologically \emph{closed} shape curves.
Namely, the curve is either the immersion of a circle, and we call it \emph{%
circular periodic}, or the immersion of a closed interval (of length
\TEXTsymbol{>} 0) and is contractible, and we call it \emph{string periodic.}
In the latter case the curve is a "string" with two end points which are
either a reversing cusp (i.e. at the Hills's boundary) or a collision point,
and in order to qualify as a periodic curve it is tacitly assumed we take
two copies of the "string" with the opposite orientation.

\begin{remark}
A shape curve consisting of a single point $p$ is an exceptional case, and
there are additional string periodic moduli curves of the fixed shape $p$.
Namely, $p$ must be a Lagrange or an Euler point, and the ray solution with
negative energy, starting at rest from $(\rho_{0},p)$ on the Hill's boundary
(cf. Section 7.4), leads directly to the triple collision point $O$. By
traversing this ray segment in both directions we obtain a string periodic
moduli curve.
\end{remark}

Next, we shall describe the distinction between periodic three-body motions
and shape periodic motions (i.e. of circular or string type). Let $%
\gamma^{\ast}$ be a closed (piecewise smooth) curve on the 2-sphere which is
the shape curve of a motion $\gamma(t),$ $t_{0}\leq t\leq t_{1}$, of
m-triangles with vanishing angular momentum, and assume $\gamma^{\ast}$ is
periodic as above. Then $\gamma(t_{0})$ and $\gamma(t_{1})$ are congruent
m-triangles and hence differ only by a rotation angle $\Delta\psi$, which we
can calculate as a line integral along the shape curve, according to the
kinematic Gauss-Bonnet theorem, see (\ref{devi}).

In particular, $\Delta\psi$ is zero if the shape curve is string periodic,
and hence the given motion $\gamma(t)$ must also be periodic. Thus the
notions of "periodic" and "shape periodic" are identical in this case. On
the other hand, a circular periodic curve $\gamma^{\ast}$ encloses a signed
area $\Delta A$ (depending on orientation and self-intersections), and the
above line integral can also be expressed as a surface integral which yields
\begin{equation}
\Delta\psi=\frac{1}{2}\Delta A   \label{angle}
\end{equation}
Therefore, the motion $\gamma(t)$ is periodic if and only if the angle (\ref%
{angle}) is a rational multiple of $2\pi$, say $\Delta\psi=(p/q)2\pi$ with $%
(p,q)=1$, and hence the number $q(t_{1}-t_{0})$ is the period of the motion.

Thus the study of periodic 3-body trajectories is completely reduced to the
study of closed shape curves on the 2-sphere, and it is a challenge to
describe or characterize the various types of these curves in terms of
simple geometric invariants. For example, due to the monotonicity theorem it
is natural to regard the number of eclipse points (counted with
multiplicity) as a measure of the complexity of the curve, hence the
simplest curves are characterized by a small number of eclipse points.

\begin{problem}
What is the minimal number of eclipse points on a (string or circular)
periodic shape curve ? What even numbers can be realized? What are those
periodic curves with a small number of eclipse points, say up to 10?
\end{problem}

The homotopy classes of closed curves inside $P=S^{2}-\left\{ \mathbf{b}_{1}%
\mathbf{,b}_{2}\mathbf{,b}_{3}\right\} $ are elements of the fundamental
group $\pi_{1}(P)$, namely the free group of two generators.

\begin{problem}
What are those homotopy classes of closed curves in $P$ which can be
represented by circular periodic shape curves ?
\end{problem}

\begin{definition}
We propose to define the \emph{chaoticity} of $\gamma^{\ast}$ to be the
following value
\begin{equation*}
ch(\gamma^{\ast})=\frac{Area(D(\gamma^{\ast}))}{Area(S^{2})}
\end{equation*}
where $D(\gamma^{\ast})\subset S^{2}$ is the closure of the set $%
\gamma^{\ast }$, and we say $\gamma^{\ast}$ is \emph{chaotic} or \emph{%
non-chaotic} if $ch(\gamma^{\ast})>0$ or $ch(\gamma^{\ast})=0$, respectively.
\end{definition}

\begin{problem}
What are the possible values of chaoticity for shape curves representing
motions with $\mathbf{\Omega}=0$ ?
\end{problem}

\begin{problem}
What are the non-chaotic shape curves other than the periodic ones ?
\end{problem}

\subsection{Triple collisions}

The works of Sundman and Siegel show that triple collisions is the only type
of \emph{essential singularity} of 3-body motions, while the binary
collisions can be regularized analytically \cite{Sund2}, \cite{Lev}. Recall
that the singularities of the Newtonian potential function $U$ in the moduli
space $\bar{M}\simeq\mathbb{R}^{3}$ consists of the triple of rays $\left\{
\overrightarrow{O\mathbf{b}}_{i},i=1,2,3\right\} $, where the base point (or
origin) $O$ represents the triple collision and the rays represent the three
types of binary collisions. Moreover, the moduli curves of triple collision
motions with total energy $h$ are exactly those geodesic curves in $(\bar
{M%
}_{h},d\bar{s}_{h}^{2})$ with the point $O$ as a limit. The following are
some pertinent problems on the geometry of such geodesics.

\begin{problem}
The existence (resp. uniqueness) problem on the shortest path in $(\bar{M}%
_{h},d\bar{s}_{h}^{2})$ linking a given point $p$ in $\bar{M}_{h}$ to the
base point $O$.
\end{problem}

By the scaling symmetry we may assume $h=0,\pm1$, where the case of $h=-1$
is most difficult and also most interesting. The existence of such a
shortest geodesic curve between $O$ and a given point $p$ in $\bar{M}_{h}$
can be proved by Hilbert's direct method when $h=0$ or $1$, whereas for the
case of $h=-1$ the existence will depend on the position of $p$ in the
Hill's region $\bar{M}_{h}$. Note that Hilbert's direct method also applies
for those $p$ with
\begin{equation*}
d(p,O)<d(p,\partial\bar{M}_{h})+d(\partial\bar{M}_{h},O)
\end{equation*}

The uniqueness problem is, however, much more interesting than the existence
problem, but it is also much more difficult and subtle. For a geodesic $\bar{%
\gamma}$ starting from $O$, the question is how far out $\bar{\gamma}$ is
the unique shortest geodesic from $O$. We remark that for points lying in
the eclipse plane, the shortest geodesic is not in the eclipse plane, and
hence the limiting shape at $O$ of the shortest geodesic must be a Lagrange
point, say $\mathbf{p}_{0}$. By the monotonicity of its shape curve $\bar{%
\gamma}$ will eventually reach the eclipse plane, but after the first
eclipse $\bar{\gamma}$ ceases to be of shortest length. Hence, the best we
can hope for is uniqueness up to the first eclipse point.

We propose to investigate first the case of $h=0$, due to the scaling
invariance of this energy level. Then the general uniqueness of a shortest
geodesic between any point $p$ and $O$ reduces to the uniqueness for eclipse
points $p$ lying at the distance $\rho=1$ from $O$, namely for points on the
eclipse circle $E^{\ast}$. Thus the problem is reduced from the moduli space
$\bar{M}$ to the shape space $M^{\ast}$, namely we ask about the uniqueness
of such shape curves between points on $E^{\ast}$ and $\mathbf{p}_{0}$.

\begin{problem}
For the case of energy level $h=0$ and for a given mass distribution, let
\emph{S} be the set of triple collision moduli curves emanating from $O$,
whose shape curve starts out from the Lagrange point $\mathbf{p}_{0}$ on the
upper hemisphere of $M^{\ast}=S^{2}$. Let \emph{S}$^{\ast}$ be the initial
arcs of the shape curves from $\mathbf{p}_{0}$ (but $\mathbf{p}_{0}$ not
included) to their first point on the equator circle $E^{\ast}$. Is the
family of curves \emph{S}$^{\ast}$ a foliation of the punctured upper
hemisphere $S_{+}^{2}-\left\{ \mathbf{p}_{0}\right\} $ ?
\end{problem}

In the case that there exists a unique shortest geodesic in $(\bar{M}_{h},d%
\bar{s}_{h}^{2})$ linking a given m-triangle $\delta$ to the collapsed
m-triangle $O$, it is certainly interesting to actually estimate its \emph{%
length, initial direction} and total \emph{rotation angle }of the triangle
in terms of its geometric invariants. See also the last part of Section 6.2.

\subsection{Binary collisions and nearby trajectories}

The base point $O$ is, of course, the only singularity for $(\bar{M},d\bar
{%
s}^{2})$. But the Riemannian manifold $(\bar{M}_{h},d\bar{s}_{h}^{2})$ has
another kind of \emph{singularity} along the triple of rays $\left\{
\overrightarrow{O\mathbf{b}}_{i},i=1,2,3\right\} $ minus the initial point $%
O $, say, of \emph{binary collision }type. One expects that understanding of
the geometry of geodesic curves in the vicinity of this type of singularity
will be an important topic in the study of the global geometry of geodesics
on $(\bar{M}_{h},d\bar{s}_{h}^{2})$.

\begin{problem}
What kind of local analysis will enable us to provide an effective control
on the local geometry of geodesic curves in the vicinity of a singular ray
of a given binary collision type ?
\end{problem}

Let us make some further remarks. In the vicinity of the ray $%
\overrightarrow {O\mathbf{b}}_{i}$ the gradient vector field $\nabla U$ is
closely approximated by the field $\nabla U_{i}$, where
\begin{equation*}
U=\sum U_{i},\text{ \ }U_{i}=\frac{m_{j}m_{k}}{r_{jk}}=\frac{(m_{j}m_{k})^{%
\frac{3}{2}}}{\sqrt{1-m_{i}}}\frac{1}{d_{i}}
\end{equation*}
and $d_{i}=d_{i}(x)=\rho\sin\sigma_{i}$ is the distance in $(\bar{M},d\bar
{%
s}^{2})$ between $x$ and the ray $\overrightarrow{O\mathbf{b}}_{i}$, cf. (%
\ref{2-11}). Thus, there is a suitable rotationally symmetric metric which
provides a good approximation of $d\bar{s}_{h}^{2}$ when we are close to
such a ray singularity. Application of Noether's theorem to this simpler
metric yields a \emph{first integral} of its geodesic equation which is
almost constant along a geodesic segment of $(\bar{M}_{h},d\bar{s}_{h}^{2})$
near $\overrightarrow{O\mathbf{b}}_{i}$. This will serve as a useful \emph{%
auxiliary} function whose analysis will provide an effective control on the
above local geometry.

\subsection{Trajectories starting at the boundary of the Hill's region}

In the case of negative energy, say $h=-1$, the variety $\bar{M}_{h}$ is the
\emph{Hill's region}, namely the proper subset of the moduli space from
which the moduli curves of the three-body motions cannot leave. The region
is enclosed by its boundary, namely Hill's surface $\partial\bar{M}_{h}$
which is the smooth surface defined by $\rho=U^{\ast}(\varphi,\theta)$, with
the (kinematic) gradient field $\nabla U$ as a normal field.

Here we shall focus on those geodesics of the metric $d\bar{s}_{h}^{2}$
starting at the surface $\partial\bar{M}_{h}$, where the metric becomes
identically zero. Hence, a curve lying on $\partial\bar{M}_{h}$ has zero
length, and a minimizing curve containing a segment on $\partial\bar{M}_{h}$
is only \emph{virtual} and cannot, of course, represent an actual trajectory
of a 3-body motion. Therefore, in the study of variational problems of this
kind one often needs a certain estimate or geometrical control of those
geodesic curves starting at $\partial\bar{M}_{h}$, that is, the moduli
curves of those 3-body motions with no kinetic energy at $t=t_{0}$. They
constitute a family $\left\{ \bar{\gamma}_{x_{0}}\right\} $ of geodesics
parametrized by their initial points $x_{0}\in\partial\bar{M}_{h}$.

Following Jacobi, it is natural to study the variational vector fields along
$\bar{\gamma}_{x_{0}}$ with respect to variations within the above family of
geodesics. These vector fields are solutions of the Jacobi equation along $%
\bar{\gamma}_{x_{0}}$ with their initial vectors belonging to $%
T_{x_{0}}(\partial\bar{M}_{h})$.

\begin{problem}
Let $x_{0}$ be a generic point of $\partial\bar{M}_{h},h=-1$, and let $\bar{%
\gamma}_{x_{0}}$ be the geodesic curve with initial point $x_{0}$. How do we
obtain an effective (that is, simple and useful) lower bound estimate of the
distance between $x_{0}$ and the first zero point of Jacobi vector fields of
the above type, in terms of the geometric invariants at $x_{0}$?
\end{problem}

\subsection{On the problem of fundamental segments}

For a fixed energy level $h=0,\pm1$, consider the family $\Sigma(h)$ of all
oriented geometric shape curves, with the exceptional ones removed, of
three-body motions with zero angular momentum. According to the monotone
m-latitude theorem the curve $\gamma^{\ast}$ can be viewed as a union of its
segments $C_{i}$ $=(p_{i},q_{i})$, between two consecutive points $p_{i}$
and $q_{i}$ of extremal m-latitude. We shall refer to them as the \emph{%
fundamental segments}. Thus the end points $p_{i},q_{i}$ lie on opposite
hemispheres, unless one of them is a binary collision point (and hence lies
on the equator circle), and moreover, the m-latitude is strictly monotonic
along the segment. Clearly, a global shape curve can be regarded as being
pieced together by such fundamental segments, and a periodic shape curve has
only a finite number of them.

Conversely, we may try to construct curves by connecting $C_{i}$ to $C_{i+1}$
in a "smooth" way. Here $C_{i}$ and $C_{i+1}$ belong to $\Sigma(h)$, so
"smooth" means their union also belongs to $\Sigma(h)$. For simplicity,
assume we are using only \emph{regular} fundamental segments $C=(p,q)$, that
is, $p$ and $q$ are regular points. Observe that $C$ is tangential to the
m-latitude circle at $p$, so its direction will be completely specified by
an index $\varepsilon=0,1$ representing "eastward" or "westward"
respectively. Thus we can associate to the starting point $p$ $%
=(\varphi,\theta)$ the following 5-tuple of numbers%
\begin{equation}
\left[ p\right] =(\varphi,\theta,\mathfrak{S}_{0},\mathfrak{S}%
_{1},\varepsilon)   \label{5-tuple}
\end{equation}
which determines $C$ completely and therefore also the 5-tuple $\left[ q%
\right] $ associated to its end point. In (\ref{5-tuple}) $\mathfrak{S}_{0},%
\mathfrak{S}_{1}$ are the Siegel numbers of $C$ at $p$, as explained in
Section 4.1 and 4.2.

Roughly speaking, the relationship between the initial data and terminal
data for a fundamental segment with regular end points provides a type of
correspondence%
\begin{equation*}
\text{ \ }\left[ p\right] =(\varphi,\theta,\mathfrak{S}_{0},\mathfrak{S}%
_{1},\varepsilon)\rightarrow\left[ q\right] =(\pi-\varphi^{\prime},\theta^{%
\prime},\mathfrak{S}_{0}^{\prime},\mathfrak{S}_{1}^{\prime
},\varepsilon^{\prime})\text{\ }
\end{equation*}
on a dense open set of $S^{2}\times\mathbb{R}^{+}\times\mathbb{R}\times$ $%
\left\{ 0,1\right\} $. Moreover, since $p$ and $q$ are points on opposite
hemispheres, let us compose the above correspondence with the reflectional
symmetry with respect to the equator circle, namely we replace $\left[ q%
\right] $ by $\left[ p^{\prime}\right] =(\varphi^{\prime},\theta^{\prime },%
\mathfrak{S}_{0}^{\prime},\mathfrak{S}_{1}^{\prime},\varepsilon^{\prime})$.
Finally, we assume $p$ (and hence also $p^{\prime}$) lies on the upper
hemisphere, thus arriving at the \emph{fundamental correspondence}%
\begin{equation}
\Theta:\mathfrak{T}_{0}\cup\mathfrak{T}_{1}\rightarrow\mathfrak{T}_{0}\cup%
\mathfrak{T}_{1}\text{, \ \ }\left[ p\right] \rightarrow\left[ p^{\prime}%
\right]   \label{dyn}
\end{equation}
\emph{\ }where the $\mathfrak{T}_{i}$ are identical copies of the
4-dimensional space $\mathfrak{T}=S_{+}^{2}\times\mathbb{R}_{+}\times
\mathbb{R}$. The correspondence is defined on a dense, open set, where it is
also invertible. In fact, with some more labour it would be possible to
extend the fundamental correspondence to include irregular points (i.e.
cusps and collisions) as well.

\begin{remark}
A periodic shape curve is the assemblage of a finite number of fundamental
segments whose initial data constitute a periodic orbit of the above
correspondence (\ref{dyn}). Namely, if $\left[ p\right] $ has even order $2k$%
, then the orbit of $\left[ p\right] $ defines $2k$ fundamental segments
which join together to a periodic curve. On the other hand, if the order is $%
2k+1$, then the end of the curve lies in the southern hemisphere, so by
running through the orbit twice the order will be $4k+2$, and the associated
curve will be periodic.
\end{remark}

Thus, the correspondence (\ref{dyn}) provides a natural way to a systematic
study of the geometry of global shape curves.

\end{document}